\documentclass [letter, 12pt]{article}
\pdfoutput=1
\usepackage{a4wide, amsmath, amsfonts, amssymb, geometry, shuffle}

\usepackage{bm}
\usepackage{fancyhdr, dsfont}
\usepackage[latin1]{inputenc}
\usepackage{graphicx} 
\usepackage{float}
\usepackage{array}
\usepackage{comment} 
\usepackage[sc,small]{caption} 
\usepackage{hyperref}
\hypersetup{
    colorlinks=true,
    linkcolor=black,
    citecolor=black,
    filecolor=black,
    urlcolor=black,
}

\usepackage{color}
\definecolor{dgreen}{rgb}{0,0.70,0.30}
\definecolor{gold}{rgb}{0.85,.66,0}
\definecolor{purple}{rgb}{1.0,0.3,0.6}

\usepackage{tikz}
\usetikzlibrary{calc} \usetikzlibrary{patterns} \usetikzlibrary{decorations.pathreplacing} \usetikzlibrary{decorations.markings} \usetikzlibrary{decorations.pathmorphing} \usetikzlibrary{positioning}

\restylefloat{figure}
\usepackage{epstopdf}

\def\beq{\begin{equation}}
\def\eeq{\end{equation}}

\def\Im{{\rm Im\,}}

\newcommand{\co}{\ , \ \ \ \ \ \ }
\newcommand{\dd}{\mathrm{d}}
\newcommand{\te}{\textrm}
\newcommand{\ap}{\alpha'}

\newcommand{\efrak}{\mathfrak e}
\newcommand{\ffrak}{\mathfrak f}

\renewcommand{\theta}{\vartheta}
               
\newcommand{\be}{\begin{equation}}
\newcommand{\ee}{\end{equation}}
\newcommand{\bea}{\begin{eqnarray}}
\newcommand{\eea}{\end{eqnarray}}

\newcommand{\ra}{\rangle}
\newcommand{\la}{\langle}
\newcommand{\ek}[2]{(e_{#1}\cdot k_{#2})}
\newcommand{\eeij}[2]{(e_{#1}\cdot e_{#2})}

\title{\vspace{1cm}\textbf{String-motivated one-loop amplitudes in \\ gauge theories with half-maximal supersymmetry} \\ \ \\} \author{\hspace{-0cm}Marcus Berg$^{\dag}$, Igor Buchberger$^{\dag}$, Oliver Schlotterer$^{\star}$\\[5mm]
\hspace{-0cm} $^{\dag}$
{\it Department of Physics, Karlstad University,
 651 88 Karlstad, Sweden} \\
\hspace{-0cm} $^{\star}$ 
 {\it Max-Planck-Institut f\"ur Gravitationsphysik, Albert-Einstein-Institut},\\[-1mm] {\it 14476 Potsdam, Germany}
}

\begin{document}

\maketitle{}
\vspace{1cm}
\begin{abstract}
We compute one-loop amplitudes in six-dimensional Yang--Mills theory with half-maximal supersymmetry from first principles: imposing gauge invariance and locality on an ansatz made from string-theory inspired kinematic building blocks yields unique expressions for the 3- and 4-point amplitudes. We check that the results are reproduced in the field-theory limit $\alpha' \rightarrow 0$ of string amplitudes in K3 orbifolds, using simplifications made in a companion string-theory paper \cite{Berg:2016wux}.
\end{abstract}

\newpage

\setcounter{tocdepth}{2}
\tableofcontents

\numberwithin{equation}{section}

\section{Introduction}

The last few years have seen significant progress on massless scattering amplitudes of string and gauge theories with less than maximal supersymmetry. 
 As an example, open-string 4-point 1-loop amplitudes with minimal supersymmetry were discussed in \cite{Bianchi:2006nf,Bianchi:2015vsa}, and the closed-string counterparts for half-maximal supergravity amplitudes can be found in \cite{Tourkine:2012vx,Ochirov:2013xba}. In a companion paper \cite{Berg:2016wux} we simplified and generalized these results, using an infrared regularization procedure due to Minahan \cite{Minahan:1987ha} to maintain manifest gauge invariance. 
 
In this paper, we will present novel representations for 1-loop 3- and 4-point amplitudes in 6-dimensional gauge theories with 8 supercharges, inspired by their string-theory ancestors from the companion paper \cite{Berg:2016wux}. In contrast to the 4-dimensional string-theory expressions in other work \cite{Tourkine:2012vx,Ochirov:2013xba, Bianchi:2015vsa}, we maintain 6-dimensional Lorentz-covariance (the maximum allowed by half-maximal supersymmetry) as we did in  \cite{Berg:2016wux}. 
We use 4-dimensional spinor helicity variables only for specific checks.  

The general philosophy of our calculational strategy will be: 
\begin{itemize}
\item String theory motivates a generic ``alphabet'' of kinematic building blocks for field-theory amplitudes. As we will see in examples, imposing locality and gauge invariance on a suitable ansatz drawn from this alphabet fixes the amplitudes we consider. Building blocks with up to two loop momenta and the systematics of their gauge variations will be discussed at general multiplicity.
\item
In general,  there is a tension between manifest locality and manifest gauge invariance. We begin from a  local representation, with crucial input from the cancellation of gauge variations of different diagrams. Then, by manipulating  integrands, we rearrange kinematic factors into gauge-invariants of the same form as in the string amplitudes from the companion paper \cite{Berg:2016wux}.
\item 
In the pure-spinor description of 10-dimensional super Yang--Mills (SYM) \cite{Witten:1985nt, Howe:1991mf, Howe:1991bx}, gauge invariance and supersymmetry are unified to BRST invariance \cite{Berkovits:2000fe}. Along with locality, this has been used to determine multiparticle amplitudes in pure-spinor superspace up to and including two loops \cite{Mafra:2010jq, Mafra:2014gja, Mafra:2015mja}. To extend this approach below maximal supersymmetry, we consider half-maximal SYM in the maximal spacetime dimension $D=6$ where 8 supercharges can be realized.
\item
A useful check is provided by comparison with \cite{Mafra:2014gja}: the $n$-point 1-loop amplitudes in our half-maximal setup
follow the same structure as the corresponding $(n{+}2)$-point amplitudes with maximal supersymmetry. 
 \end{itemize}
This last feature is inherited from the structure of the string integrands, where comparison of the pure-spinor superspace results in \cite{maxsusy} with the orbifold amplitudes in \cite{Berg:2016wux} reveals the same $+2$ offset in multiplicity. 
The counting is uniform with the relevant
string compactifications: at complex dimension 2, we find the first nontrivial Calabi-Yau manifold (K3),
that breaks half the supersymmetry. At complex dimension 3, supersymmetry is broken to a quarter (${\cal N}=1$ in 4-dimensional counting), and the multiplicity offset in 1-loop amplitudes is +2 in the parity-even and +3 in the parity-odd sector, respectively. This means the parity-even part of our results, written in dimension-agnostic variables, applies universally to gauge-theory amplitudes with ${\cal N}=2$ and ${\cal N}=1$ supersymmetry. The parity-odd contributions 
to 4-dimensional ${\cal N}=1$ amplitudes, on the other hand, are quite different from the present 6-dimensional results with half-maximal supersymmetry. The methods of this work could be applied there too, but we postpone this to the future. 

This paper is mostly about gauge-theory amplitudes, but it is of great interest
to pursue the analogous calculations for supergravity. In particular, it is interesting to test to what extent the
Bern--Carrasco--Johansson (BCJ) duality \cite{Bern:2008qj} holds in our calculations, and whether the requisite supergravity
(1-loop) amplitudes with 16 supercharges can be obtained from the double-copy construction \cite{Bern:2010ue}. We 
report our results on this in section \ref{sect4}, where we find that the 3-point function in half-maximal gauge theory
satisfies the BCJ duality, but --- in contrast to the 4-dimensional expressions in \cite{Carrasco:2012ca, Johansson:2014zca} --- our representation of the 4-point function does not naturally lend itself to the duality.  

For comparison with the literature, we consider compactification on $T^2$ from 6 to 4 dimensions,
and specialize to a 4-dimensional helicity basis, finding a perfect match with known results. The match involves a single  free numerical factor that depends on the field content of the specific model. Many consistent string models
contain exotic matter, whereas much of the work on field-theory amplitudes does not, so a completely general match goes beyond the scope of this work. Instead, in appendix \ref{appE} we describe simplified string models that are by themselves inconsistent (in particular when restoring
couplings to a gravitational sector), but can usefully be compared to existing work on amplitudes.

\section{Kinematic building blocks for 1-loop}
\label{sect2}

In this section, we introduce a system of kinematic building blocks for 1-loop amplitudes of half-maximal SYM. The overall guiding principle is invariance under linearized gauge transformations, that can be compactly implemented through the Grassmann (and thereby nilpotent) operator
\beq
\delta \equiv \sum_{i=1}^n \omega_i k_i^m\frac{\partial}{\partial e^m_i}  \ ,
\label{BG11}
\eeq
with vector index $m=0,1,\ldots,D{-}1$ and $n$ being the number of external legs.
The fermionic bookkeeping variables $\omega_{i}$ keep track of unphysical longitudinal polarizations $e_i \rightarrow k_i$ in the $i^{\te{th}}$ external leg:
\beq
\delta e_i^m = k_i^m \omega_i \ .
\label{BG11a}
\eeq
We will follow the ideas of Berends and Giele \cite{Berends:1987me} to construct multiparticle generalizations of the polarization vector $e^m$ and its gauge-invariant linearized field strength,
\beq
f_i^{mn} \equiv k_{i}^m e_{i}^n - k_{i}^n e_{i}^m \co \delta f_i^{mn} =0 \ .
\label{BG11b}
\eeq
The multiparticle variables of this section are designed to represent tree-level subdiagrams with an off-shell leg and therefore transform covariantly under (\ref{BG11}). They provide a suitable starting point to obtain both local and gauge-invariant expressions for the 1-loop amplitudes under investigation.

\subsection{Local multiparticle polarizations}
\label{sect21}

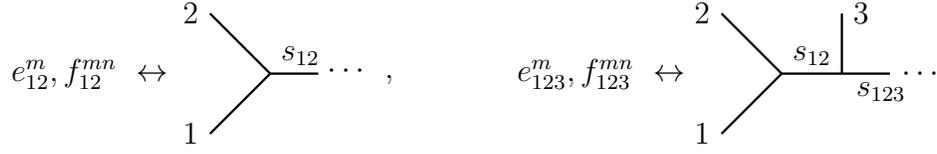
\begin{figure}[h]
\begin{center}
\begin{tikzpicture} [scale=0.8, line width=0.30mm]
\begin{scope}[xshift=-8.5cm] 
\draw (-3,0) node{$e^{m}_{12},f^{mn}_{12} \  \leftrightarrow  $};
\draw (0,0) -- (-1,1) node[left]{$2$};
\draw (0,0) -- (-1,-1) node[left]{$1$};
\draw (0,0) -- (0.8,0);
\draw (0.5,0.3) node{$s_{12}$};
\draw (1.5, 0) node{${\cdots} \  \ , $};
\end{scope}
\draw (-3,0) node{$e^{m}_{123},f^{mn}_{123} \  \leftrightarrow  $};
\draw (0,0) -- (-1,1) node[left]{$2$};
\draw (0,0) -- (-1,-1) node[left]{$1$};
\draw (0,0) -- (1.8,0);
\draw (0.5,0.3) node{$s_{12}$};
\draw (1,0) -- (1,1) node[right]{$3$};
\draw (1.65,-0.3) node{$s_{123}$};
\draw (2.3, 0) node{${\cdots}  $};
%
\end{tikzpicture}
\caption{Cubic-vertex subdiagrams with an off-shell leg $\cdots$ can be represented by local multiparticle polarizations $e^{m}_{12},f^{mn}_{12}$ and $e^{m}_{123},f^{mn}_{123}$, respectively.}
\label{f:loc}
\end{center}
\end{figure}

In order to attach the tree-level subdiagrams in figure \ref{f:loc} to a graph of arbitrary loop order, we define local 2- and 3-particle generalizations of polarizations and field strengths. In the conventions for multiparticle momenta and Mandelstam invariants where
\beq
s_{12} \equiv (k_1\cdot k_2) \co
s_{12\ldots p} \equiv \frac{1}{2} (k_{12\ldots p})^2 \co
k_{12\ldots p}^m \equiv k_1^m + k_2^m +\ldots+k_p^m \ ,
\label{mand}
\eeq
the 2-particle polarization and field strength
\begin{align}
e^{m}_{12}  &\equiv  e_2^m (k_2\cdot e_1) -  e_1^m (k_1\cdot e_2) +\frac{1}{2}(k_1^m - k_2^m) (e_1 \cdot e_2)
\label{BG31} \\
f^{mn}_{12}&\equiv k_{12}^m e_{12}^n - k_{12}^n e_{12}^m
-s_{12}\big( e_1^m e_2^n - e_1^n e_2^m \big)  
\label{BG32}
\end{align}
can be used to relate the factorization limit of $n$-point amplitudes on a 2-particle channel $\sim (k_i{+}k_j)^{-2}$ 
to an $(n{-}1)$-point amplitude with one gluon polarization replaced by $e^m_{ij}$. Their 3-particle counterparts read
\begin{align}
e^{m}_{123}  &\equiv  e_3^m (k_3\cdot e_{12}) -  e_{12}^m (k_{12}\cdot e_3) +\frac{k_{12}^m - k_{3}^m}{2} (e_{12} \cdot e_3) + \frac{s_{12}}{2} \big( e_2^m (e_1\cdot e_3) -  e_1^m (e_2\cdot e_3) \big)
\label{BG33} \\
f^{mn}_{123}&\equiv k_{123}^m e_{123}^n - k_{123}^n e_{123}^m
-(s_{13}\!+\!s_{23})\big( e_{12}^m e_3^n - e_{12}^n e_3^m \big)  -
s_{12} \big( e_1^m e_{23}^n- e_1^n e_{23}^m
-(1\!\leftrightarrow\! 2) \big) \ ,
\label{BG34}
\end{align}
and the combinations of $e^{m}_{ijl} $ and $f^{mn}_{ijl}$ that we will encounter in the next section capture the polarization dependence of 3-particle factorization channels $\sim (k_i{+}k_j{+}k_l)^{-2}$.

\noindent
These definitions can be motivated by a resummation of Feynman diagrams \cite{Berends:1987me} or through OPEs of vertex operators in string theory -- see \cite{Mafra:2014oia} for a supersymmetric derivation in the pure-spinor formalism and appendix \ref{appA} for the bosonic RNS counterpart. The propagators of the diagrams in figure \ref{f:loc} are cancelled by 
numerators containing the Mandelstam invariants (\ref{mand}). They appear in both the definition of $f^{mn}_{12},f^{mn}_{123}$ and in the action of the gauge variation (\ref{BG11}),
\begin{align}
\delta e^{m}_{12}  &= k_{12}^m \omega_{12} + s_{12}(\omega_1 e_2^m - \omega_2 e_1^m)
\notag \\
\delta f^{mn}_{12}&= s_{12}(\omega_1 f_2^{mn} - \omega_2 f_1^{mn})  \ ,
\label{gBG32} \\
\delta e^{m}_{123}  &= k_{123}^m \omega_{123} + (s_{13}+s_{23}) (\omega_{12} e_3^m - \omega_3 e_{12}^m) + s_{12}(\omega_{1} e_{23}^m - \omega_{23} e_{1}^m - \omega_{2} e_{13}^m + \omega_{13} e_{2}^m)
\notag \\
\delta f^{mn}_{123}&= (s_{13}+s_{23}) (\omega_{12} f_3^{mn} - \omega_3 f_{12}^{mn}) + s_{12}(\omega_{1} f_{23}^{mn} - \omega_{23} f_{1}^{mn} - \omega_{2} f_{13}^{mn} + \omega_{13} f_{2}^{mn}) \ ,
\notag
\end{align}
which will play a central role in this work. Given that the right-hand side is entirely furnished by multiparticle polarizations and multiparticle gauge scalars
\begin{align}
\omega_{12} &\equiv \frac{1}{2} \big[ \omega_2 (k_2 \cdot e_1) - \omega_1 (k_1 \cdot e_2) \big]
\co
\omega_{123} \equiv \frac{1}{2} \big[ \omega_3 (k_3 \cdot e_{12}) - \omega_{12} (k_{12} \cdot e_3) \big] \ ,
\label{om3}
\end{align}
the gauge algebra of the $e_{12\ldots p}^m, \ f_{12\ldots p}^{mn}$ and $\omega_{12\ldots p}$ is said to be {\em covariant}. Nilpotency of the gauge variation (\ref{BG11}) can be checked from the covariant transformation of the fermionic gauge scalars $\omega_{12\ldots p}$,
\beq
\delta \omega_{12}  = s_{12}\omega_{1} \omega_{2} \co
\delta \omega_{123}  = (s_{13}+s_{23}) \omega_{12}\omega_3 + s_{12}(\omega_{1} \omega_{23} - \omega_2 \omega_{13} ) \ .
\label{gom}
\eeq
Note that the gauge algebra (\ref{gom}) of multiparticle gauge scalars resembles the BRST variation\footnote{BRST invariance in pure-spinor superspace powerfully combines the supersymmetry of 10-dimensional SYM with gauge invariance of the bosonic superfield components \cite{Berkovits:2000fe}.} of multiparticle vertex operators $V_{12\ldots p}$ in the pure-spinor superstring \cite{Mafra:2014oia}.

\subsection{Berends--Giele currents}
\label{sect22}

\begin{figure}[h]
\begin{center}
\begin{tikzpicture} [scale=0.8, line width=0.30mm]
\draw (-8,0) node{$\efrak^{m}_{123},\ffrak^{mn}_{123} \ \, \ \leftrightarrow  $};
\draw (-4.5,0.5) -- (-5,1) node[left]{$3$};
\draw (-4.7,0) -- (-5.4,0) node[left]{$2$};
\draw (-4.5,-0.5) -- (-5,-1) node[left]{$1$};
\draw[fill=gray] (-4,0) circle(0.7cm);
\draw (-3.3,0)--(-2.5,0);
\draw (-2, 0) node{${\ldots} $};
\draw (-1.3,0) node{$=$};
\scope[xshift=1cm]
\draw (0,0) -- (-1,1) node[left]{$2$};
\draw (0,0) -- (-1,-1) node[left]{$1$};
\draw (0,0) -- (1.8,0);
\draw (0.5,0.3) node{$s_{12}$};
\draw (1,0) -- (1,1) node[right]{$3$};
\draw (1.65,-0.3) node{$s_{123}$};
\draw (2.7, 0) node{${\cdots} \  \ + $};
\scope[xshift=-0.4cm]
\draw (5.5,0) -- (4.5,1) node[left]{$3$};
\draw (5.5,0) -- (4.5,-1) node[left]{$2$};
\draw (5.5,0) -- (7.3,0);
\draw (6,-0.3) node{$s_{23}$};
\draw (6.5,0) -- (6.5,-1) node[right]{$1$};
\draw (7,0.3) node{$s_{123}$};
\draw (7.8, 0) node{${\ldots} $};
\endscope
\draw (5.5,1) node{};
\endscope
\end{tikzpicture}
\caption{Berends--Giele currents $\efrak^m_{12\ldots p}$ and $\ffrak^{mn}_{12\ldots p}$ combine multiparticle 
polarizations with appropriate propagators so as to reproduce the cubic-vertex subdiagrams in a color-ordered 
$(p{+}1)$-point tree amplitude with an off-shell leg $\cdots$.}
\label{f:bg}
\end{center}
\end{figure}
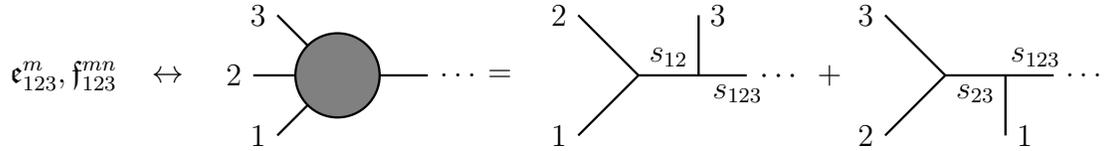

In order to simplify the recursive definition and the gauge algebra of the above multiparticle polarizations $e_{12\ldots p}^m$ and $f_{12\ldots p}^{mn}$, it is convenient to change basis to Berends--Giele currents\footnote{We use the $\mathfrak{Fraktur}$ typeface to distinguish the non-local Berends--Giele currents $\efrak^{m}_{12\ldots p}, \ \ffrak^{mn}_{12\ldots p} \sim s^{1-p}$ from the local multiparticle polarizations $e^m_{12}$, $f^{mn}_{12}, \ldots$ \ in eqs.\ (\ref{BG31}) to (\ref{BG34}).} \cite{Berends:1987me},
\begin{align}
\efrak_{12}^m &\equiv \frac{e_{12}^{m}}{s_{12}} \co \ffrak_{12}^{mn} \equiv \frac{f_{12}^{mn}}{s_{12}}  \notag \\
\efrak_{123}^m &\equiv \frac{e_{123}^{m}}{s_{12} s_{123}}+\frac{e_{321}^{m}}{s_{23} s_{123}}
\label{BG81}  \\
\ffrak_{123}^{mn} &\equiv \frac{f_{123}^{mn}}{s_{12} s_{123}}+\frac{f_{321}^{mn}}{s_{23} s_{123}} \ . \notag
\end{align}
As one can see from the 3-particle instances $\efrak_{123}^m$ and $\ffrak_{123}^{mn}$, the cubic graphs in figure \ref{f:loc} are combined according to a color-ordered 4-point amplitude with one off-shell leg, see figure \ref{f:bg}.

\noindent
The 2- and 3-particle instances (\ref{BG81}) can be reproduced from the compact recursion \cite{Berends:1987me, Mafra:2015vca}
\begin{align}
\efrak^{m}_{P}  &=  {1 \over 2 s_P} \sum_{XY=P}  \bigl[ \efrak_{Y}^m (k^Y\cdot  \efrak^{X})
+ \efrak^{Y}_n  \ffrak_X^{mn}
- (X \leftrightarrow Y)\bigr] 
\label{BG01} \\
\ffrak^{mn}_P &= k_P^m \efrak_P^n - k_P^n \efrak_P^m
- \sum_{XY=P}\!\!\big( \efrak_X^m \efrak_Y^n - \efrak_X^n \efrak_Y^m \big)  \ ,
\label{BG02}
\end{align}
with multiparticle labels $P=12\ldots p$ and initial conditions $\efrak^{m}_{1}\equiv e^m_1$, $\ffrak^{mn}_{1}\equiv f^{mn}_1$.  The summation $\sum_{XY=P}$ means to deconcatenate (i.e.\ split up) the word $P = 12\ldots p$ referring to external particles $1,2,\ldots p$ into non-empty words $X = 12\ldots j$ and $Y=j{+}1\ldots p$ with $j=1,2,\ldots,p{-}1$. As an example, for $p=4$ (four letters), the deconcatenation parts of (\ref{BG02}) include $\sum_{XY=1234} \efrak_X^m \efrak_Y^n = \efrak_1^m \efrak_{234}^n+\efrak_{12}^m \efrak_{34}^n+\efrak_{123}^m \efrak_4^n$.

The same kind of deconcatenation pattern arises when translating the gauge variations (\ref{gBG32}) and (\ref{gom}) to the Berends--Giele framework,
\begin{align}
\delta \efrak^m_P &= k^m_P \Omega_P + \sum_{XY=P} \big[ \Omega_X \efrak^m_Y  - (X \leftrightarrow Y) \big] \label{BG09} \\
\delta \ffrak^{mn}_P&= \sum_{XY=P} \big[ \Omega_X \ffrak^{mn}_Y  - (X \leftrightarrow Y) \big]  \ ,
\label{BG10}
\end{align}
for instance
\beq
\delta  \efrak^m_{12} =k_{12}^m \Omega_{12} + (\Omega_1 \efrak^m_{2} - \Omega_2 \efrak^m_{1}) \co
\delta \ffrak^{mn}_{123} = \Omega_1 \ffrak^{mn}_{23} + \Omega_{12} \ffrak^{mn}_{3} - \Omega_{23} \ffrak^{mn}_{1} - \Omega_3 \ffrak^{mn}_{12} \ .
\label{ex10}
\eeq
We have introduced Berends--Giele currents $\Omega_P$ associated with the multiparticle gauge scalars (\ref{om3}),
\beq
\Omega_1 \equiv \omega_1 \co \Omega_{12} \equiv \frac{\omega_{12}}{s_{12}} \co \Omega_{123} \equiv \frac{\omega_{123}}{s_{12}s_{123}}+\frac{\omega_{321}}{s_{23}s_{123}} \ ,
\label{BG08a}
\eeq
which are reproduced from the recursion
\beq
\Omega_P = \frac{1}{2 s_P} \sum_{XY=P} \big[ \Omega_Y (k_Y \cdot \efrak_X) - (X \leftrightarrow Y) \big] \ ,
\label{BG08}
\eeq
and translate the gauge variations (\ref{gom}) into the simple form
\beq
\delta \Omega_P =  \sum_{XY=P} \Omega_X \Omega_Y \ .
\label{gBG08}
\eeq
As exemplified by $\delta f_{123}^{mn}$ in (\ref{gBG32}) and $\delta \ffrak_{123}^{mn}$ in (\ref{ex10}), the absence of additional Mandelstam variables (\ref{mand}) on the right-hand side simplifies the gauge variation of Berends--Giele currents as compared to their local constituents. Nevertheless, the all-multiplicity pattern of local gauge variations $\delta e^m_{12\ldots p},\delta f^{mn}_{12\ldots p}$ and $\delta \omega_{12\ldots p}$ is well-understood from \cite{Mafra:2011kj, Mafra:2014oia}.

\subsection{Tree-level building blocks}
\label{sect23}

Based on arguments in pure-spinor superspace \cite{Lee:2015upy}, tree-level amplitudes of YM theories in arbitrary dimension have been expressed in terms of the kinematic structure \cite{Mafra:2015vca}
\beq
\mathfrak{M}_{A,B,C}\equiv \frac{1}{2} \efrak^m_A \ffrak^{mn}_B \efrak^n_C + \te{cyc}(A,B,C)
\ .
\label{tree1}
\eeq
These building blocks are totally antisymmetric in $A,B,C$ and represent cubic diagrams where Berends--Giele currents labelled by $A,B$ and $C$ are connected through a cubic vertex. By the gauge-variations (\ref{BG09}) and (\ref{BG10}) as well as
\beq
k_P^m \ffrak^{mn}_P = \sum_{XY=P} (\efrak_X^m \ffrak^{mn}_Y - \efrak_Y^m \ffrak^{mn}_X) \ ,
\label{tree5}
\eeq
they transform covariantly in terms of multiparticle gauge scalars (\ref{BG08})\footnote{The same gauge algebra holds in presence of fermions, see \cite{Mafra:2015vca} for the supersymmetric completion of $\mathfrak{M}_{A,B,C}$ via 10-dimensional gauginos.},
\beq
\delta \mathfrak{M}_{A,B,C}= \sum_{XY=A} \big( \Omega_X  \mathfrak{M}_{Y,B,C} - \Omega_Y  \mathfrak{M}_{X,B,C} + \Omega_B  \mathfrak{M}_{X,Y,C} - \Omega_C  \mathfrak{M}_{X,Y,B}
\big) + \te{cyc}(A,B,C)
\ ,
\label{tree2}
\eeq
for instance $\delta \mathfrak{M}_{12,3,4}=\Omega_1  \mathfrak{M}_{2,3,4} - \Omega_2  \mathfrak{M}_{1,3,4} + \Omega_3  \mathfrak{M}_{1,2,4} - \Omega_4  \mathfrak{M}_{1,2,3}$ in a 4-point context.

As shown in \cite{Mafra:2015vca}, the Berends--Giele formula \cite{Berends:1987me} for color-ordered tree amplitudes,
\beq
A^{\te{tree}}(1,2,\ldots,n) = s_{12\ldots n-1} (\efrak_{12\ldots n-1} \cdot e_n)
\label{tree3}
\eeq
is supersymmetrized by the pure-spinor superspace formula of \cite{Mafra:2010jq}. Based on efficient manipulations in pure-spinor superspace\footnote{BRST integration by parts of their 10-dimensional ancestors in pure-spinor superspace straightforwardly relates
\[
\sum_{XY=P}\mathfrak{M}_{X,Y,Q}= \sum_{XY=Q} \mathfrak{M}_{P,X,Y}\ , \ \ \ \ \te{e.g.} \ \ \ \
\mathfrak{M}_{12,3,4} = \mathfrak{M}_{34,1,2}
\ , \ \ \ \  \mathfrak{M}_{123,4,5}=\mathfrak{M}_{12,3,45}+\mathfrak{M}_{1,23,45} \ .
\]
}, the expression in (\ref{tree3}) can be converted to manifestly cyclic formulae for $n$-point amplitudes such as
\begin{align}
A^{\te{tree}}(1,2,3,4) &= \frac{1}{2}(\mathfrak{M}_{12,3,4} + \mathfrak{M}_{23,4,1}+\mathfrak{M}_{34,1,2} + \mathfrak{M}_{41,2,3})  \notag \\
A^{\te{tree}}(1,2,3,4,5) &= \mathfrak{M}_{12,3,45} + \te{cyc}(1,2,3,4,5) 
\label{tree4} \\
A^{\te{tree}}(1,2,3,4,5,6) &= \frac{1}{3}\mathfrak{M}_{12,34,56} + \frac{1}{2}(\mathfrak{M}_{123,45,6}+\mathfrak{M}_{123,4,56}) +   \te{cyc}(1,2,3,4,5,6) \ , \notag 
\end{align}
which only require currents of multiplicity $\leq \lfloor \frac{n}{2} \rfloor$ as anticipated in \cite{Berends:1989hf}. It is easy to check gauge invariance of (\ref{tree4}) via (\ref{tree2}), and manifestly cyclic expressions at higher multiplicity can be found in \cite{Mafra:2010jq} in pure-spinor superspace.

\subsection{Parity-even 1-loop building blocks}
\label{sect24}

The Berends--Giele organization also applies to loop amplitudes: The uplifts of $\efrak^m_P$ and $\ffrak_P^{mn}$ to pure-spinor superspace \cite{Lee:2015upy} have been used to construct BRST invariant and local expressions for 5- and 6-point 1-loop amplitudes \cite{Mafra:2014gja} as well as 2-loop 5-point amplitudes \cite{Mafra:2015mja} in 10-dimensional SYM. To extend the method beyond maximal supersymmetry, we shall now introduce kinematic building blocks for 1-loop amplitudes in 6 dimensions with half-maximal supersymmetry. In absence of the no-triangle property of maximal SYM \cite{Bern:1994zx}, we expect loop integrals of bubble and triangle topology in the half-maximal setup.

\begin{figure}[h]
\begin{center}
\begin{tikzpicture} [scale=0.8, line width=0.30mm]
\draw (0,0) -- (-1,1) node[left]{$2$};
\draw (0,0) -- (-1,-1) node[left]{$1$};
\draw (0,0) -- (4,0);
\draw (0.8,0) -- (0.8,1) node[above]{$3$};
\draw (1.6,0) -- (1.6,1) node[above]{$4$};
\draw (2.4,0.5)node{$\ldots$};
\draw (3.2,0) -- (3.2,1) node[above]{$p$};
\draw (4,0) .. controls (4.5,1) and (6.5,1) .. (7,0);
\draw (4,0) .. controls (4.5,-1) and (6.5,-1) .. (7,0);
\draw (5.5,-0.75)node{$<$}node[above]{$\ell$};
\draw (7,0) -- (10.2,0);
\draw (7.8,0) -- (7.8,1) node[above]{$p+1$};
\draw (8.6,0.5)node{$\ldots$};
\draw (9.4,0) -- (9.4,1) node[above]{$n-2$};
\draw (10.2,0) -- (11.2,1) node[right]{$n-1$};
\draw (10.2,0) -- (11.2,-1) node[right]{$n$};
\end{tikzpicture}
\caption{Cubic diagrams of bubble topology with kinematic numerator $T_{A,B}$ at ${A=12\ldots p}$ and $B=n,n{-}1 \ldots p{+}1$, where $\ell$ denotes the loop momentum.}
\label{f:bub}
\end{center}
\end{figure}
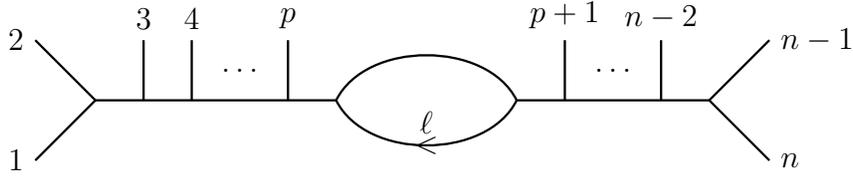

Since we will be interested in both local and gauge-invariant amplitude representations, we start by introducing local 1-loop building blocks before giving their Berends--Giele counterparts based on $\efrak^m_P$ and $\ffrak_P^{mn}$ in section \ref{sect26}. As motivated by the string-theory discussion of \cite{Berg:2016wux}, suitable kinematic numerators for bubble diagrams as in figure \ref{f:bub} are given by
\beq
T_{A,B} \equiv - \frac{1}{2} f^{mn}_{A} f^{mn}_{B} = T_{B,A} \ .
\label{BG35}
\eeq
The multiparticle labels $A$ and $B$ in the subscripts of the
multiparticle field strengths $f^{mn}_{12\ldots p}$, see (\ref{BG32}) and (\ref{BG34}),
refer to the tree-level subdiagrams seen in figure \ref{f:bub}. Their gauge variations in (\ref{gBG32}) imply covariant transformation for $T_{A,B}$ in (\ref{BG35}) such as
\begin{align}
\delta T_{1,2} &= 0 \co \delta T_{12,3} = s_{12}( \omega_1 T_{2,3} - \omega_2 T_{1,3}) \notag 
 \\
\delta T_{12,34} &= s_{12}( \omega_1 T_{2,34} - \omega_2 T_{1,34}) + s_{34}( \omega_3 T_{12,4} - \omega_4 T_{12,3}) \label{gBG35} \\ 
\delta T_{123,4}  &= (s_{13}+s_{23}) (\omega_{12} T_{3,4} - \omega_3 T_{12,4}) + s_{12} (\omega_1 T_{23,4} - \omega_{23} T_{1,4} - \omega_2 T_{13,4} + \omega_{13} T_{2,4}) \notag \ .
\end{align}
As will become clearer from the examples in sections \ref{sect31} and \ref{sect34}, loop momenta $\ell_m$ in the numerators of triangle diagrams and higher $n$-gons require vectorial and tensorial kinematic building blocks to contract with. This leads us to define generalizations of (\ref{BG35}),
\begin{align}
T^m_{A,B,C} \equiv e^m_A T_{B,C} + (A\leftrightarrow B,C) 
\ , \ \ \ \ \, T^{mn}_{A,B,C,D} \equiv 2e^{(m}_A e^{n)}_B T_{C,D} + (AB\leftrightarrow AC,AD,BC,BD,CD)  \ ,
\label{BG35a}
\end{align}
with (anti-)symmetrization conventions determined by $2e^{(m}_A e^{n)}_B=e_A^me_B^n +e_A^n e_B^m$.
Again, the covariant gauge variations (\ref{gBG32}) of local multiparticle polarizations propagate to the transformation of (\ref{BG35a}), e.g.
\begin{align}
\delta T^m_{1,2,3} &= \omega_1 k_1^m T_{2,3} + \omega_2 k_2^m T_{1,3} + \omega_3 k_3^m T_{1,2}   \notag \\
\delta T^m_{12,3,4} &= \omega_{12} k_{12}^m T_{3,4} + \omega_3 k_3^m T_{12,4} + \omega_4 k_4^m T_{12,3} + s_{12}(\omega_1 T^m_{2,3,4} - \omega_2 T^m_{1,3,4} )
\label{BG35b} \\
\delta T^{mn}_{1,2,3,4} &= 2 \omega_1 k_1^{(m} T^{n)}_{2,3,4} +2 \omega_2 k_2^{(m} T^{n)}_{1,3,4} +2 \omega_3 k_3^{(m} T^{n)}_{1,2,4} +2 \omega_4 k_4^{(m} T^{n)}_{1,2,3}  \ .
\notag
\end{align}
In section \ref{sect3}, we will identify combinations of scalar, vector and tensor building blocks whose gauge variation cancels $\ell$-dependent propagators $(\ell - k_{12\ldots p})^2$, i.e.\ which qualify as triangle- and box numerators.

\subsection{Parity-odd 1-loop building blocks}
\label{sect25}

The running of chiral fermions in the loop introduces Levi-Civita tensors into the integrands of multiplicity $\geq 3$. This requires a parity-odd completion of the above building blocks whose form is inspired by the contribution of worldsheet fermions with odd spin structures in the RNS superstring \cite{Polchinski:1998rr,Green:1987mn}
\beq
E^m_{A|B,C} \equiv
\frac{i}{4}\epsilon^{m}{}_{npqrs} e_A^n f_B^{pq} f_C^{rs} = E^m_{A|C,B} \ .
\label{locG09}
\eeq
The lack of symmetry under $A\leftrightarrow B$ or $A\leftrightarrow C$ (represented by the vertical bar $A|\ldots$ in the subscript) is an artifact of the asymmetric superghost pictures in the string computation \cite{Berg:2016wux}. Throughout this work, we will choose reference leg 1 to be part of $A$ in each term.

In contrast to the variations (\ref{BG35}) and (\ref{BG35b}) of the parity-even building blocks, the gauge algebra of (\ref{locG09}) now relies on momentum conservation: Only by imposing $k_A^m\!+\!k_B^m \!+\! k_C^m=0$, one can show that
\begin{align}
\delta E^m_{1|2,3} &= 0 \co \delta E^m_{12|3,4} = s_{12} (\omega_1 E^m_{2|3,4} - \omega_2 E^m_{1|3,4})
\label{locG10} \\
\delta E^m_{1|23,4} &= s_{23}(\omega_2 E^m_{1|3,4}  - \omega_3 E^m_{1|2,4} + \omega_1 E^m_{3|2,4} - \omega_1 E^m_{2|3,4}  )  \ .\notag
\end{align}
In analogy to the parity-even building blocks (\ref{BG35a}), the parity-odd vector (\ref{locG09}) allows for tensorial generalizations such as
\beq
E^{mn}_{A|B,C,D} \equiv 2 e_B^{(m} E^{n)}_{A|C,D} + 2 e_C^{(m} E^{n)}_{A|B,D} +2 e_D^{(m} E^{n)}_{A|B,C}  \ .
\label{locG11}
\eeq
Once the gauge variation of this tensor building block is simplified via
\beq
2\eta^{mn} \epsilon^{abcdef} = \eta^{ma} \epsilon^{nbcdef} + \eta^{na} \epsilon^{mbcdef} -
 \eta^{mb} \epsilon^{nacdef} - \eta^{nb} \epsilon^{macdef} +(ab \leftrightarrow cd,ef)
\label{odd13}
\eeq
based on ``overantisymmetrization'' $ \epsilon^{[abcdef} \eta^{g]n}=0$ in 6 dimensions, the trace component contributes to an anomalous gauge variation:
\beq
\delta E^{mn}_{1|2,3,4} =  \eta^{mn} \omega_1 Y_{2,3,4} +  2 \big[ k_{2}^{(m} ( \omega_2 E^{n)}_{1|3,4} - \omega_1 E^{n)}_{2|3,4} )  + (2\leftrightarrow 3,4) \big] \ .
\label{locG12}
\eeq
The scalar building block
\beq
Y_{A,B,C} \equiv
\frac{i}{4}\epsilon_{mnpqrs} f_A^{mn} f_B^{pq} f_C^{rs}
\label{locG13}
\eeq
represents the chiral box anomaly specific to 6 dimensions \cite{Chen:2014eva}, that we will encounter  in the 4-point 1-loop amplitude.

\subsection{Berends--Giele building blocks at 1-loop}
\label{sect26}

Recombining the local multiparticle polarizations into Berends--Giele currents $\efrak^m_P, \ffrak^{mn}_P$ in (\ref{BG01}) and (\ref{BG02}) leads to simplified gauge variations (\ref{BG09}) and (\ref{BG10}). Accordingly, the Berends--Giele versions
\begin{align}
{\cal T}_{A,B} &\equiv -\frac{1}{2} \ffrak^{mn}_A\ffrak^{mn}_B \co
{\cal T}^m_{A,B,C} \equiv \efrak^m_A {\cal T}_{B,C}+\efrak^m_B {\cal T}_{A,C}+\efrak^m_C {\cal T}_{A,B}\notag
\\
{\cal T}^{mn}_{A,B,C,D}&\equiv 2 \efrak^{(m}_A \efrak^{n)}_{B} {\cal T}_{C,D}
+ (AB\leftrightarrow AC,AD,BC,BD,CD) \label{ct1}
\end{align}
of the local 1-loop building blocks in (\ref{BG35}) and (\ref{BG35a}) obey a gauge algebra with deconcatenation rules and multiparticle gauge scalars $\Omega_P$ defined in (\ref{BG08}),
\begin{align}
\delta {\cal T}_{A,B} &\equiv \sum_{XY=A} (\Omega_X {\cal T}_{Y,B} - \Omega_Y {\cal T}_{X,B} ) + (A\leftrightarrow B)
\notag \\
\delta {\cal T}^m_{A,B,C} &\equiv \Omega_A  k_A^m {\cal T}_{B,C}
+ \sum_{XY=A} (\Omega_X {\cal T}^m_{Y,B,C} - \Omega_Y {\cal T}^m_{X,B,C} ) 
+ (A\leftrightarrow B,C) \label{BG15} \\
\delta  {\cal T}^{mn}_{A,B,C,D} &\equiv  2\Omega_A k_A^{(m}  {\cal T}^{n)}_{B,C,D}
+ \sum_{XY=A} (\Omega_X {\cal T}^{mn}_{Y,B,C,D} - \Omega_Y {\cal T}^{mn}_{X,B,C,D} ) 
+ (A\leftrightarrow B,C,D) \ ,
\notag 
\end{align}
bypassing the Mandelstam invariants in (\ref{gBG35}) and (\ref{BG35b}). Similarly, adjusting the local parity-odd building blocks in (\ref{locG09}) and (\ref{locG11}) to Berends--Giele currents,
\begin{align}
{\cal E}^m_{A|B,C} &\equiv
\frac{i}{4}\epsilon^{m}{}_{npqrs} \efrak_A^n \ffrak_B^{pq} \ffrak_C^{rs} = {\cal E}^m_{A|C,B} 
\label{Giele09} \\
{\cal E}^{mn}_{A|B,C,D} &\equiv 2 \efrak_B^{(m} {\cal E}^{n)}_{A|C,D} + 2 \efrak_C^{(m} {\cal E}^{n)}_{A|B,D} +2 \efrak_D^{(m} {\cal E}^{n)}_{A|B,C}  \ ,
\end{align}
translates the gauge variations in (\ref{locG10}) and (\ref{locG12}) into the following deconcatenation rules:
\begin{align}
\delta {\cal E}^m_{A|B,C} &= \sum_{XY=A} (\Omega_X {\cal E}^m_{Y|B,C}-\Omega_Y {\cal E}^m_{X|B,C}) +  \sum_{XY=B} \big[ \Omega_X {\cal E}^m_{A|Y,C}-\Omega_Y {\cal E}^m_{A|X,C} + \Omega_A ( {\cal E}^m_{Y|X,C} -  {\cal E}^m_{X|Y,C})\big]
\notag \\
&+  \sum_{XY=C} \big[ \Omega_X {\cal E}^m_{A|B,X}-\Omega_Y {\cal E}^m_{A|B,X} + \Omega_A ( {\cal E}^m_{Y|B,X} -  {\cal E}^m_{X|B,Y})
\big] \notag
\\
\delta {\cal E}^{mn}_{A|B,C,D}  &= \eta^{mn} \Omega_A {\cal Y}_{B,C,D} + \sum_{XY=A} (\Omega_X {\cal E}^{mn}_{Y|B,C,D} - \Omega_Y {\cal E}^{mn}_{X|B,C,D} )  \label{odd9}  \\
&+ \Big[  \sum_{XY=B} (\Omega_X {\cal E}^{mn}_{A|Y,C,D} -  \Omega_Y {\cal E}^{mn}_{A|X,C,D} )
+ \Omega_A \sum_{XY=B} ( {\cal E}^{mn}_{Y|X,C,D} -  {\cal E}^{mn}_{X|Y,C,D} )
\notag \\
& \ \ \ \ + 2k_B^{(m} (\Omega_B {\cal E}^{n)}_{A|C,D} - \Omega_A {\cal E}^{n)}_{B|C,D} )
+(B\leftrightarrow C,D) \Big]  \notag
\end{align}
with the obvious Berends--Giele version of the anomaly building block (\ref{locG13}):
\beq
{\cal Y}_{A,B,C} \equiv
\frac{i}{4}\epsilon_{mnpqrs} \ffrak_A^{mn} \ffrak_B^{pq} \ffrak_C^{rs} \ .
\label{odd15}
\eeq
We recall that the gauge algebra (\ref{odd9}) relies on momentum conservation. Note that the above gauge variations take a similar form as seen in the BRST algebra of maximally supersymmetric 1-loop building blocks in \cite{Mafra:2014oia} and \cite{Mafra:2014gsa}. In particular, the generalization of the BRST covariant building blocks to tensors of arbitrary rank \cite{Mafra:2014gsa} can be easily adapted to the half-maximal framework, and the resulting definition and gauge algebra of ${\cal T}^{m_1 m_2\ldots m_r}_{A_1,A_2,\ldots,A_{r+2}}$ or ${\cal E}^{m_1 m_2\ldots m_r}_{A_1|A_2,\ldots,A_{r+2}}$ will be explored in the future.

\subsection{Gauge-invariant and pseudo-invariant 1-loop building blocks}
\label{sect27}

The covariant transformations (\ref{BG15}) and (\ref{odd9}) are suitable to construct gauge-invariant combinations of Berends--Giele building blocks. The simplest parity-even examples
\begin{align}
\delta{\cal T}_{1,2}= \delta( {\cal T}_{1,23} + {\cal T}_{12,3} - {\cal T}_{13,2} )=0
\label{invE}
\end{align}
turn out to exhibit the same pattern of combining different multiparticle labels as seen in the parity-odd sector:
\begin{align}
\delta{\cal E}^m_{1|2,3}= \delta( {\cal E}^m_{1|23,4} + {\cal E}^m_{12|3,4} - {\cal E}^m_{13|2,4} ) = 0\,.
\label{invO}
\end{align}
While the parity-even gauge algebra (\ref{BG15}) along with momentum conservation allows for invariants with additional free vector indices,
\begin{align}
\delta( {\cal T}^m_{1,2,3} + k_2^m {\cal T}_{12,3} + k_3^m {\cal T}_{13,2}) &=0 
\label{invEa}  \ ,
\end{align}
the anomalous term $\eta^{mn} \Omega_A {\cal Y}_{B,C,D} $ in the variation (\ref{odd9}) of ${\cal E}^{mn}_{A|B,C,D}$ 
complicates the construction of
tensor invariants in the parity-odd sector. Hence, we follow the terminology of \cite{Mafra:2014gsa} to relax the requirement of gauge invariance such that anomaly kinematics (\ref{odd15}) is admitted: Kinematic factors  whose gauge variation  can be expressed in terms of $\Omega_A$ and ${\cal Y}_{B,C,D} \sim \epsilon_{mnpqrs} \ffrak_B^{mn}\ffrak_C^{pq}\ffrak_D^{rs}$ will be referred to as {\em pseudo-invariant}. Then, one can view the combination of different tensor ranks in
\begin{align}
\delta( {\cal E}^{mn}_{1|2,3,4} + 2k_2^{(m} {\cal E}^{n)}_{12|3,4} +2k_3^{(m} {\cal E}^{n)}_{13|2,4} + 2k_4^{(m} {\cal E}^{n)}_{14|2,3} ) &=  \eta^{mn} \Omega_1 {\cal Y}_{2,3,4}
\label{invOa}
\end{align}
as following the same pattern to achieve pseudo-invariance as seen in (\ref{invEa}). Even though the special role of the first slot $A$ in ${\cal E}^m_{A|B,C}$ and ${\cal E}^{mn}_{A|B,C,D}$ causes the parity-even and parity-odd gauge algebras (\ref{BG15}) and (\ref{odd9}) to differ in their details, the examples in (\ref{invE}) and (\ref{invO}) as well as (\ref{invEa}) and (\ref{invOa}) suggest that the construction of gauge-(pseudo-)invariants follows the same rules. Accordingly, we introduce a unifying notation for combinations of parity-even and parity-odd building blocks
\begin{align}
M_{A,B} &\equiv {\cal T}_{A,B} =  -\frac{1}{2} \ffrak^{mn}_A\ffrak^{mn}_B \notag\\
M^m_{A|B,C} &\equiv {\cal T}^m_{A,B,C}+ {\cal E}^m_{A|B,C} = \big[ \efrak^m_A M_{B,C} +  (A\leftrightarrow B,C) \big]+
\frac{i}{4}\epsilon^{m}{}_{npqrs} \efrak_A^n \ffrak_B^{pq} \ffrak_C^{rs} \label{uni} \\
M^{mn}_{A|B,C,D} &\equiv {\cal T}^{mn}_{A,B,C,D}+ {\cal E}^{mn}_{A|B,C,D} = \big[ \efrak_B^{m} {\cal E}^{n}_{A|C,D}+ \efrak_B^{n} M^{m}_{A|C,D} + (B\leftrightarrow C,D) \big]+ \efrak_A^{n} {\cal T}^{m}_{B,C,D}   \ ,
\notag
\end{align}
whose relative coefficients are part of the string-theory input and which yield a compact form for
the gauge-(pseudo-)invariants which have been identified in the string computations of \cite{Berg:2016wux}: The scalar invariants
\begin{align}
C_{1|2} &\equiv M_{1,2}
\label{BG2pt}
\\
C_{1|23} &\equiv M_{1,23} + M_{12,3}  - M_{13,2}   
\label{BG17}
\\
C_{1|234} &\equiv M_{1,234} + M_{123,4}  + M_{412,3} + M_{341,2} + M_{12,34} + M_{41,23}  
\label{BG19} 
\end{align}
and vector invariants
\begin{align}
C^m_{1|2,3} &\equiv M^m_{1|2,3} + k_2^m M_{12,3} + k_3^m M_{13,2}
\label{BG18}
\\
C^m_{1|23,4} &\equiv M^m_{1|23,4} + M^m_{12|3,4}  - M^m_{13|2,4}  - k_2^m M_{132,4} + k_3^m M_{123,4} - k_4^m (M_{41,23} + M_{412,3} - M_{413,2}) 
\label{BG20}
\end{align}
are constructed from the same combinations of multiparticle labels as the maximally supersymmetric BRST invariants $C_{1|2,3,4}, C_{1|23,4,5}, C_{1|234,5,6}$ as well as $C^m_{1|2,3,4,5},C_{1|23,4,5,6}^m$ defined in section 5 of \cite{Mafra:2014oia}. The tensor pseudo-invariant
\begin{align}
C^{mn}_{1|2,3,4} &\equiv M^{mn}_{1|2,3,4} +  2\big[ k_2^{(m} M^{n)}_{12|3,4}+(2\leftrightarrow 3,4) \big] - 2 \big[ k_2^{(m} k_3^{n)} M_{213,4}+ (23\leftrightarrow 24,34)\big] \ ,
\label{BGtens}
\end{align}
on the other hand, resembles the 6-point tensor $C^{mn}_{1|2,3,4,5,6}$ in pure-spinor superspace defined in (3.14) of \cite{Mafra:2014gsa} -- see \cite{maxsusy} for its appearance in closed-string amplitudes. From the gauge algebras (\ref{BG15}) and (\ref{odd9}), it is straightforward to check that
\beq
\delta C_{1|A}=0 \co \delta C^m_{1|A,B}=0
\co
 \delta C^{mn}_{1|2,3,4} =2 i \omega_1\eta^{mn} 
 \epsilon(k_2,e_2,k_3,e_3,k_4,e_4)  = \omega_1\eta^{mn} {\cal Y}_{2,3,4} \ ,
\label{BG21var}
\eeq
using momentum conservation for the vectors and the tensor. In addition to the tensor (\ref{BGtens}), one can construct scalar pseudo-invariants from the additional building block
\begin{align}
{\cal J}_{1|2|3,4}  &\equiv (e_2)_m (e^m_1 M_{3,4} + {\cal E}^m_{1|3,4}) + \frac{1}{2} \big[ (e_2 \cdot e_3) M_{1,4} + (3\leftrightarrow 4) \big]
\label{Jdef} \\
\delta {\cal J}_{1|2|3,4}  &= \Omega_1 {\cal Y}_{2,3,4} + k_2^m (\Omega_2 M^{m}_{1|3,4} - \Omega_1 M^m_{2|3,4}) + \big[ s_{23}( \Omega_{23} M_{1,4} - \Omega_1 M_{23,4})
+  (3\leftrightarrow 4) \big]\ .
\label{Jvar}
\end{align}
Its covariant gauge variation (\ref{Jvar}) based on momentum conservation\footnote{Note that the derivation of (\ref{Jvar}) is based on $(k_{34}\cdot e_2)  {\cal T}_{3,4}=k_2^m {\cal T}^m_{2,3,4} + s_{23}  {\cal T}_{23,4} + s_{24}  {\cal T}_{24,3}$.} implies that
\beq
P_{1|2|3,4} \equiv {\cal J}_{1|2|3,4} + k_2^m M^m_{12|3,4} + s_{23} M_{123,4} + s_{24} M_{124,3} 
\label{BG24}
\eeq
is pseudo-invariant as well:
\beq
\delta P_{1|2|3,4} =2 i \omega_1 \epsilon(k_2,e_2,k_3,e_3,k_4,e_4) = \omega_1 {\cal Y}_{2,3,4} \ .
\label{BG24var}
\eeq
Its composition from $M_{A,B}$ and $M^m_{A|B,C}$ follows the patterns of the BRST pseudo-invariant $P_{1|2|3,4,5,6}$ in (5.22) of \cite{Mafra:2014gsa}. More generally, the recursion introduced in \cite{Mafra:2014oia, Mafra:2014gsa} allows to construct BRST {(pseudo-)invariants} at arbitrary multiplicity and tensor rank, including comparable generalizations of $P_{1|2|3,4}$ in (\ref{BG24}). The master recursion for an arbitrary number of multiparticle slots in section 8 of \cite{Mafra:2014gsa} can be used to obtain (pseudo-)invariants of higher multiplicity $n\geq 5$ in the half-maximal setup.

\section{Constructing 1-loop field-theory amplitudes}
\label{sect3}

In this section, we propose manifestly local expressions for the 3- and 4-point amplitudes of half-maximal SYM in $D=6$. The kinematic factors of individual diagrams are constructed from the string-theory motivated family of building blocks introduced in the previous section, and designed to produce cancellations between their gauge variations when assembling the overall amplitude. To manifest gauge invariance at the level of the integrand, we pick the convention for shifting the loop momentum $\ell$ such that the only $\ell$-dependent propagators are inverse to $\ell^2,(\ell{-}k_1)^2,(\ell{-}k_{12})^2, \ldots,(\ell{-}k_{12\ldots n-1})^2$, i.e.\ they refer to momenta of the form
\beq
\ell_{12\ldots p} \equiv \ell-k_{12\ldots p} \co p=0,1,2,\dots,n{-}1 \ ,
\label{loc70}
\eeq
where $k_{12\ldots p}$ is defined in \eqref{mand}.
Locality is implemented by drawing all  cubic-vertex diagrams 
that are compatible with color-ordering of the external legs and free of tadpole subgraphs\footnote{Tadpole subgraphs are incompatible with the string prescription that motivates our choice of building blocks: The tadpoles have $n{-}1$ propagators with external momenta only, that cannot arise from the maximum number of $n{-}2$ kinematic poles $s^{-1}_{i\ldots j}$ admitted by the singularity structure of the string-theory integrands in \cite{Berg:2016wux}. We note that this is not directly related to the general consistency relations
known as {\it tadpole cancellation} in string theory (see e.g. the textbooks \cite{Polchinski:1998rr,Ibanez:2012zz}) --- we have
not (yet) demanded tadpole cancellation in the models of appendix \ref{appE}.}. Following the spirit of the duality between color and kinematics \cite{Bern:2008qj}, the quartic vertices of the SYM Feynman rules are absorbed into the kinematic factors of the cubic graphs. Since quartic vertices arise from the gauge-invariant completion of the SYM Lagrangian, checking gauge invariance of the amplitudes is sufficient to make sure the quartic-vertex kinematics is appropriately captured. The general structure of a $D$-dimensional $n$-point 1-loop amplitude in the cubic-graph expansion reads \cite{Bern:2010ue}
\begin{align}
A^{\te{1-loop}}_{{\cal N}}(1,2,\ldots,n) &= \int \frac{ \dd^D\ell }{(2\pi)^D}  \sum_{i \in \Gamma_{12\ldots n}} \frac{  N_i(\ell) }{ \prod_{\alpha=1}^{n}   p^2_{\alpha,i}(\ell)} \ .
\label{cubgrph} 
\end{align}
The summation range $\Gamma_{12\ldots n}$ selects cubic 1-loop graphs $i$ that are compatible with the
cyclic ordering $1,2,\ldots,n$ of the color-stripped single-trace amplitude in (\ref{cubgrph}). Any graph $i$ is associated with (possibly loop-momentum dependent) internal momenta $p_{1,i}(\ell),p_{2,i}(\ell),\ldots,p_{n,i}(\ell)$ from its edges $\alpha$, and the design of their kinematic numerators $N_i(\ell)$ is guided by the gauge variation (\ref{BG11}):
\beq
\te{{\em Each  term  of} $\delta N_i(\ell)$ {\em must  cancel  a  propagator, i.e.\ contain  a  factor  of} $p^2_{\alpha,i}(\ell) \, , \ \ \alpha=1,2,\ldots,n$} \ .
\label{loc71}
\eeq
By locality of the numerators, this is a necessary condition for gauge invariance of the overall integrand. 
To ensure it is also sufficient, it remains to check that all the contributions from $\delta N_i(\ell)$ with fewer propagators cancel between diagrams. The global delta function imposing momentum conservation $k_1+k_2+\ldots+k_n$ in (\ref{cubgrph}) is left implicit. Finally, in slight abuse of notation, the specification of $4{\cal N}$ supercharges through the subscript ${\cal N}$ of $A^{\te{1-loop}}_{{\cal N}}(\ldots)$ follows the 4-dimensional counting of supersymmetries although the amplitudes constructed in this section live in 6 dimensions.

\subsection{The local form of the 3-point amplitude}
\label{sect31}

At the 3-point level, our ansatz for a cubic-graph expansion for half-maximal 1-loop amplitudes without tadpoles involves three bubble-diagrams and one triangle:
\begin{align}
A^{\te{1-loop}}_{{\cal N}=2}(1,2,3) &= \int \frac{ \dd^D\ell }{(2\pi)^D} \, \bigg\{ \frac{ T_{1,23} }{s_{23} \ell^2 \ell_{1}^2 } + \frac{ T_{12,3} }{s_{12} \ell^2 \ell_{12}^2 } + \frac{ T_{31,2} }{s_{13} \ell_{1}^2 \ell_{12}^2 } +
 \frac{ N_{1|2,3}(\ell) }{ \ell^2 \ell_{1}^2 \ell_{12}^2 } \bigg\}  \ ,
\label{BG30} 
\end{align}
see (\ref{loc70}) for the $\ell$-dependent propagators.
The bubble numerators in (\ref{BG30}) have already been identified with the scalar building blocks $T_{A,B}$ in (\ref{BG35}). By their gauge algebra (\ref{gBG35}) and the absence of tadpoles, this is a canonical choice compatible with the general principle (\ref{loc71}). The triangle numerator $N_{1|2,3}(\ell)$ is initially left undetermined, but the requirement to cancel the gauge variation of the bubbles,
\beq
\delta \Big( \frac{ T_{1,23} }{s_{23} \ell^2 \ell_{1}^2 } + \frac{ T_{12,3} }{s_{12} \ell^2 \ell_{12}^2 } + \frac{ T_{31,2} }{s_{13} \ell_{1}^2 \ell_{12}^2 }  \Big) = \frac{ \omega_1 T_{2,3}(\ell_1^2-\ell^2) 
+ \omega_2 T_{1,3}(\ell_{12}^2-\ell_1^2) 
+ \omega_3 T_{1,2}(\ell^2-\ell_{12}^2) }{\ell^2 \ell_1^2 \ell_{12}^2} \ ,
\label{loc72}
\eeq
with $\ell_{12\ldots p}^2 = \ell^2 - 2 (\ell \cdot k_{12\ldots p}) + k_{12\ldots p}^2$ fixes its gauge variation to be
\beq
\delta N_{1|2,3}(\ell) =2(\ell \cdot k_1) \omega_1 T_{2,3} +2  (\ell \cdot k_2)   \omega_2 T_{1,3}+2  (\ell \cdot k_3)   \omega_3 T_{1,2} \ ,
\label{loc73}
\eeq
using the vanishing of Mandelstam invariants (\ref{mand}) in 3-particle momentum phase space\footnote{We keep kinematic identities  covariant and dimension-agnostic (except in section \ref{sect5}, for checks in $D=4$). 
Hence, we will not use the common strategy of
factorizing $s_{12}= \frac{1}{2}(k_3^2 -k_2^2 - k_1^2)=0$ into 4-dimensional spinor brackets $\langle 12\rangle$ and $[12]$,
one of which is taken to be non-zero for complex momenta \cite{Elvang:2010kc}. Instead,
 in the next section \ref{sect32} we will introduce
a $D$-dimensional infrared regularization to track the cancellation of the vanishing 3-particle $s_{ij}$ in intermediate steps.}
\beq
s_{12}= \frac{1}{2} \big[ (k_1+k_2)^2 - k_1^2 - k_2^2 \big]= \frac{1}{2} \big[k_3^2 - k_1^2 - k_2^2 \big]=0 \ .
\label{loc74}
\eeq
In view of the gauge algebra (\ref{BG35b}), the minimal solution to (\ref{loc73}) is  $2\ell_m T^m_{1,2,3}$. However, we
are led to  the nonminimal solution
\beq
N_{1|2,3}(\ell)  =
2\ell_m (T^m_{1,2,3}+E^m_{1|2,3}) + T_{12,3} + T_{13,2} + T_{1,23} \; , 
\label{loc75}
\eeq
which will appear in 4-point gauge variations and play out with the BCJ duality. Moreover, (\ref{loc75}) resembles the structure of the maximally supersymmetric pentagon numerator in (4.5) of \cite{Mafra:2014gja}. We have allowed 
for the parity-odd term $\ell_m E^m_{1|2,3}\sim \epsilon(\ell,e_1,k_2,e_2,k_3,e_3)$ defined by (\ref{locG09}),
for chiral fermions to run in the triangular loop. 

The freedom to choose nonminimal solutions might seem to be at odds with the ``fixing'' of the 
amplitudes claimed in the introduction. In fact, the parity-even extensions of the triangle numerator by $T_{ij,k}$
in \eqref{loc75} vanish as detailed in section \ref{sect32} below, but their inclusion parallels certain non-vanishing contributions to the triangle numerators in the 4-point amplitude, see eqs.\ (\ref{exota}) to (\ref{exot}). The parity-odd term $\ell_m E^m_{1|2,3}$ is local and gauge-invariant by itself, but the string-motivated combination $\ell_m M^m_{A|B,C}$, defined in (\ref{uni}) at generic multiplicity,
tells us that $\ell_m T^m_{1,2,3}$ should appear only in the combination $\ell_m (T^m_{1,2,3}+E^m_{1|2,3})$.
As a check, 
this term can be directly calculated from string theory as in \cite{Berg:2016wux}.

\subsection{Infrared regularization}
\label{sect32}

In (\ref{loc74}), we see 
the usual vanishing $s_{ij}=0$ of 3-particle Mandelstam invariants 
for massless external states. This threatens  to introduce singular propagators 
of the form ``$1/0$'' in the bubble terms in (\ref{BG30}). Fortunately, their numerators are
\begin{align}
T_{12,3} &=-\frac{1}{2}f_{12}^{mn} f_3^{mn} = (k_{12}^m e_{12}^n  - s_{12}e_1^m e_2^n)(k_3^n e_3^m - k_3^m e_3^n) \notag \\
&=  -(s_{13}+s_{23})(e_{12} \cdot e_3) - s_{12}(e_1 \cdot e_3)(k_3 \cdot e_2)+s_{12}(e_2 \cdot e_3)(k_3 \cdot e_1)\label{loc77} \\
&= s_{12} \big[ (e_2 \!\cdot \!e_3)(k_2 \!\cdot \!e_1) - (e_1\! \cdot \!e_3)(k_1\! \cdot \!e_2) + \tfrac{1}{2} (e_1 \!\cdot\! e_2)(k^m_1 - k^m_2) e_3^m  - (e_1 \!\cdot\! e_3)(k_3\! \cdot \! e_2)+(e_2 \! \cdot \! e_3)(k_3 \! \cdot \! e_1)  \big]\notag \\
&= s_{12} (e_1 \cdot e_2)(k_1 \cdot e_3)
\notag 
\end{align}
using transversality $(k_3 \cdot e_3)= - (k_{12}\cdot e_3) = 0$ and no Mandelstam identity other than $s_{12}+s_{13}+s_{23}=0$. This identifies the bubble contribution $\frac{ T_{12,3}}{s_{12}}$ as a ``$0/0$'' indeterminate, that  requires an infrared regularization procedure
to resolve. 

\begin{figure}[h]
\begin{center}
\begin{tikzpicture} [scale=0.8, line width=0.30mm]
\scope[xshift=3.2cm]
\draw (0,0) -- (-1,1) node[left]{$2$};
\draw (0,0) -- (-1,-1) node[left]{$1$};
\endscope
\draw (3.2,0) -- (4,0);
\draw (3.6,-0.5)node{$s_{12}$};
\draw (4,0) .. controls (4.5,1) and (6.5,1) .. (7,0);
\draw (4,0) .. controls (4.5,-1) and (6.5,-1) .. (7,0);
\draw (5.5,-0.75)node{$<$}node[above]{$\ell$};
\draw (7,0) -- (7.8,0) node[right]{$3$};
\draw (5.5,1.5)node{$T_{12,3}= s_{12}(e_1 \cdot e_2)(k_1 \cdot e_3)$};
\draw[<->] (9.8,0.25)--(11,0.25);
\scope[xshift=10cm]
\scope[xshift=4cm]
\draw (0,0) -- (-1,1) node[left]{$2$};
\draw (0,0) -- (-1,-1) node[left]{$1$};
\endscope
\draw (4,0) .. controls (4.5,1) and (6.5,1) .. (7,0);
\draw (4,0) .. controls (4.5,-1) and (6.5,-1) .. (7,0);
\draw (5.5,-0.75)node{$<$}node[above]{$\ell$};
\draw (7,0) -- (7.8,0) node[right]{$3$};
\draw (5.9,1.5)node{$M_{12,3}=(e_1 \cdot e_2)(k_1 \cdot e_3)$};
\endscope
\end{tikzpicture}
\caption{The singular propagator $s_{12}^{-1}$ in the one-mass bubble diagram is compensated by the formally vanishing numerator $T_{12,3}= s_{12}(e_1 \cdot e_2)(k_1 \cdot e_3)$. If the 
diagram on the left is a ``snail'', then the diagram on the right is a ``shy snail''.
All our snails are shy.}
\label{f:3bub}
\end{center}
\end{figure}
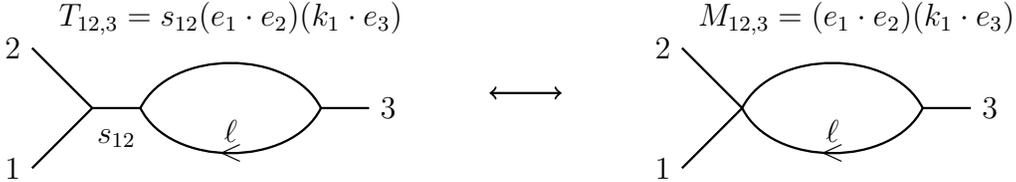

The problem of singularities $\sim s_{12\ldots n-1}^{-1}$ in the phase space of $n$ massless particles 
also arises in string amplitudes in orbifolds with half-maximal supersymmetry. In \cite{Berg:2016wux},
where we constructed the string-theory input for the SYM amplitudes in this paper, we used the following
proposal by Minahan in 1987 \cite{Minahan:1987ha}, 
that we referred to as {\em minahaning}.  Infrared singularities are regularized by a lightlike ``deformation''  momentum $p^m$ perpendicular to all the polarization vectors,
that deforms 3-particle momentum conservation to
\beq
k_1^m+k_2^m+k_3^m=p^m \co p^2=0 \ .
\label{loc78}
\eeq
This allows the Mandelstam invariants in the 3-point function  to be nonzero in intermediate steps. 
For instance, we have
\beq
k_3 \cdot p = k_3 \cdot (k_1+k_2+k_3)=k_1\cdot k_3+
k_2\cdot k_3=-s_{12}  \ ,
\label{Mina5}
\eeq
so $s_{12}$ is linear in the deformation $p$. (In string theory, the virtue
of this particular regularization procedure is that it preserves modular invariance of 1-loop amplitudes
while keeping the external states on-shell.)
Then, by (\ref{loc77}), the dependence on $p^m$ automatically drops out from the bubble contributions, and the singular propagator is cancelled as visualized in figure \ref{f:3bub},
\beq
M_{12,3} = \frac{ T_{12,3} }{s_{12}  } = \frac{ (k_3\cdot p) (e_1 \cdot e_2)(k_1 \cdot e_3) }{(k_3 \cdot p)} = (e_1 \cdot e_2)(k_1 \cdot e_3) \ .
 \label{loc79}
 \eeq
This casts the 3-point amplitude (\ref{BG30}) into the following form,
\begin{align}
A^{\te{1-loop}}_{{\cal N}=2}(1,2,3) &= \int \frac{ \dd^D\ell }{(2\pi)^D} \, \bigg\{ \frac{ (e_2 \cdot e_3)(k_2 \cdot e_1)}{\ell^2 \ell_{1}^2 } + \frac{ (e_1 \cdot e_2)(k_1 \cdot e_3) }{ \ell^2 \ell_{12}^2 } + \frac{ (e_1 \cdot e_3)(k_3 \cdot e_2) }{ \ell_{1}^2 \ell_{12}^2 }  \label{rBG30}  \\
&\! \! \! \! \! \! \! \! \! \! \! \! \! \! \! \! \! \! \! \! +
 \frac{2 \ell_m \big[
e_1^m(k_2 \cdot e_3)(k_3 \cdot e_2) +    
e_2^m(k_3 \cdot e_1)(k_1 \cdot e_3) +    
e_3^m(k_1 \cdot e_2)(k_2 \cdot e_1) +    
  i\epsilon^{m}(e_1,k_2,e_2,k_3,e_3)  
 \big] }{ \ell^2 \ell_{1}^2 \ell_{12}^2 } \bigg\}  \ ,
\notag
\end{align}
where we have dropped any term $\sim s_{ij}$ in the triangle numerator (\ref{loc75}). 
Note that at the level of the integrand, we are treating the one-mass bubbles (``shy snails'' of figure \ref{f:3bub}) on equal footing with triangle diagrams. (This could also be
natural for computing effective actions: wavefunction renormalization and  gauge coupling corrections
are related by Ward identities.) 


One might worry whether the deformation (\ref{loc78}) of the kinematic phase space interferes with the gauge algebra, since we used momentum conservation  in section \ref{sect26} and \ref{sect27} to identify gauge-invariants. 
This worry is unfounded --- indeed we used momentum conservation 
for our vectorial and tensorial building blocks, but only scalar building blocks are minahaned. 
For instance, we use $n$-particle
momentum conservation  to cast $\delta E^m_{1|23,4}$ in (\ref{locG10}) into covariant form and 
to rewrite $\delta T^m_{1,2,3}$ in (\ref{invEa}) as $k_2^m (\omega_2 T_{1,3} - \omega_1 T_{2,3}) + (2\leftrightarrow 3) $ where the gauge-invariant completion is more evident. The gauge algebra for the scalars $M_{A,B}\sim\ffrak_A^{mn} \ffrak_B^{mn}$, where minahaning is required in case of single-particle slots $A$ or $B$, is not tied to any phase-space constraints. 

In more general terms,
the vector and tensor building blocks seen in section \ref{sect2} are built from products of at least 3 Berends--Giele currents, see (\ref{ct1}) and (\ref{Giele09}). Hence, the associated external propagators contain at most $n{-}2$ massless momenta, and the naively singular propagators of the scalars $M_{12\ldots n-1,n}$ are bypassed. To summarize: in the sector of the gauge algebra that relies on momentum conservation, we are in fact free to set the deformation vector $p$ in (\ref{loc78}) to zero from the outset.

As expected, the representation (\ref{rBG30}) of the 3-point amplitude integrates to zero in dimensional regularization: the scale-free bubble integral vanishes by cancellation between infrared and ultraviolet divergences, and the triangle contributions with tensor structure $\ell^m \rightarrow k^m_j$  vanish upon integration, on kinematic grounds. While the main emphasis of this work is  on  integrands and their systematic construction via gauge invariance and locality, we will also study the integrated expressions 
 as consistency checks.

\subsection{The gauge-invariant form of the 3-point amplitude}
\label{sect33}

By the minahaning prescription (\ref{loc78}) explained in the previous section, the bubble contributions $\frac{ T_{12,3}}{s_{12}}$ are well-defined expressions (\ref{loc79}) in the 3-particle phase space. With the choice of triangle numerator in (\ref{loc75}), the 3-point integrand in (\ref{BG30}) is gauge-invariant due to interplay of the triangle with the bubbles. Alternatively, we can make gauge invariance manifest at the level of individual diagrams, as follows. Eliminate the second and third bubble in (\ref{BG30}) by adding
\beq
0= \frac{ T_{12,3}}{s_{12}\ell^2 \ell_1^2 \ell_{12}^2} \Big[ \ell_{12}^2 - \ell_1^2 + 2 (\ell \cdot k_2) \Big]
+ \frac{ T_{13,2}}{s_{13}\ell^2 \ell_1^2 \ell_{12}^2} \Big[ \ell^2 - \ell_{12}^2 + 2 (\ell \cdot k_3) \Big]
\label{loc81}
\eeq
to the integrand, setting $s_{12}=0$ in the brackets $[\ldots]$. With the definitions (\ref{BG17}) and (\ref{BG18}) of the scalar and vectorial gauge-invariants $C_{1|23}$ and $C_{1|2,3}^m$, we arrive at the alternative representation
\beq
A^{\te{1-loop}}_{{\cal N}=2}(1,2,3) = \int \frac{ \dd^D\ell }{(2\pi)^D} \, \bigg\{ \frac{ C_{1|23} }{\ell^2 \ell_{1}^2 } +
 \frac{2 \ell_m  C^m_{1|2,3}+ s_{23} C_{1|23} }{ \ell^2 \ell_{1}^2 \ell_{12}^2 } \bigg\}  
\label{BG30a}
\eeq
with manifest gauge invariance. (We also included the vanishing scalar triangle $ s_{23} C_{1|23}$ to make contact with the maximally supersymmetric pentagon in (5.5) of \cite{Mafra:2014gja}.) To compare (\ref{BG30a}) with
the manifestly local expression (\ref{rBG30}), we write out polarizations and momenta:
\begin{align}
A^{\te{1-loop}}_{{\cal N}=2}(1,2,3) &= \int \frac{ \dd^D\ell }{(2\pi)^D} \, \bigg\{ \frac{ (e_2 \cdot e_3)(k_2 \cdot e_1)+ (e_1 \cdot e_3)(k_3 \cdot e_2)+ (e_1 \cdot e_2)(k_1 \cdot e_3)}{\ell^2 \ell_{1}^2 }   \notag \\
&\! \! \! \! \! \! \! \! \! \! \! \! \! \! \! \! \! \! \! \! +
 \frac{2 \ell_m \big[
e_1^m(k_2 \cdot e_3)(k_3 \cdot e_2) +    
e_2^m(k_3 \cdot e_1)(k_1 \cdot e_3) +    
e_3^m(k_1 \cdot e_2)(k_2 \cdot e_1) 
 \big] }{ \ell^2 \ell_{1}^2 \ell_{12}^2 } 
\label{BG30b}  \\
&\! \! \! \! \! \! \! \! \! \! \! \! \! \! \! \! \! \! \! \! +
 \frac{2 \ell_m \big[
k_2^m (e_1\cdot e_2)(k_1 \cdot e_3)+
k_3^m (e_1\cdot e_3)(k_1 \cdot e_2)    +
  i\epsilon^{m}(e_1,k_2,e_2,k_3,e_3)  
 \big] }{ \ell^2 \ell_{1}^2 \ell_{12}^2 } \bigg\}  \ .
\notag
\end{align}
We see that the gauge-invariant form (\ref{BG30b}) of the 3-point amplitude happens to also have manifest locality, but
as emphasized earlier, amplitudes at higher multiplicity generically exhibit a tension between locality and gauge invariance. At 4 points, for instance, the gauge-invariant triangle ``numerators'' such as $\ell_m C^{m}_{1|23,4}$ that will appear in section \ref{sect35} involve kinematic poles (say $ s_{12}^{-1}$) that do not match the propagator structure of the triangle diagram under discussion (say $(s_{23} \ell^2 \ell_1^2 \ell_{123}^2)^{-1}$ along with $\ell_m C^{m}_{1|23,4}$). When all diagrams of the amplitude (\ref{cubgrph}) are assembled, those superficially non-local contributions will collapse to local expressions, as is guaranteed from the manifestly local starting point of our construction.

The gauge-invariant bubble coefficient in (\ref{BG30b}) can be recognized as the 3-point tree,
\beq
C_{1|23} = (e_2 \cdot e_3)(k_2 \cdot e_1)+ (e_1 \cdot e_3)(k_3 \cdot e_2)+ (e_1 \cdot e_2)(k_1 \cdot e_3) = A^{\te{tree}}(1,2,3) \ .
\label{loc82}
\eeq
It arises as the leading UV-divergence of (\ref{BG30a}) when performing the $\dd^D \ell$-integral in (\ref{BG30b}) in $D\geq 4$ dimensions, for which we introduce the shorthand $|_{\rm UV}$, as in
\beq
A^{\te{1-loop}}_{{\cal N}=2}(1,2,3) \, \big|_{\te{UV}} = C_{1|23}  = A^{\te{tree}}(1,2,3) \ .
\label{UV3}
\eeq

\subsection{The local form of the 4-point amplitude}
\label{sect34}

The 4-point analogue of the ansatz (\ref{BG30}) in terms of cubic diagrams without tadpoles reads
\begin{align}
A^{\te{1-loop}}_{{\cal N}=2}&(1,2,3,4) =  \int \frac{ \dd^D\ell }{(2\pi)^D} \, \bigg\{
\frac{ M_{12,34} }{ \ell^2 \ell_{12}^2} + \frac{ M_{41,23} }{ \ell_{1}^2 \ell_{123}^2} + \frac{M_{123,4}}{ \ell^2 \ell_{123}^2}
+ \frac{M_{1,234}}{ \ell^2 \ell_{1}^2} + \frac{M_{341,2}}{ \ell^2_{12} \ell_{1}^2} + \frac{M_{412,3}}{ \ell^2_{12} \ell_{123}^2}  
 \notag \\
& \ \ \ \ \ \ \ \ +\frac{ N_{12|3,4}(\ell)}{s_{12} \ell^2 \ell_{12}^2 \ell_{123}^2}+ \frac{ N_{1|23,4}(\ell) }{ s_{23} \ell^2 \ell_{1}^2   \ell_{123}^2} +  
\frac{ N_{1|2,34}(\ell)}{s_{34}\ell^2 \ell_{1}^2 \ell_{12}^2 } + \frac{ N_{41|2,3}(\ell)}{s_{14} \ell_{1}^2 \ell_{12}^2 \ell_{123}^2} +  \frac{ N^{\te{box}}_{1|2,3,4}(\ell) }{\ell^2 \ell_{1}^2 \ell_{12}^2 \ell_{123}^2} \bigg\} \ .
\label{4ptans}
\end{align}
The propagators in the first line are associated with bubble diagrams,
and the numerators $N_{A|B,C}(\ell)$ and $N^{\te{box}}_{1|2,3,4}(\ell) $ of the triangles and the box, respectively, will be inferred from gauge invariance.

\subsubsection{Bubbles}

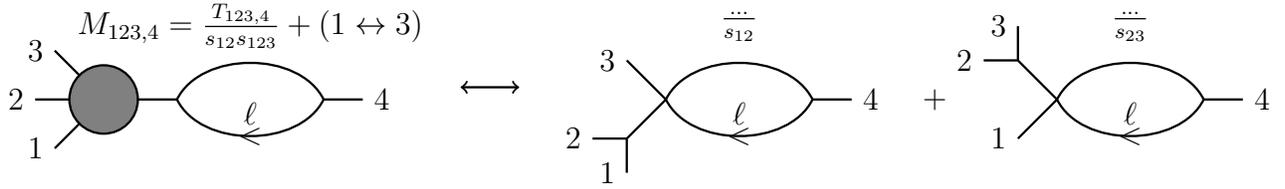
\begin{figure}[h]
\begin{center}
\begin{tikzpicture} [scale=0.65, line width=0.30mm]
\draw (2,0.5) -- (1.5,1) node[left]{$3$};
\draw (1.8,0) -- (1.1,0) node[left]{$2$};
\draw (2,-0.5) -- (1.5,-1) node[left]{$1$};
\draw[fill=gray] (2.5,0) circle(0.7cm);
\draw (3.2,0) -- (4,0);
\draw (4,0) .. controls (4.5,1) and (6.5,1) .. (7,0);
\draw (4,0) .. controls (4.5,-1) and (6.5,-1) .. (7,0);
\draw (5.5,-0.75)node{$<$}node[above]{$\ell$};
\draw (7,0) -- (7.8,0) node[right]{$4$};
\draw (5.5,1.5)node{$M_{123,4} = \frac{ T_{123,4} }{s_{12}s_{123}} + (1\leftrightarrow 3)$};
\draw[<->] (9.8,0.25)--(11.0,0.25);
\scope[xshift=18cm]
\draw (3.2,0.8) -- (3.2,1.5) node[left]{$3$};
\draw (3.2,0.8) -- (2.5,0.8) node[left]{$2$};
\draw (4,0) -- (3.2,0.8);
\draw (4,0) -- (3.2,-0.8) node[left]{$1$};
\draw (4,0) .. controls (4.5,1) and (6.5,1) .. (7,0);
\draw (4,0) .. controls (4.5,-1) and (6.5,-1) .. (7,0);
\draw (5.5,-0.75)node{$<$}node[above]{$\ell$};
\draw (7,0) -- (7.8,0) node[right]{$4$};
\draw (5.5,1.5)node{$\frac{ \ldots }{s_{23}} $};
\endscope
\scope[xshift=10cm]
\draw (3.2,-0.8) -- (3.2,-1.5) node[left]{$1$};
\draw (3.2,-0.8) -- (2.5,-0.8) node[left]{$2$};
\draw (4,0) -- (3.2,-0.8);
\draw (4,0) -- (3.2,0.8) node[left]{$3$};
\draw (4,0) .. controls (4.5,1) and (6.5,1) .. (7,0);
\draw (4,0) .. controls (4.5,-1) and (6.5,-1) .. (7,0);
\draw (5.5,-0.75)node{$<$}node[above]{$\ell$};
\draw (7,0) -- (7.8,0) node[right]{$4$};
\draw (5.5,1.5)node{$\frac{  \ldots }{s_{12}} $};
\draw (9.5,0)node{$+$};
\endscope
\end{tikzpicture}
\caption{The kinematic factors $T_{123,4}$ and $T_{321,4}$ of an external bubble compensate the singular propagator $s_{123}^{-1}$ upon minahaning. The leftover poles in $s_{12}$ and $s_{23}$ correspond to tree-level subdiagrams whose local numerators in the ellipsis can be assembled from~(\ref{red49}).
}
\label{f:4bub}
\end{center}
\end{figure}

Our ansatz for the bubbles in the 4-point amplitude (\ref{4ptans}) is again based on the scalar building block $T_{A,B}$ in (\ref{BG35}). At 4 points, the tree-level subgraphs connected to an external bubble as depicted in figure \ref{f:4bub} involve 3 external legs and 2 propagators. 
There are two pole channels $\sim(s_{12}s_{123})^{-1}$ and $\sim(s_{23}s_{123})^{-1}$ admitted by the cyclic ordering,
that are combined into the Berends--Giele current $f^{mn}_{123} \rightarrow \ffrak^{mn}_{123}$ in (\ref{BG81}). This is why the bubble kinematics in (\ref{4ptans}) is chosen as ${\cal T}_{A,B}=M_{A,B}$, see (\ref{uni}).

Similarly to our previous discussion around (\ref{loc79}), the propagators of an external bubble include a 3-particle Mandelstam invariant $s_{123}$ that naively vanishes in 4-particle phase space, but 
this is compensated by a zero of the numerators $T_{123,4}$ and $T_{321,4}$. In more detail, we extend the minahaning procedure
from section \ref{sect32} to 4 points: a lightlike deformation momentum $\sum_{j=1}^4 k_j^m = p^m$ amounts to using no Mandelstam identity other than $\sum_{i<j}^4 s_{ij}=0$. Given the cancellation of $\frac{ s_{12}+s_{13}+s_{23}}{s_{123}}=1=-\frac{ s_{14}+s_{24}+s_{34}}{s_{123}}$ in intermediate steps (which can also be found in \cite{Johansson:2014zca} in the context of the same diagrams), we ultimately arrive at finite expressions like \cite{Berg:2016wux}
\begin{align}
M_{123,4} &=   (e_1\cdot e_3)(e_2\cdot e_4) - \frac{1}{2}(e_1\cdot e_2)(e_3\cdot e_4)-\frac{1}{2}(e_1\cdot e_4)(e_2\cdot e_3) \notag \\
&\ \ \ \ + (e_1\cdot e_2) \Big[ \frac{ (k_2\cdot e_3) - (k_1\cdot e_3) }{2s_{12}}(k_3\cdot e_4) - \frac{ (k_1\cdot e_4) (k_2\cdot e_3)}{s_{23}} \Big]   \notag \\
&\ \ \ \ + (e_2\cdot e_3) \Big[ \frac{ (k_2\cdot e_1) - (k_3\cdot e_1) }{2 s_{23}}(k_1\cdot e_4) - \frac{ (k_2\cdot e_1) (k_3\cdot e_4)}{s_{12}} \Big] \notag \\
&\ \ \ \ + (e_1\cdot e_3) \Big[ \frac{ (k_1\cdot e_2)(k_3\cdot e_4) }{s_{12}} +  \frac{ (k_1\cdot e_4)(k_3\cdot e_2) }{s_{23}}  \Big]\ , \label{red49}
\end{align} 
that still exhibit the 2-particle propagators $s_{12}^{-1}, s_{23}^{-1}$ and can be visualized through the collapsed propagators  in figure \ref{f:4bub}. The same kind of regularization procedure was instrumental for the 4-point 4-loop amplitude of ${\cal N}=4$ SYM \cite{Bern:2012uf}.\footnote{Another related topic is the ''cancelled propagator
argument" in string theory (see for example fig.\ 9.9 in \cite{Polchinski:1998rq}): an external bubble in 
a gauge-boson amplitude at 1-loop could cause a mass shift in the effective action, which would  naively
interfere with gauge invariance. For recent related work, see e.g.\ \cite{Sen:2016gqt} and references therein.}

The remaining bubble topology with 2 external legs on each side
and propagators of the form $\sim (s_{12} s_{34})^{-1}$ does not need regularization. Still, one can identify a global prefactor of $s_{12}=s_{34}$ in the numerator $T_{12,34}$, cancelling one of the propagators \cite{Berg:2016wux}, 
\begin{align}
M_{12,34} &= \frac{1}{s_{12}} \Big[ s_{12}(e_1\cdot e_4)(e_2\cdot e_3)-s_{12}(e_1\cdot e_3)(e_2\cdot e_4) +(s_{13}-s_{23})(e_1\cdot e_2)(e_3\cdot e_4)    \label{red48} \\
&\!\!\!\!\! + (e_1\cdot e_2) \big( (k_1\cdot e_3)(k_2\cdot e_4) - (k_2\cdot e_3)(k_1\cdot e_4) \big)  + (e_3\cdot e_4) \big( (k_4\cdot e_2)(k_3\cdot e_1) - (k_4\cdot e_1)(k_3\cdot e_2) \big) 
\Big] \ .
\notag
   \end{align}
Upon insertion into (\ref{4ptans}), this pinpoints the bubble contributions to the 4-point amplitude in its local form. Their gauge variation
\begin{align}
&
\delta\left(
\frac{ M_{12,34} }{ \ell^2 \ell_{12}^2} + \frac{ M_{41,23} }{ \ell_{1}^2 \ell_{123}^2} + \frac{M_{123,4}}{ \ell^2 \ell_{123}^2}
+ \frac{M_{1,234}}{ \ell^2 \ell_{1}^2} + \frac{M_{341,2}}{ \ell^2_{12} \ell_{1}^2} + \frac{M_{412,3}}{ \ell^2_{12} \ell_{123}^2}\right)\notag \\
& \hspace{1cm} = \ 
\frac{ 
\omega_{12} T_{3,4} ( \ell_{12}^2-\ell^2 )+ \omega_3 T_{12,4} (  \ell_{123}^2-\ell_{12}^2) + \omega_{4} T_{12,3}( \ell^2-\ell_{123}^2 )
}{s_{12} \ell^2 \ell_{12}^2 \ell_{123}^2} \notag \\
&\hspace{1.5cm} + \
\frac{ 
\omega_{23} T_{1,4} (\ell_{123}^2-\ell^2_1 )+ \omega_1 T_{23,4} ( \ell_{1}^2-\ell^2 ) + \omega_{4} T_{23,1}(\ell^2-\ell_{123}^2 ) 
}{ s_{23} \ell^2 \ell_{1}^2   \ell_{123}^2} \label{boxvar} \\
&\hspace{1.5cm} + \
 \frac{
\omega_{34} T_{1,2} (\ell^2-\ell^2_{12} )+ \omega_1 T_{2,34} (\ell_{1}^2-\ell^2 ) + \omega_{2} T_{1,34}( \ell_{12}^2-\ell_{1}^2)
}{s_{34}\ell^2 \ell_{1}^2 \ell_{12}^2 } \notag  \\
&\hspace{1.5cm} + \
\frac{ 
\omega_{14}T_{2,3}(\ell_{123}^2 - \ell_{1}^2)+\omega_2T_{41,3}(\ell_{12}^2-\ell_{1}^2 )+\omega_3T_{41,2}( \ell_{123}^2-\ell_{12}^2) }{s_{14} \ell_{1}^2 \ell_{12}^2 \ell_{123}^2} \,,
\notag
\end{align}
is compatible with (\ref{loc71}) and takes the right form to conspire with the triangle diagrams.

\subsubsection{Parity-even triangles}

\begin{figure}[h]
\begin{center}
\begin{tikzpicture} [scale=0.8, line width=0.30mm]
\draw (0,0) -- (-1,1) node[left]{$1$};
\draw (0,0) -- (-1,-1) node[left]{$4$};
\draw (0,0) -- (1,0);
\draw (2.7,1)--(2.7,-1);
\draw(2.7,1) -- (3.1,1.6)node[right]{$2$};
\draw(2.7,-1) -- (3.1,-1.6)node[right]{$3$};
\draw[->](2.7,-1)--(1.85,-0.5);
\draw(1,0)--(1.85,-0.5);
\draw(1.5,-0.9)node{$\ell+k_4$};
\draw(2.7,1)--(1.85,0.5);
\draw[->](1,0)--(1.85,0.5);
\draw(1.5,0.9)node{$\ell-k_1$};
\end{tikzpicture}
\caption{The ``special'' triangle diagram where the vector part $\sim T^m_{41,2,3}$ of the numerator contracts with the shifted loop-momentum $\ell \rightarrow \ell + k_4$.}
\label{f:tri}
\end{center}
\end{figure}
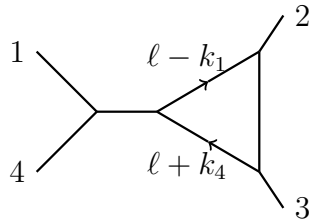

By analogy with the 3-point expression (\ref{loc75}), the numerators of the 4 triangle diagrams in (\ref{4ptans}) will have both parity-even and parity-odd contributions, to be denoted by $N^{\rm even}_{12|3,4}(\ell) $ and $N^{\rm odd}_{12|3,4}(\ell) $, respectively:
\beq
N_{A|B,C}(\ell) = N^{\rm even}_{A|B,C}(\ell)  + N^{\rm odd}_{A|B,C}(\ell) 
\label{evodd}
\eeq
The requirement (\ref{loc71}) on the triangle numerators' gauge variation interlocks the scalar and vectorial parts. For instance, the parity-even expressions (with $T^m_{A,B,C}$ defined by (\ref{BG35a})) 
\begin{align}
N^{\rm even}_{12|3,4}(\ell) &= 2\ell_m T^m_{12,3,4} + T_{123,4} + T_{124,3} + T_{12,34} 
\label{exota}
\\
N^{\rm even}_{1|23,4}(\ell) &= 2\ell_m T^m_{1,23,4} - T_{231,4} + T_{1,234} + T_{14,23}
\\
N^{\rm even}_{1|2,34}(\ell) &= 2 \ell_m T^m_{1,2,34} + T_{12,34} - T_{341,2} - T_{342,1} 
\\
N^{\rm even}_{41|2,3}(\ell) &= 2(\ell_m + k_m^4) T^m_{41,2,3} + T_{412,3} + T_{413,2} + T_{41,23} - s_{14} \big[ e_4^m T^m_{1,2,3} + (e_1\cdot e_4) T_{2,3}
\big]
\label{exot}
\end{align}
extending the pattern of (\ref{loc75}),
provide the right interplay between scalar and vector contributions  to produce differences of the inverse propagators $\ell^2_{12\ldots j}$ and $s_{ij}$ in their gauge variation:
\begin{align}
\delta N^{\rm even}_{12|3,4}(\ell)&= s_{12} \big[ \omega_1 N^{\rm even}_{2,3,4}(\ell) + \omega_{13} T_{2,4} + \omega_{14} T_{2,3} - \omega_2 N^{\rm even}_{1,3,4}(\ell)- \omega_{23} T_{1,4} - \omega_{24} T_{1,3} \big]  \notag \\
& \ \ \ \ + \omega_{12} T_{3,4} (\ell^2 - \ell_{12}^2)
+ \omega_3 T_{12,4} (\ell_{12}^2 - \ell_{123}^2) + \omega_{4} T_{12,3}(\ell_{123}^2 - \ell^2) \ ,
\notag \\
\delta N^{\rm even}_{1|23,4}(\ell)&=s_{23} \big[ \omega_2 N^{\rm even}_{1,3,4} (\ell) + \omega_{12} T_{3,4} + \omega_{24} T_{1,3}  - \omega_3 N^{\rm even}_{1,2,4} (\ell) - \omega_{13} T_{2,4} - \omega_{34} T_{1,2} \big]  \notag \\
& \ \ \ \ + \omega_{23} T_{1,4} (\ell^2_1 - \ell_{123}^2)
+ \omega_1 T_{23,4} (\ell^2 - \ell_{1}^2) + \omega_{4} T_{23,1}(\ell_{123}^2 - \ell^2) \ ,\label{BG38} \\
\delta N^{\rm even}_{1|2,34}(\ell)&=s_{34} \big[ \omega_3 N^{\rm even}_{1,2,4} (\ell) + \omega_{13} T_{2,4} + \omega_{23} T_{1,4}  - \omega_4 N^{\rm even}_{1,2,3}(\ell) - \omega_{14} T_{2,3} - \omega_{24} T_{1,3} \big]  \notag \\
& \ \ \ \ + \omega_{34} T_{1,2} (\ell^2_{12} - \ell^2)
+ \omega_1 T_{2,34} (\ell^2 - \ell_{1}^2) + \omega_{2} T_{1,34}(\ell_{1}^2 - \ell_{12}^2) \ .
\notag
\end{align}
The shorthand $N^{\rm even}_{i,j,k}(\ell)$ on the right-hand sides refers to the parity-even triangle numerator in (\ref{loc75}):
\beq
N^{\rm even}_{i,j,k}(\ell) \equiv
2\ell_m T^m_{i,j,k} + T_{ij,k} + T_{i,jk} + T_{ik,j} \ .
\label{offtri}
\eeq
The exceptional terms $2k_4^m T^m_{41,2,3}-s_{14}e_4^m T^m_{1,2,3} - s_{14} (e_1 \cdot e_4)T_{2,3}$ in the fourth triangle numerator (\ref{exot}) without any counterparts in  $N^{\rm even}_{12|3,4}(\ell) ,N^{\rm even}_{1|23,4}(\ell) $ and $N^{\rm even}_{1|2,34}(\ell) $ can be justified as follows: The gauge variation of the naive ansatz $2\ell_m  T^m_{41,2,3} + T_{412,3} + T_{413,2} + T_{41,23}$ for $N^{\rm even}_{41|2,3}(\ell) $,
\begin{align}
 \delta&\big[ 2\ell_m  T^m_{41,2,3} + T_{412,3} + T_{413,2} + T_{41,23} \big] = 2\omega_{41}T_{2,3}\big[ (\ell \cdot k_{14}) - s_{14} \big] + 2 \omega_2 T_{41,3} \big[ (\ell \cdot k_{2}) + s_{14} \big]  \notag
 \\
&+ 2 \omega_3 T_{41,2}(\ell \cdot k_3)  + s_{14} \big[ \omega_{42} T_{1,3} + \omega_{43} T_{1,2} - \omega_{12}T_{3,4} - \omega_{13} T_{2,4} + \omega_4 N^{\rm even}_{1,2,3}(\ell) - \omega_1 N^{\rm even}_{4,2,3} (\ell)\big] \label{BG38d}  
\end{align}
does not satisfy the necessary condition (\ref{loc71}) for gauge invariance. However, the addition of $2k_4^m T^m_{41,2,3}$
is easily seen to complete the first line of (\ref{BG38d}) to be expressible via differences of $\ell^2_{12\ldots j}$ and can be motivated by a diagrammatic argument: In our conventions for the shifts of the integration variable, $\ell$ is the momentum in the $n$-gon edge between the external legs 1 and $n$. For the ``special'' triangle graph with legs 1 and 4 forming a tree-level subdiagram, the momentum in the analogous adjacent edge is $\ell+k_4$ rather than $\ell$, see figure \ref{f:tri}. Hence, it is not surprising that our analysis driven by gauge invariance points towards a numerator of the form $N^{\rm even}_{41|2,3}(\ell) = 2(\ell_m+k_m^4) T^{m}_{41,2,3}+\ldots$. The remaining terms of the ellipsis --- specifically the subtraction of $s_{14} \big[ e_4^m T^m_{1,2,3} + (e_1\cdot e_4) T_{2,3} \big]$ --- can be inferred by demanding the variation to follow the structure of (\ref{BG38}),
\begin{align}
 \delta N^{\rm even}_{41|2,3}(\ell) &= s_{41} \big[ \omega_4 N^{\rm even}_{1,2,3}(\ell)  -  \omega_1 N^{\rm even}_{2,3,4}(\ell) +  \omega_{24} T_{1,3}+ \omega_{34} T_{1,2}  - \omega_{12} T_{3,4} - \omega_{13} T_{2,4} \big] \notag \\
 & \ \ \ \ + \omega_{41} T_{2,3} (\ell_{123}^2 - \ell_{1}^2) + \omega_2 T_{41,3} (\ell_{1}^2 - \ell_{12}^2) + \omega_3 T_{41,2} (\ell_{12}^2 - \ell_{123}^2) \ .
\label{BG39}
\end{align}

The appearance of $\ell^2_{12\ldots j}$ in the triangles' gauge variation cancels the contribution (\ref{boxvar}) from the bubbles. The remaining terms $\sim s_{ij}$ in the above $\delta N^{\rm even}_{A|B,C}$ need to conspire with the box graph. In particular, the sign change of $\omega_{24} T_{1,3}+ \omega_{34} T_{1,2}$ and the conversion $N^{\rm even}_{4,2,3} (\ell) \rightarrow N^{\rm even}_{2,3,4} (\ell)$ going from (\ref{BG38d}) to (\ref{BG39}) is essential to render the desired gauge variation of the box numerator linear in $\ell$: Only the relative minus sign in $\ell_{12\ldots j}^2 - \ell_{12\ldots j-1}^2$ makes the quadratic piece $\sim \ell^2$ disappear.

\subsubsection{Parity-odd triangles}

The parity-odd part of the triangles can be reconstructed by demanding all the $N_{A|B,C}(\ell)$ to be expressible in terms of the string-theory motivated ``parity-(odd$+$even)'' building blocks $M^m_{A|B,C}$ and ${\cal J}_{1|i|j,k}$ defined in (\ref{uni}) and (\ref{Jdef}), respectively, 
\begin{align}
N^{\rm odd}_{12|3,4} (\ell)&= 2\ell_m E^m_{12|3,4} \co N^{\rm odd}_{1|23,4} (\ell)= 2\ell_m E^m_{1|23,4} \co N^{\rm odd}_{1|2,34}(\ell) = 2\ell_m E^m_{1|2,34} \notag
\\
N^{\rm odd}_{41|2,3} (\ell)&=2(\ell_m+k_m^4) E^m_{41|2,3}  - 2 s_{41} e_4^m E^m_{1|2,3} \ ,
\label{odd41}
\end{align}
see (\ref{locG09}) for the definition of $E^m_{A|B,C}$.
For example, combining (\ref{odd41}) with (\ref{exot}) leads to the following expression for the special triangle numerator,
\beq
N_{41|2,3} (\ell)=2(\ell_m+k_m^4) s_{41} M^m_{41|2,3}  - 2 s_{41} {\cal J}_{1|4|2,3}+ T_{412,3} + T_{413,2} + T_{41,23} \ ,
\eeq 
which shares its structure with the maximally supersymmetric pentagon numerator in section 4.4.2 of \cite{Mafra:2014gja}, where the tree-level subdiagram involving legs 1 and 6 singles out $\ell+k_6$ in the same manner. The overall parity-odd gauge variation of the triangles (\ref{odd41}) is given by 
\begin{align}
&\delta\left(\frac{ N^{\rm odd}_{12|3,4} (\ell) }{s_{12} \ell^2 \ell_{12}^2 \ell_{123}^2} + \frac{N^{\rm odd}_{1|23,4} (\ell)}{ s_{23} \ell^2 \ell_{1}^2 \ell_{123}^2} + \frac{ N^{\rm odd}_{1|2,34}(\ell) }{  s_{34} \ell^2 \ell_1^2 \ell_{12}^2}
+ \frac{N^{\rm odd}_{41|2,3} (\ell) }{s_{14} \ell_1^2 \ell_{12}^2 \ell_{123}^2 } \right) = \frac{2 }{\ell^2 \ell_1^2 \ell_{12}^2 \ell_{123}^2}  \notag \\
& \ \ \ \ \times \Big[
 \ell_m (\omega_1 E_{2|3,4}^m - \omega_2 E_{1|3,4}^m)(\ell^2_1 - \ell^2_{12})
+ \ell_m (\omega_1 E_{3|2,4}^m - \omega_3 E_{1|2,4}^m)(\ell^2_{12} - \ell^2_{123}) \notag \\
& \ \ \ \ \ \ \ \ 
+ \ell_m (\omega_1 E_{4|2,3}^m - \omega_4 E_{1|2,3}^m)(\ell^2_{123} - \ell^2) 
 -2 \ell^2 \omega_1 k_4^m E^m_{4|2,3}
\Big] \ ,
\label{doddtri}
\end{align}
where the last term $\sim  -2 \ell^2 \omega_1 k_4^m E^m_{4|2,3}$ will be seen to play an important role for the 6-dimensional gauge anomaly.

\subsubsection{The box numerator}

The $\ell$-dependent part of the box numerator can be readily written down by promoting the constituents of the triangle in (\ref{loc75}) to have another free vector index (while extending the combinatorics to 4 legs),
\begin{align}
N^{\te{box}}_{1|2,3,4}(\ell)  &= 2 \ell_m \ell_n M^{mn}_{1|2,3,4} + 2 \ell_m \big[ s_{12}M^m_{12|3,4} + s_{23}M^m_{1|23,4} + {\rm cyc}(2,3,4) \big] + N^{\te{scal}}_{1|2,3,4}
\label{BG360}  \ ,
\end{align}
see (\ref{uni}) for the tensor building block. Assuming that the scalar contributions $ N^{\te{scal}}_{1|2,3,4}$ to the box numerator are parity-even, we can extract the entire parity-odd gauge variation from (\ref{BG360}) and find
\begin{align}
\delta N^{\te{box}}_{1|2,3,4}(\ell) \, \big|_{\rm odd} &= 2\ell^2 \omega_1 Y_{2,3,4} + 
2\ell_m (\omega_2 E_{1|3,4}^m - \omega_1 E_{2|3,4}^m)(\ell^2_1 - \ell^2_{12})
 \label{BG350}  \\
&\! \! \! \! \! \! \! \! \! \! \! \! \! + 2\ell_m (\omega_3 E_{1|2,4}^m - \omega_1 E_{3|2,4}^m)(\ell^2_{12} - \ell^2_{123})
+ 2\ell_m (\omega_4 E_{1|2,3}^m - \omega_1 E_{4|2,3}^m)(\ell^2_{123} - \ell^2) 
\notag \ ,
\end{align}
using (\ref{locG12}) and (\ref{locG09}) for $\delta E^{mn}_{1|2,3,4}$ and $\delta E^m_{A|B,C}$, respectively. With the relation
\beq
Y_{2,3,4}= 2i \epsilon(k_2,e_2,k_3,e_3,k_4,e_4)=2k_4^mE^m_{4|2,3} \ ,
\label{YYY}
\eeq
it is easy to verify that (\ref{BG350}) naively cancels the entire parity-odd variation of the triangles in (\ref{doddtri}), but in the next section we will see the expected box anomaly. It remains to find a scalar completion $N^{\te{scal}}_{1|2,3,4}$ of the $\ell$-dependent parity-even building blocks $\ell_m T^m_{A,B,C}$ and $\ell_m \ell_n T^{mn}_{1,2,3,4}$ in (\ref{BG360}). The expression for $N^{\te{scal}}_{1|2,3,4}$ will be designed to cancel the gauge variations (\ref{BG38}) and (\ref{BG39}) of the triangles which are not yet accounted for by the bubbles (\ref{boxvar}):
\begin{align}
\delta N^{\te{box}}_{1|2,3,4}(\ell) \, \big|_{\rm even} &= \omega_1  N^{\rm even}_{2,3,4} (\ell)(\ell^2 - \ell_{1}^2)
+ \omega_2 N^{\rm even}_{1,3,4}(\ell) (\ell^2_{1} - \ell_{12}^2)
+ \omega_3  N^{\rm even}_{1,2,4}(\ell )(\ell^2_{12} - \ell_{123}^2) \notag \\
&
+ \omega_4  N^{\rm even}_{1,2,3} (\ell)(\ell^2_{123} - \ell^2) + ( \omega_{13} T_{2,4} - \omega_{24} T_{1,3}) (\ell^2 - \ell_{1}^2 + \ell_{12}^2 - \ell_{123}^2)
\label{BG37} \\
&+(\omega_{12}T_{3,4} - \omega_{34} T_{1,2}) (\ell^2 - \ell_{12}^2) +  (\omega_{23}T_{1,4} - \omega_{14} T_{2,3}) (\ell_{1}^2 - \ell_{123}^2) \ . \notag 
\end{align}
One can check that this is accomplished by the following local expression for the scalar box in (\ref{BG360}):
\begin{align}
N^{\te{scal}}_{1|2,3,4} &=  T_{12,34} + T_{13,24} + T_{14,23} + \frac{2}{3} \big[ T_{123,4}+ T_{321,4} + T_{234,1} + T_{432,1} + T_{134,2} + T_{431,2} + T_{124,3} + T_{421,3} \big]
\notag \\
&  + \frac{1}{3} \big[ (k_2^m {-} k_1^m) T^m_{12,3,4} + (k_3^m {-} k_2^m) T^m_{1,23,4} +{\rm cyc}(2,3,4) \big] - \frac{1}{6} \, T^{mn}_{1,2,3,4} \big[ k_1^m k_1^n + (1\leftrightarrow 2,3,4) \big]  \ . \label{BG361}
\end{align}
In analogy with the maximally supersymmetric hexagon numerator in section 4.4.3 of \cite{Mafra:2014gja}, the scalar part (\ref{BG361}) of the box $N^{\te{box}}_{1|i,j,k}(\ell)$ involves a combination of $T_{A,B}$ which depends on the ordering $i,j,k$. The second line\footnote{Note that the gauge variation of the second line of (\ref{BG361}) is given by
\[
\frac{1}{3} \big[ T_{12,4}\omega_3(s_{23}-s_{13}) + T_{12,3} \omega_4 (s_{24}-s_{14}) \big] + (12\leftrightarrow 13,14,23,24,34)
\ .
\]
}, on the other hand, is permutation invariant in 2,3,4, and the parity-odd contributions cancel when representing the last line via $T^m_{1i,j,k}\rightarrow s_{1i} M^m_{1i|j,k}, \ 
T^m_{1,ij,k}\rightarrow s_{ij} M^m_{1|ij,k}$ and
$T^{mn}_{1,2,3,4}\rightarrow M^{mn}_{1|2,3,4}$ in terms of the ``parity-(odd$+$even)''  building blocks (\ref{uni}).

\subsubsection{The box anomaly}

The anomaly kinematics $Y_{2,3,4}=2i \epsilon(k_2,e_2,k_3,e_3,k_4,e_4)$ from both $\delta N_{41|2,3}(\ell)$ and $\delta N^{\te{box}}_{1|2,3,4}(\ell)$ deserves particular attention since this is where the tensor trace $\delta(\ell_m \ell_n M_{1|2,3,4}^{mn}) =\ell^2 \omega_1 Y_{2,3,4} + \ldots$ conspires with the special triangle with $k_{41}$ in a massive corner (i.e.\ where the $\ell^{-2}$ propagator is absent):
\beq
\delta A^{\te{1-loop}}_{{\cal N}=2}(1,2,3,4) = 2 \omega_1 Y_{2,3,4}  \int \frac{ \dd^D\ell }{(2\pi)^D} \, \bigg\{-
 \frac{ 1}{ \ell_{1}^2 \ell_{12}^2 \ell_{123}^2} +  \frac{ \eta_{mn} \ell^m \ell^n }{\ell^2 \ell_{1}^2 \ell_{12}^2 \ell_{123}^2} \bigg\} \ .
\label{anom0}
\eeq
Naively, one would be tempted to set (\ref{anom0}) to zero since the integrand appears to vanish. As is well known, 
dimensional regularization reveals a logarithmic divergence that requires a refined analysis. One can show  with conventional (see e.g.\ \cite{Bern:1995db} or section 5.1 of \cite{Chen:2014eva}) or worldline techniques (see e.g.\ section 4.5 of \cite{Mafra:2014gja})   that tensor $n$-gon integrals in $D=2n-2$ dimensions give rise to the following rational terms
\beq
\int \dd^D\ell \, \bigg\{
\frac{ \eta_{pq} \ell^p \ell^q }{\ell^2 (\ell-k_1)^2   \ldots (\ell-k_{12\ldots n-1})^2} 
\, -\, \frac{1}{(\ell-k_1)^2  \ldots (\ell-k_{12\ldots n-1})^2} 
\bigg\}  \, \Big|_{D=2n-2}
= \frac{   \pi^{n-1}  }{(n-1)!} \ ,
\label{anom}
\eeq
when combined with an appropriate scalar $(n{-}1)$-gon. Hence, we identify the following anomalous gauge variation in the above 4-point amplitude,
\beq
 A^{\te{1-loop}}_{{\cal N}=2}(1,2,3,4) \, \big|_{e_1 \rightarrow k_1}= \frac{ \pi^3 }{3(2\pi)^6} \,
Y_{2,3,4}
= \frac{i  }{96 \pi^3}\, \epsilon(k_2,e_2,k_3,e_3,k_4,e_4) \ .
\label{anom2}
\eeq
As can be ``discovered'' from the field-theory perspective (see e.g.\ \cite{Chen:2014eva}),
this anomaly can be cancelled by contributions due to additional fields in the gravitational sector,
once the relations between couplings in the gauge and gravitational sectors are suitably tuned. 
From the string-theory point of view, this is the Green-Schwarz mechanism
generalized to $D=6$ (see e.g.\  \cite{Anastasopoulos:2006cz}). 
The additional states are  $p$-form fields, possibly on collapsed cycles of the K3 orbifold,
but in string theory no couplings need to be adjusted:
the coupling relations suitable for anomaly cancellation arise
from the same open-string loop diagrams as those that gave rise to the anomaly
 (e.g.\ diagrams discussed in \cite{Berg:2016wux}, 
but in the long-cylinder limit instead of the field-theory limit discussed later in this paper). 

\subsection{The gauge-invariant form of the 4-point amplitude}
\label{sect35}

To make pseudo-invariance of the 4-point amplitude (\ref{4ptans}) manifest, one can repeat the procedure of section \ref{sect33} and perform algebraic rearrangements of the integrand similar to (\ref{loc81}). The guiding principle is to eliminate those cubic diagrams where the reference leg 1 is involved in a non-trivial tree-level subdiagram\footnote{This scheme of eliminating cubic diagrams descends from string theory, where integration by parts allows to eliminate the worldsheet origin of the associated propagator structures, see e.g.\ \cite{maxsusy, Berg:2016wux}.}, i.e.\ where either $\ell^2$ or $\ell_1^2$ is absent. However, the above discussion of the box anomaly suggests that the special triangle with propagators $\ell^2_1\ell^2_{12}\ell^2_{123}$ is an exception. In the process of these rearrangements, the kinematic building blocks $T_{A,B}$, $M^{m}_{A|B,C}$, $M^{mn}_{A|B,C,D}$ and ${\cal J}_{1|2|3,4}$ are assembled into the gauge (pseudo-)invariants $C_{1|A}$, $C^{m}_{1|A,B}$, $C^{mn}_{1|A,B,C}$ and $P_{1|2|3,4}$ introduced in section \ref{sect27}. By tedious but straightforward manipulations, one can show that the integrand of (\ref{4ptans}) agrees with
\begin{align}
A^{\te{1-loop}}_{{\cal N}=2}(1,2,3,4) &=  \int \frac{ \dd^D\ell }{(2\pi)^D} \, \bigg\{ \frac{ C_{1|234} }{\ell^2 \ell_{1}^2} + \frac{ 2\ell_m C^m_{1|23,4} {+} s_{34} C_{1|234} {-} s_{24} C_{1|324}}{\ell^2 \ell_{1}^2 \ell_{123}^2}+ \frac{ 2\ell_m C^m_{1|2,34} {+} s_{23} C_{1|432} {-} s_{24} C_{1|342}}{\ell^2 \ell_{1}^2 \ell_{12}^2}\notag \\
& - \frac{2 P_{1|4|2,3} }{\ell_{1}^2 \ell_{12}^2 \ell_{123}^2} + \frac{2 \ell_m \ell_n C^{mn}_{1,2,3,4} + 2 \ell_m (s_{23} C^m_{1|23,4}  +s_{24} C^m_{1|24,3}  +s_{34} C^m_{1|2,34}   )+ C_{1|2|3|4}^{\te{scal}} }{\ell^2 \ell_{1}^2 \ell_{12}^2 \ell_{123}^2}  \bigg\}
\label{BG27}
\end{align}
upon insertion of all the numerators, with the following shorthand for the gauge-invariant scalar box:
\begin{align}
C_{1|2|3|4}^{\te{scal}} &\equiv
\frac{1}{3}(4 s_{23} s_{34} C_{1|234} - 2s_{23} s_{24} C_{1|324} - 2s_{24} s_{34} C_{1|243}) -\frac{1}{6} C^{mn}_{1|2,3,4} \sum_{j=1}^4 k_j^m k_j^n \notag \\
&+\frac{1}{3} \big[ s_{23} (k_3^m - k_2^m) C_{1|23,4}^m + s_{24} (k_4^m - k_2^m) C_{1|24,3}^m + s_{34} (k_4^m - k_3^m) C_{1|34,2}^m \big] \ .
\label{BG28}
\end{align}
This representation generalizes the form (\ref{BG30a}) of the 3-point amplitude and confines the leading UV contribution to the single bubble in the first term such that, in analogy with (\ref{UV3}),
\beq
A^{\te{1-loop}}_{{\cal N}=2}(1,2,3,4) \, \big|_{\te{UV}}  = C_{1|234} = A^{\te{tree}}(1,2,3,4) \ ,
\label{UV4}
\eeq 
see (\ref{t8}) below for  the tree amplitude expressed  in momenta and polarization vectors. The anomalous gauge variation is carried by $P_{1|4|2,3}$ and $\ell_m \ell_n C^{mn}_{1|2,3,4} $, see (\ref{BG21var}), so one can immediately reproduce the anomaly (\ref{anom0}) from the representation in (\ref{BG27}). However, the latter obscures locality through the propagators $\sim s_{ij}^{-1}$ that were absorbed into numerators, such as $\ell_m C^m_{1|23,4}$ present in the box and one triangle. 
For a complete picture of the 4-point amplitude and its symmetry properties, one should consider both the manifestly local representation (\ref{4ptans}) and the manifestly gauge pseudo-invariant representation (\ref{BG27}). Note that the latter form of the 4-point amplitude shares the structure of the maximally supersymmetric 6-point amplitude in (5.10) of \cite{Mafra:2014gja}. Accordingly, cyclic symmetry of the integrated expression under $(1,2,3,4) \rightarrow (4,1,2,3)$ modulo anomaly can be verified along the lines of section 5.4 of \cite{Mafra:2014gja}.

\subsubsection{An empirical invariantization map}

Following the same steps as for the maximally supersymmetric setup explained in section 5.1 of \cite{Mafra:2014gja}, there is a systematic and intuitive mapping from the above local representations to the manifestly gauge (pseudo-)invariant expressions in (\ref{BG30a}), (\ref{BG27}) and (\ref{BG28}). Whenever the reference leg 1 enters a kinematic building block through a single-particle slot $A=1$, it signals a (pseudo-)invariant according to
\beq
M_{A,B} \rightarrow \delta_{A,1} C_{1|B} \co 
M^m_{A|B,C} \rightarrow   \delta_{A,1} C^m_{1|B,C} \co 
M^{mn}_{A|B|C,D}\rightarrow  \delta_{A,1}  C^{mn}_{1|B,C,D} \co 
{\cal J}_{1|4|2,3}\rightarrow P_{1|4|2,3} \ .
\label{inv1}
\eeq
Building blocks with leg 1 in a multiparticle slot (say $A=12$ or $A=123$), on the other hand, are absorbed into the (pseudo-)invariant completions of the cases with $A=1$ and therefore mapped to zero through the Kronecker delta $\delta_{A,1}$ in the empirical ``invariantization'' prescription (\ref{inv1}). This formal map is checked to reproduce the above manifestly gauge (pseudo-)invariant amplitude representations from the local ones in (\ref{BG30}) and (\ref{4ptans}).

\subsubsection{A simplified representation}

In contrast to the maximally supersymmetric 6-point amplitude in \cite{Mafra:2014gja}, the present 4-point context turns out to admit additional simplifications. The BCJ relations \cite{Bern:2008qj} among different permutations of $C_{1|234} \sim A^{\te{tree}}(1,2,3,4)$ imply the vanishing of the scalar triangle numerators, and additional on-shell relations detailed in appendix \ref{sectB} cast the scalar box numerator into a compact form:
\beq
s_{34} C_{1|234} - s_{24} C_{1|324}= s_{23} C_{1|432} - s_{24} C_{1|342} =0 \co 
C_{1|2|3|4}^{\te{scal}} = - \frac{1}{3}  s_{23} s_{34} C_{1|234} \ .
\label{bcjtriag}
\eeq
Upon insertion into (\ref{BG27}), we arrive at the following simplified and manifestly pseudo-invariant form of the 4-point amplitude with half-maximal supersymmetry,
\begin{align}
A^{\te{1-loop}}_{{\cal N}=2}(1,2,3,4) &=  \int \frac{ \dd^D\ell }{(2\pi)^D} \, \bigg\{ \frac{ C_{1|234} }{ \ell^2 \ell_{1}^2} + \frac{2 \ell_m C^m_{1|23,4} }{\ell^2 \ell_{1}^2 \ell_{123}^2} + \frac{ 2\ell_m C^m_{1|2,34} }{\ell^2 \ell_{1}^2 \ell_{12}^2}  - \frac{2 P_{1|4|2,3} }{\ell_{1}^2 \ell_{12}^2 \ell_{123}^2}
&\label{BG27simp} \\
&\! \! \! \! \! \! \! \! \! \! \! \! \! \! \! \! \! \! \! \!  \! \! \! \!  + \frac{2 \ell_m \ell_n C^{mn}_{1,2,3,4} +2 \ell_m (s_{23} C^m_{1|23,4}  +s_{24} C^m_{1|24,3}  +s_{34} C^m_{1|2,34}   ) -\frac{1}{3}  s_{23} s_{34} C_{1|234}}{\ell^2 \ell_{1}^2 \ell_{12}^2 \ell_{123}^2}  \bigg\}   \ .
\notag
\end{align}
Its integrand can be regarded as the main result of this work, and the integrated expression in $D=4$ dimensions is discussed in section \ref{sec:spinhel} to demonstrate agreement with results in the literature. Note that from the above construction via locality and gauge invariance, we are free to add any multiple of the maximally supersymmetric amplitude $A^{\te{1-loop}}_{{\cal N}=4}(1,2,3,4)$ made of a box diagram with permutation invariant and local numerator $s_{23} s_{34} C_{1|234}$. In section \ref{sec:run}, we will see explicitly that this freedom 
is equivalent to the freedom of adjusting the field content of the theory, i.e.\ the number of supermultiplets that run in the loop and their gauge-group representations.

The diversity of gauge theories increases rapidly when reducing from maximal to  half-maximal supersymmetry
(running of gauge couplings,
variety of supermultiplets and representations). We would like to highlight that all this additional complexity of the 4-point 1-loop amplitude in general dimensions is compactly captured by the kinematic building blocks in \eqref{BG27simp}.
As we discuss further in the conclusions, upon dimensional reduction of this theory to $D=4$, the parity-even
part of {\it minimally} supersymmetric gauge theory is also given by this expression.

\section{Supergravity from the duality between color and kinematics}
\label{sect4}

A major virtue of the cubic-graph organization of gauge-theory amplitudes is that 
it often admits the construction of supergravity amplitudes at various loop orders by double-copy \cite{Bern:2010ue}. For this to work, the kinematic constituents must mirror all the properties of the color factors \cite{Bern:2008qj}, in other words they should satisfy the BCJ duality between color and kinematics\footnote{See \cite{Carrasco:2015iwa} for a recent pedagogical account.}. In this section, it will be demonstrated that the 3-point gauge-theory amplitude presented in sections \ref{sect31} and \ref{sect33} obeys the BCJ duality, so we can infer the related half-maximal supergravity amplitude. However,  in the formulation
of the 4-point amplitude we gave in sections \ref{sect34} and \ref{sect35}, the duality is not manifest
and further work is needed.

\subsection{Review of the BCJ duality and double-copy construction}
\label{sect41}

\begin{figure}[h]
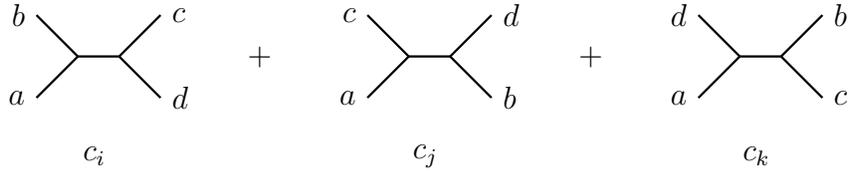

\begin{center}
 \tikzpicture[scale=1.1, line width=0.30mm]
 \scope[yshift=-0.5cm, xshift=0.0cm]
 \draw  (2,0.5) -- (1.5,1) node[left]{$b$};
 \draw   (2,0.5) -- (1.5,0) node[left]{$a$};
 \draw   (2,0.5) -- (2.5,0.5) ;
 \draw   (2.5,0.5) -- (3,1) node[right]{$c$};
 \draw   (2.5,0.5) -- (3,0) node[right]{$d$};
 \draw (2.2, -0.7) node{$c_i$};
 \draw (4.2,0.5) node{$+$};
 \endscope
 \scope[yshift=-0.5cm, xshift=4.0cm]
 \draw   (2,0.5) -- (1.5,1) node[left]{$c$};
 \draw   (2,0.5) -- (1.5,0) node[left]{$a$};
 \draw   (2,0.5) -- (2.5,0.5) ;
 \draw   (2.5,0.5) -- (3,1) node[right]{$d$};
 \draw   (2.5,0.5) -- (3,0) node[right]{$b$};
 \draw (4.2,0.5) node{$+$};
 \draw (2.2, -0.7) node{$c_j$};
 \endscope
 \scope[yshift=-0.5cm, xshift=8.0cm]
 \draw   (2,0.5) -- (1.5,1) node[left]{$d$};
 \draw   (2,0.5) -- (1.5,0) node[left]{$a$};
 \draw   (2,0.5) -- (2.5,0.5) ;
 \draw   (2.5,0.5) -- (3,1) node[right]{$b$};
 \draw   (2.5,0.5) -- (3,0) node[right]{$c$};
 \draw (2.2, -0.7) node{$c_k$};
 \endscope
 \endtikzpicture
\caption{Jacobi identities imply the vanishing of the color factors associated to the above triplet of cubic
 graphs, $c_i + c_j + c_k = 0$. The legs $a$, $b$, $c$ and $d$ may represent arbitrary subdiagrams with the same momenta in the external edges of the graphs $i,j,$ and $k$. According to the BCJ duality, their corresponding kinematic numerators $N_i$ can be chosen such that $N_i + N_j + N_k=0$.}
\label{f:bcj}
\end{center}
\end{figure}

The BCJ duality between color and kinematics is based on the dictionary between cubic graphs and structure constants $f^{abc}$ of an arbitrary gauge group. The color representative $c_i$ of a graph is obtained by dressing each cubic vertex with a factor of $f^{abc}$ and by contracting the color indices $a,b,c$ across the internal lines. Then, the Jacobi identity
\beq
f^{abe}f^{cde} + f^{bce}f^{ade} + f^{cae} f^{bde} = 0
\label{jac1}
\eeq
universal to any Lie algebra relates triplets of graphs depicted in figure \ref{f:bcj} in the sense that their color factors add up to zero, $c_i + c_j + c_k=0$. According to the BCJ conjecture\footnote{Although the most general form of BCJ duality and the double copy construction remain conjectures, they are supported by a steadily growing list of examples up to and including 4 loops \cite{Bern:2012uf, Bern:2013uka, Bern:2014sna}. Also there are examples without any supersymmetry such as \cite{Boels:2013bi, Bern:2013yya, Nohle:2013bfa, Chiodaroli:2013upa, Johansson:2014zca, Mogull:2015adi}, and the 4-dimensional version of the half-maximal 1-loop amplitudes under investigation has been cast into BCJ form in \cite{Carrasco:2012ca, Johansson:2014zca}.}, gauge-theory amplitudes can be represented\footnote{Ambiguities for local cubic-graph representations of gauge-theory amplitudes arise from the freedom to assign the contributions from quartic gluon vertices to different cubic diagrams. Redistributions as required by the BCJ duality are often referred to as ``generalized gauge freedom'' 
\cite{Bern:2008qj, Bern:2010yg, Bern:2010ue}, and a concrete non-linear gauge transformation to
implement such rearrangements of tree-level diagrams was identified in \cite{Lee:2015upy}.}
such that for each such triplet of graphs $(i,j,k)$, the corresponding triplets of kinematic weights $N_{i},N_j,N_k$ (or {\em numerators} for short) depending on polarizations and (external and internal) momenta sum to zero as well, $N_i + N_j + N_k=0$. At loop level where the numerators may depend on loop momenta $\ell$, such kinematic Jacobi identities are understood to hold for any value of $\ell$. A gauge-theory amplitude is said to obey BCJ duality if the numerators $N_i$ of all the cubic graphs $i$ are antisymmetric under flips of their vertices and if they satisfy all the kinematic Jacobi identities.

According to the double-copy conjecture, the integrands of gauge-theory amplitudes that obey the BCJ duality are converted to supergravity integrands once the color factors $c_i$ are replaced by another copy of the kinematic numerators $\tilde N_i$ for each cubic graph $i$, i.e. \cite{Bern:2010ue}
\beq
{\cal M}^{\te{$g$-loop}}_{{\cal N}+ \widetilde{\cal N}} = {\cal A}^{\te{$g$-loop}}_{{\cal N}}  \, \big|_{c_i \rightarrow \tilde N} \ .
\label{dbcpy}
\eeq
The additional kinematic numerators $\tilde N_i$ do not need to come from the same theory, 
in fact they can even violate the kinematic Jacobi relations. As indicated by the subscripts in (\ref{dbcpy}), the supersymmetries ${\cal N}$ and  $\widetilde{\cal N}$ of the gauge theories with numerators $N_i$ and $\tilde N_i$ add up to yield the amount of supersymmetry of the supergravity amplitude ${\cal M}^{\te{$g$-loop}}_{{\cal N}+ \widetilde{\cal N}} $. As for the state dependence of $N_i$ and $\tilde N_i$, 
the supergravity spectrum emerges as the tensor product of the gauge-theory states, e.g.\ graviton polarization tensors follow from the traceless parts of the gluon polarizations $e^{(m} \tilde e^{n)}$, and the ${\cal N}=8$ supergravity multiplet arises as a double copy of the ${\cal N}=4$ SYM multiplet. 

The standard double-copy realization of pure ${\cal N}=4$ supergravity 
is as $({\cal N}=0) \times ({\cal N}=4)$ SYM, an asymmetric  (``heterotic'') realization (see e.g.\ \cite{Bern:2011rj}). In this work, we have in mind 
the symmetric double copy $({\cal N}=2) \times ({\cal N}=2)$ SYM, that gives ${\cal N}=4$ supergravity coupled to ${\cal N}=4$ matter
multiplets with maximum spin 1 and $\frac{3}{2}$, respectively. String constructions of these matter-coupled supergravities 
leave some freedom to tune the matter
content, see e.g.\ \cite{Tourkine:2012vx} and references therein.


\subsection{BCJ duality and double copy of the 3-point amplitude}
\label{sect42}

The local representation (\ref{BG30}) of the 3-point amplitude will now be shown to obey the BCJ duality. There are 3 inequivalent classes of kinematic Jacobi relations $N_i + N_j + N_k=0$ to check:
\begin{itemize}
\item Symmetry of the bubbles versus absence of tadpoles: As depicted in figure \ref{f:bcj1}, the symmetry of bubble numerators $T_{A,B}=T_{B,A}$ and the absence of tadpoles is consistent with the BCJ duality.

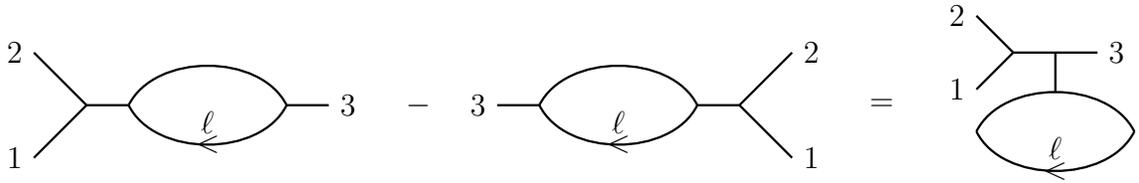
\begin{figure}[h]
\begin{center}
\begin{tikzpicture} [scale=0.7, line width=0.30mm]
\scope[xshift=3.2cm]
\draw (0,0) -- (-1,1) node[left]{$2$};
\draw (0,0) -- (-1,-1) node[left]{$1$};
\endscope
\draw (3.2,0) -- (4,0);
%
\draw (4,0) .. controls (4.5,1) and (6.5,1) .. (7,0);
\draw (4,0) .. controls (4.5,-1) and (6.5,-1) .. (7,0);
\draw (5.5,-0.75)node{$<$}node[above]{$\ell$};
\draw (7,0) -- (7.8,0) node[right]{$3$};
\draw(9.5,0) node{$-$};
\scope[xshift=7.8cm]
\draw (7.8,0) -- (8.8,1) node[right]{$2$};
\draw (7.8,0) -- (8.8,-1) node[right]{$1$};
\draw (7,0) -- (7.8,0);
%
\draw (4,0) .. controls (4.5,1) and (6.5,1) .. (7,0);
\draw (4,0) .. controls (4.5,-1) and (6.5,-1) .. (7,0);
\draw (5.5,-0.75)node{$<$}node[above]{$\ell$};
\draw (4,0) -- (3.2,0) node[left]{$3$};
\endscope
\draw(18.3,0) node{$=$};
\scope[xshift=16.1cm, yshift=-0.5cm]
\draw(5.5,0.75)--(5.5,1.5);
\draw(5.5,1.5) -- (6.3,1.5)node[right]{$3$};
\draw(5.5,1.5) -- (4.7,1.5);
\draw (4.7,1.5) -- (4,0.8)node[left]{$1$};
\draw (4.7,1.5) -- (4,2.2)node[left]{$2$};
\draw (4,0) .. controls (4.5,1) and (6.5,1) .. (7,0);
\draw (4,0) .. controls (4.5,-1) and (6.5,-1) .. (7,0);
\draw (5.5,-0.75)node{$<$}node[above]{$\ell$};
\endscope
\end{tikzpicture}
\caption{Kinematic Jacobi relation $T_{12,3}-T_{3,12}=0$ relating the antisymmetrization of bubbles to a tadpole diagram with vanishing numerator.}
\label{f:bcj1}
\end{center}
\end{figure}

\item The formally vanishing scalar admixtures to the triangle numerators $N_{1|2,3}(\ell)$ in (\ref{loc75}) yield bubble numerators upon antisymmetrization in 2,3, consistent with the triplet of diagrams in figure \ref{f:bcj2}.

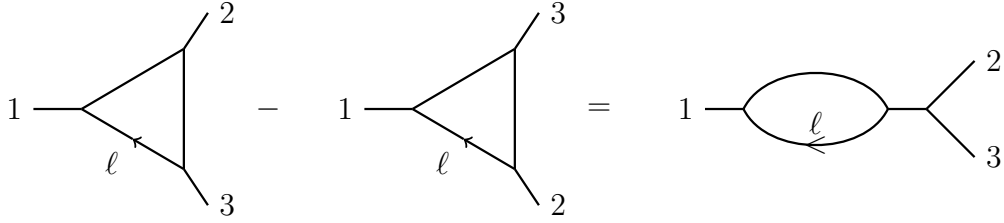
\begin{figure}[h]
\begin{center}
\begin{tikzpicture} [scale=0.8, line width=0.30mm]
\draw (1,0) -- (0.2,0)node[left]{$1$};
\draw (2.7,1)--(2.7,-1);
\draw(2.7,1) -- (3.1,1.6)node[right]{$2$};
\draw(2.7,-1) -- (3.1,-1.6)node[right]{$3$};
\draw[->](2.7,-1)--(1.85,-0.5);
\draw(1,0)--(1.85,-0.5);
\draw(1.5,-0.9)node{$\ell$};
\draw(2.7,1)--(1,0);
\draw(4.1,0)node{$-$};
\scope[xshift=5.5cm]
\draw (1,0) -- (0.2,0)node[left]{$1$};
\draw (2.7,1)--(2.7,-1);
\draw(2.7,1) -- (3.1,1.6)node[right]{$3$};
\draw(2.7,-1) -- (3.1,-1.6)node[right]{$2$};
\draw[->](2.7,-1)--(1.85,-0.5);
\draw(1,0)--(1.85,-0.5);
\draw(1.5,-0.9)node{$\ell$};
\draw(2.7,1)--(1,0);
\endscope
\draw(9.6,0)node{$=$};
\scope[scale=0.8,xshift=11cm]
\draw (7.8,0) -- (8.8,1) node[right]{$2$};
\draw (7.8,0) -- (8.8,-1) node[right]{$3$};
\draw (7,0) -- (7.8,0);
%
\draw (4,0) .. controls (4.5,1) and (6.5,1) .. (7,0);
\draw (4,0) .. controls (4.5,-1) and (6.5,-1) .. (7,0);
\draw (5.5,-0.75)node{$<$}node[above]{$\ell$};
\draw (4,0) -- (3.2,0) node[left]{$1$};
\endscope
\end{tikzpicture}
\caption{Kinematic Jacobi relation $N_{1|2,3}(\ell) - N_{1|3,2}(\ell)=2 T_{1,23}$ relating the antisymmetrization of two triangles to a bubble diagram.}
\label{f:bcj2}
\end{center}
\end{figure}

\item Antisymmetrizing a triangle numerator in legs 1,2 is also is consistent with the bubble numerators under the BCJ duality. Since our shift conventions for the loop momentum fix $\ell$ to reside in the edge next to leg 1, the momentum routing in figure \ref{f:bcj3} requires particular care, and the numerator of the second triangle takes $\ell-k_2$ instead of $\ell$ as its argument.

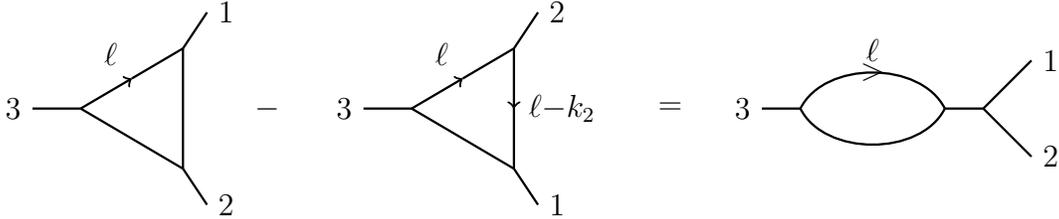
\begin{figure}[h]
\begin{center}
\begin{tikzpicture} [scale=0.8, line width=0.30mm]
\draw (1,0) -- (0.2,0)node[left]{$3$};
\draw (2.7,1)--(2.7,-1);
\draw(2.7,1) -- (3.1,1.6)node[right]{$1$};
\draw(2.7,-1) -- (3.1,-1.6)node[right]{$2$};
\draw(2.7,1)--(1.85,0.5);
\draw[->](1,0)--(1.85,0.5);
\draw(1.5,0.9)node{$\ell$};
\draw(2.7,-1)--(1,0);
\draw(4.1,0)node{$-$};
\scope[xshift=5.5cm]
\draw (1,0) -- (0.2,0)node[left]{$3$};
\draw (2.7,1)--(2.7,-1);
\draw(2.7,1) -- (3.1,1.6)node[right]{$2$};
\draw(2.7,-1) -- (3.1,-1.6)node[right]{$1$};
\draw(2.7,1)--(1.85,0.5);
\draw[->](1,0)--(1.85,0.5);
\draw(1.5,0.9)node{$\ell$};
\draw(2.7,-1)--(1,0);
\draw[->] (2.7,0.01)--(2.7,0);
\draw(3.5,0)node{$\ell{-}k_2$};
\endscope
\draw(10.8,0)node{$=$};
\scope[scale=0.8,xshift=12.2cm]
\draw (7.8,0) -- (8.8,1) node[right]{$1$};
\draw (7.8,0) -- (8.8,-1) node[right]{$2$};
\draw (7,0) -- (7.8,0);
%
\draw (4,0) .. controls (4.5,1) and (6.5,1) .. (7,0);
\draw (4,0) .. controls (4.5,-1) and (6.5,-1) .. (7,0);
\draw (5.5,0.75)node{$>$}node[above]{$\ell$};
\draw (4,0) -- (3.2,0) node[left]{$3$};
\endscope
\end{tikzpicture}
\caption{Kinematic Jacobi relation $N_{1|2,3}(\ell) - N_{1|3,2}(\ell{-}k_2)=2 T_{12,3}$ relating the antisymmetrization of triangles involving the reference leg 1 to a bubble diagram.}
\label{f:bcj3}
\end{center}
\end{figure}

\end{itemize}
Given that the representation (\ref{BG30}) of the gauge-theory amplitude satisfies all these Jacobi identities, it is qualified to enter the double-copy construction. By converting the color factors of the bubble- and triangle diagrams to additional kinematics, see (\ref{dbcpy}), one obtains
\begin{align}
{\cal M}^{\te{1-loop}}_{{\cal N}+\tilde {\cal N}=4}(1,2,3)=
\int \frac{ \dd^D\ell }{(2\pi)^D} \, \bigg\{ \frac{ |2T_{1,23}|^2 }{2 s_{23} \ell^2 \ell_{1}^2 } + \frac{ |2T_{12,3}|^2 }{2s_{12} \ell^2 \ell_{12}^2 } + \frac{|2 T_{31,2}|^2 }{2s_{13} \ell_{1}^2 \ell_{12}^2 } +
 \frac{ |N_{1|2,3}(\ell)|^2 }{ \ell^2 \ell_{1}^2 \ell_{12}^2 } + \frac{ |N_{1|3,2}(\ell)|^2 }{ \ell^2 \ell_{1}^2 \ell_{13}^2 } \bigg\}  
 \label{sug3}
\end{align}
after combining both orientations of the triangle. However, by the formally vanishing bubble numerator $T_{12,3} =s_{12}(e_1\cdot e_2) (e_3\cdot k_1)$ derived in (\ref{loc77}), the $\ell$-independent part of the bubble contributions yields
\beq
\frac{ |T_{12,3}|^2 }{s_{12}} = \frac{| -(p\cdot k_3) (e_1\cdot e_2) (e_3\cdot k_1) |^2 }{-(p\cdot k_3)} = - (p\cdot k_3) |(e_1\cdot e_2) (e_3\cdot k_1) |^2 \rightarrow 0
 \label{sug3a}
\eeq
upon minahaning (\ref{Mina5}) with a deformation momentum $p^m$
in intermediate steps. Likewise, the scalar parts $\sim T_{ij,k}$ of the triangle numerator $N_{1|2,3}(\ell)$ in (\ref{loc75}) vanish by the same argument, whereas (\ref{sug3}) simplifies to
\begin{align}
{\cal M}^{\te{1-loop}}_{{\cal N}+\tilde {\cal N}=4}(1,2,3)= 4
\int \frac{ \dd^D\ell }{(2\pi)^D} \,  \ell_m \ell_n (T^m_{1,2,3}{+}E^m_{1|2,3})( \tilde T^n_{1,2,3} {+}\tilde E^n_{1|2,3}) \, \bigg\{ 
 \frac{1}{ \ell^2 \ell_{1}^2 \ell_{12}^2 } + \frac{ 1 }{ \ell^2 \ell_{1}^2 \ell_{13}^2 } \bigg\}   \ .
 \label{sug3b}
\end{align}
Given that all of $T^m_{1,2,3},E^m_{1|2,3}, \tilde T^n_{1,2,3},\tilde E^n_{1|2,3}$ are perpendicular to the external momenta,
the only contribution to the integrated expression has tensor structure $\ell^m \ell^n\rightarrow \eta^{mn}$ and leaves a no-scale integral.

Of course, the same discussion applies to the manifestly gauge-invariant representation of the SYM numerators given in section \ref{sect33}, i.e.\ to the image of the numerators under the invariantization map (\ref{inv1}). In the alternative form
\begin{align}
{\cal M}^{\te{1-loop}}_{{\cal N}+\widetilde{\cal N}=4}(1,2,3)= 4
\int \frac{ \dd^D\ell }{(2\pi)^D} \, \ell_m \ell_n C^m_{1|2,3} \tilde C^n_{1|2,3} \, \bigg\{ 
 \frac{ 1}{ \ell^2 \ell_{1}^2 \ell_{12}^2 } + \frac{ 1 }{ \ell^2 \ell_{1}^2 \ell_{13}^2 } \bigg\}   \ ,
 \label{sug3c}
\end{align}
the UV-divergence due to the trace component $\ell_m \ell_n \rightarrow \eta_{mn}$ of the tensor integral 
\beq
{\cal M}^{\te{1-loop}}_{{\cal N}+\widetilde{\cal N}=4}(1,2,3) \, \big|_{\rm UV}= 
C^m_{1|2,3} \tilde C^m_{1|2,3}
 \label{sug3d}
\eeq
manifestly agrees with the low-energy limit of the corresponding string computation \cite{Berg:2016wux}, see \cite{Gregori:1997hi, Berg:2016wux} for a discussion of the components with different numbers of gravitons, B-fields and dilatons. For 3 gravitons, the counterterm associated with (\ref{sug3d}) is an operator $R^2$ which is on-shell equivalent to the Gauss-Bonnet term, and 
therefore yields vanishing amplitudes in strictly 4 dimensions. Likewise, the parity-odd contribution to (\ref{sug3d}) obviously drops out in $D=4$ for any state configuration.


\subsection{Deviations from the duality in the 4-point amplitude}
\label{sect43}

Since the 4-point gauge-theory amplitude (\ref{BG27simp}) has been constructed from the same principles as its BCJ-satisfying 3-point counterpart, it is tempting to hope for the duality to hold also at 4 points. However, 
in our representation one can identify a simple counterexample among the kinematic Jacobi identities that renders the 4-point supergravity amplitude inaccessible to naive\footnote{We note that two new approaches to double-copy constructions of gravity amplitudes have been developed since the first preprint version of this article \cite{loopKLT, Bern:2017yxu} which do not require a BCJ representation of the gauge-theory input.}
double copy of the present building blocks. Just like the 3-point BCJ discussion was identical for the manifestly local and the manifestly gauge-invariant representation,  for our
 4-point arguments we could
choose  either (\ref{4ptans}) or  (\ref{BG27simp}). For convenience we pick the latter.

\begin{figure}[h]
\begin{center}
\begin{tikzpicture} [scale=0.8, line width=0.30mm]
\draw (1,0) -- (0.2,0)node[left]{$4$};
\draw (2.7,1)--(2.7,-1);
\draw(2.7,1) -- (3.1,1.6)node[right]{$1$};
\draw(2.7,-1) -- (3.1,-1.6);
\draw(3.1,-1.6) -- (3.5,-1.8)node[right]{$2$};
\draw(3.1,-1.6) -- (3.0,-2.1)node[left]{$3$};
\draw(2.7,1)--(1.85,0.5);
\draw[->](1,0)--(1.85,0.5);
\draw(1.5,0.9)node{$\ell$};
\draw(2.7,-1)--(1,0);
\draw(4.1,0)node{$-$};
\scope[xshift=5.5cm]
\draw (1,0) -- (0.2,0)node[left]{$4$};
\draw (2.7,1)--(2.7,-1);
\draw(3.1,1.6) -- (3.5,1.8)node[right]{$3$};
\draw(3.1,1.6) -- (3.0,2.1)node[left]{$2$};
\draw(2.7,1) -- (3.1,1.6);
\draw(2.7,-1) -- (3.1,-1.6)node[right]{$1$};
\draw(2.7,1)--(1.85,0.5);
\draw[->](1,0)--(1.85,0.5);
\draw(1.5,0.9)node{$\ell$};
\draw(2.7,-1)--(1,0);
\draw[->] (2.7,0.01)--(2.7,0);
\draw(3.5,0)node{$\ell{-}k_{23}$};
\endscope
\draw(10.8,0)node{$=$};
\scope[scale=0.8,xshift=12.2cm]
\draw (7.8,0) -- (7.8,1) node[above]{$1$};
\draw (7,0) -- (8.6,0);
\draw (8.6,0) -- (9.4,0.8) node[right]{$2$};
\draw (8.6,0) -- (9.4,-0.8) node[right]{$3$};
%
\draw (4,0) .. controls (4.5,1) and (6.5,1) .. (7,0);
\draw (4,0) .. controls (4.5,-1) and (6.5,-1) .. (7,0);
\draw (5.5,0.75)node{$>$}node[above]{$\ell$};
\draw (4,0) -- (3.2,0) node[left]{$4$};
\endscope
\end{tikzpicture}
\caption{Kinematic Jacobi relation between the antisymmetrization of triangles involving the reference leg 1 to a bubble diagram. The vector part of the second triangle numerator is contracted into the momentum $\ell-k_{23}$ of the edge adjacent to leg 1. The numerator representation in (\ref{BG27simp}) fails to obey this kinematic Jacobi relation.}
\label{f:bcj4}
\end{center}
\end{figure}
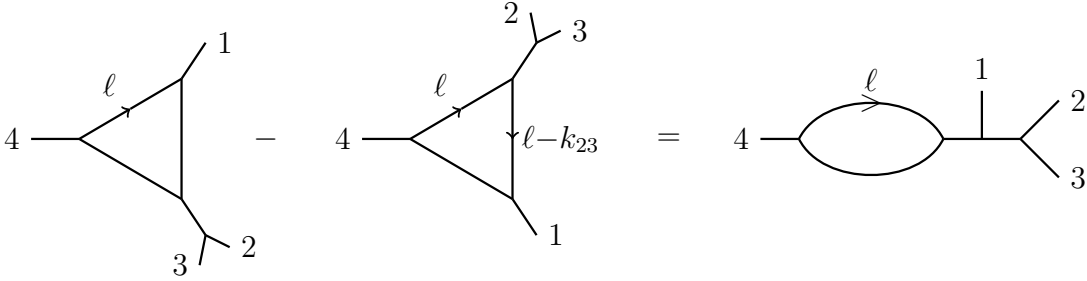

As depicted in figure \ref{f:bcj4}, the antisymmetrization of two triangles with massive momentum $k_2+k_3$ in one corner should reproduce a bubble numerator that vanishes in the parametrization of (\ref{BG27simp}). However, we obtain 
\beq
\ell_m C^m_{1|23,4} - (\ell_m - k_m^{23}) C^m_{1|23,4} = k_m^{23} C^m_{1|23,4} = P_{1|2|3,4} - P_{1|3|2,4} \neq 0
\label{bcj9}
\eeq
instead of the vanishing bubble numerator from the analogous antisymmetrization in figure \ref{f:bcj4}. This
is an obstacle to BCJ duality and prevents us from immediately
writing down a double-copy expression for the supergravity amplitude.

This obstacle was anticipated from  the analogous closed-string amplitude in our companion paper  \cite{Berg:2016wux}. 
It closely parallels the obstacle observed in the representation of the maximally supersymmetric 6-point amplitude in section 6.3 of \cite{Mafra:2014gja}. Both here and there, the mismatch in kinematic Jacobi identities boils down to pseudo-invariant kinematic factors as seen in (\ref{bcj9}). It is tempting to speculate that there is a tension between anomalies and the BCJ duality \cite{Mafra:2014gja}, but ongoing work \cite{WIP1} and the recent results of \cite{WIP2} indicate that there is no direct connection.


\section{Comparison with 4-dimensional results}
\label{sect5}

So far, we have discussed half-maximal SYM amplitudes in the highest dimensions $D=6$ where 8 supercharges can be consistently realized. This section is devoted to their dimensional reduction to $D=4$, where the parity-odd sector of gluon amplitudes including their box anomaly drops out and their dimension-agnostic parity-even integrands\footnote{We explicitly recall this feature in section \ref{sect6} by performing the field-theory limit on the string-theory results of \cite{Berg:2016wux}.} are expressed in terms of ($D=4$) spinor-helicity variables. In the manifestly gauge-invariant framework of section \ref{sect3}, the initially 6-dimensional kinematic factors will be specialized as 
\be\label{6to4}
C_{\dots}^{\dots} \rightarrow {\cal C}_{\dots}^{\dots}\equiv C_{\dots}^{\dots}\big|^{D=4}_{\rm even} \,, \hspace{1cm}
P_{1|i|j,k} \rightarrow {\cal P}_{1i|jk}\equiv P_{1|i|j,k}\big|^{D=4}_{\rm even} \,.
\ee
with explicit spinor-helicity expressions for the 4-dimensional objects ${\cal C}_{\dots}^{\dots}$ and ${\cal P}_{\dots}$. The resulting 4-dimensional 4-point amplitudes with ${\cal N}=2$ supersymmetry will be shown to match the expressions in the literature \cite{Bern:1994cg, Bern:1995db, Carrasco:2012ca, Johansson:2014zca}.

\subsection{Spinor-helicity expressions versus polarization vectors}
\label{sect5_1}

In this section, we collect spinor-helicity expressions for the parity-even gauge-invariant building blocks (\ref{6to4}) in $D=4$.  Conventions for the relevant momentum spinors and $\sigma$-matrices are summarized in appendix \ref{sectY}. As expected from supersymmetric Ward identities \cite{Grisaru:1976vm, Grisaru:1977px}, only the MHV components of the scalar, vectorial and tensorial kinematic factors in (\ref{6to4}) are non-zero, e.g.
\beq
{\cal C}_{1^-|2^-3^-}= {\cal P}_{1^-2^-|3^-4^+}   = {\cal C}^{m}_{1^-|2^-3^-,4^-} ={\cal C}^{mn}_{1^-|2^-,3^-,4^-} = 0 \ .
\label{nonMHV}
\eeq
In the MHV sector, one obtains 3-point helicity components such as
\begin{align}
{\cal C}_{1^-|2^-3^+}=\,\frac{1}{\sqrt{2}}\frac{\la 12\ra^3}{\la13\ra \la23\ra}\,, \ \ \ \ \ \ 
{\cal C}_{1^+|2^-3^-}=\,\frac{1}{\sqrt{2}}\frac{\la 23\ra^3}{\la12\ra \la13\ra}\,, \ \ \ \ \ \ 
{\cal C}_{1^+|2^+3^-}=\,\frac{1}{\sqrt{2}}\frac{ [12]^3}{[13] [23]}\,,
\label{3ch}
\end{align}
as well as
\begin{align}
{\cal C}_{1^-|2^-,3^+}^m= \frac{1}{2\sqrt{2}}  \frac{\la12\ra^2 }{\la 1 3\ra}  [ 3|\sigma^m |1\ra\,, \ \ \ \ \ \
{\cal C}_{1^-|2^+,3^+}^m=\frac{1}{2\sqrt{2}}\frac{[23]^3 }{[12][13]} \Big(  [3|\sigma^m |3\ra - [2|\sigma^m |2\ra \Big) \,.
\label{Cm_mpp}
\end{align}
At 4 points, we have the following inequivalent cases for scalar kinematic factors:
\begin{align}\label{C_mmpp}
{\cal C}_{1^-|2^-3^+ 4^+} &
= -\frac{1}{2}\frac{\la 12\ra^2[34]^2}{s_{12}s_{23}}= 
2\frac{\la12 \ra^3}{\la23\ra\la34\ra \la41\ra}\,, \\
{\cal P}_{1^-2^-|3^+4^+}  &=
-\frac{1}{2}\frac{\la 12\ra^2[34]^2}{s_{12}} 
\ , \ \ \ \ \ \ 
{\cal P}_{1^-2^+|3^-4^+}  ={\cal P}_{1^-2^+|3^+4^-}  = 0 \, ,
\label{4ch} 
\end{align}
for vector kinematic factors we have:
\begin{align}
\label{Cm_mppm}
{\cal C}^m_{1^-|2^+3^+,4^-}=\,&\,\frac{1}{2} 
\bigg( \frac{\la 14\ra^3}{\la23\ra\la24\ra\la34\ra} [1|\sigma^m|4\ra
+\frac{\la 14\ra^3}{\la13\ra\la12\ra\la23\ra} [4|\sigma^m|1\ra\,
 \bigg)\,,
\\
\label{Cm_mmpp}
{\cal C}^m_{1^-|2^-3^+,4^+}=\,
&\,\frac{1}{2} \bigg(
2\frac{\la12\ra^3}{\la14\ra\la23\ra\la34\ra}[4|\sigma^m|4\ra 
- \frac{ \la 12\ra^3}{\la14\ra\la24\ra\la34\ra} [3|\sigma^m|4\ra 
-\frac{\la12\ra^3}{\la13\ra\la23\ra\la34\ra}[4|\sigma^m|3\ra 
 \bigg)\,,
\end{align}
and for tensor kinematic factors:
\begin{align}
\label{Cmn_mmpp}
{\cal C}^{mn}_{1^-|2^-,3^+,4^+}=\,-&
\frac{1}{2}\bigg(\frac{\la12\ra^2}{\la 13\ra\la 14\ra}[3|\sigma^{(m}|1\rangle [4|\sigma^{n)}|1\ra\,+
\frac{\la12\ra^2}{\la 23\ra\la 24\ra}[3|\sigma^{(m}|2\rangle [4|\sigma^{n)}|2\ra\,
\bigg)\\
&\,+\,
\frac{1}{2}\frac{\la12\ra^3}{\la23\ra\la24\ra\la 34\ra}\Big( [3|\sigma^{(m}|3\ra - [4|\sigma^{(m}|4 \ra \Big) [1|\sigma^{n)}|2\ra\,
.\notag
\end{align}
Note that the expressions in (\ref{3ch}) and (\ref{C_mmpp}) can be reproduced from the Parke-Taylor formula \cite{Parke:1986gb} for the corresponding tree-level amplitudes (\ref{UV3}) and (\ref{UV4}). Moreover, the spinor-helicity results in (\ref{4ch}) identify $s_{ij} {\cal P}_{ij| kl}$ with the gluon components of the kinematic variables $\kappa_{ij} + \kappa_{kl}$ in \cite{Johansson:2014zca}.

It is worth stressing the crucial difference between the helicity-agnostic expressions underlying the above components of ${\cal C}_{1|234}$ and those underlying ${\cal P}_{12|34}$. Using Lorentz traces over linearized field strengths $f_{i}^{mn} \equiv k_i^m e_i^n - k_i^n e_i^m$,
\begin{align}
t(1,2) & \equiv
(e_1\cdot k_2) (e_2 \cdot k_1) - (e_1\cdot e_2) (k_1\cdot k_2)
\label{red28} \\
t(1,2,3,4) &\equiv (e_1 \cdot k_2) (e_2\cdot k_3)(e_3 \cdot k_4)(e_4\cdot k_1)  - \te{antisymmetrization in all} \ (k_j\leftrightarrow e_j) \ , \notag
\end{align}
it was found in the string-theory companion paper \cite{Berg:2016wux} that
\begin{align}
{\cal C}_{1|234} &=    \frac{1}{s_{12} s_{23}} \, \Big[   t(1,2) t(3,4)  - t(1,2,3,4) + {\rm cyc}(2,3,4) \Big]
\label{polC}
\\
{\cal P}_{12|34} &=  \frac{1}{s_{12}} \, \Big[ t(1,3)t(2,4) + t(1,4)t(2,3) - t(1,2)t(3,4) - t(1,4,2,3) \Big] \ ,
\label{polP}
\end{align}
where only ${\cal C}_{1|234}$
is expressible in terms of the famous $t_8$-tensor from maximal supersymmetry,
\beq
t_8(1,2,3,4) \equiv \frac{1}{2} \, \Big[
 t(1,2) t(3,4) -  t(1,2,3,4) + {\rm cyc}(2,3,4) \Big] 
 = \frac{1}{2} \, s_{12} s_{23} C_{1|234}  \ .
\label{t8}
\eeq
By contrast, the object  ${\cal P}_{12|34}$  of (\ref{polP}) only preserves 8 supercharges.\footnote{This can be seen by 
relating ${\cal P}_{12| 34}$ to the $\kappa_{ij}$-variables in (4.8) and (4.9) of \cite{Johansson:2014zca}, for example.}
In  (\ref{polP}), the dependence of ${\cal P}_{12|34}$ on the polarization vectors involves tensor structures $\sim \prod_{i=1}^4 (e_i \cdot k_{j_i})$, i.e.\ terms without any contraction of the type\footnote{The absence of tensor structures $\prod_{i=1}^n (e_i \cdot k_{j_i})$, i.e.\ the omnipresence of at least one factor of $(e_i\cdot e_j)$ in $n$-point tree-level amplitudes of the open superstring has been stressed and exploited in \cite{Barreiro:2012aw, Barreiro:2013dpa}. An investigation of (possibly non-supersymmetric) gauge-invariant scalar kinematic factors including $\prod_{i=1}^n (e_i \cdot k_{j_i})$ can be found in \cite{Boels:2016xhc}.} $(e_i\cdot e_j)$. An important difference between  ${\cal P}_{12|34}$ and the $t_8$-tensor (\ref{t8}) is that expressions of the form $\prod_{i=1}^4 (e_i \cdot k_{j_i})$ in ${\cal P}_{12|34} $ cannot be eliminated via momentum conservation and transversality $(e_i \cdot k_i)=0$. Note also that the symmetries
\beq
{\cal P}_{12|34}  = {\cal P}_{21|34}  = {\cal P}_{12|43} = {\cal P}_{34|12}   
\label{symP}
\eeq
are manifest in (\ref{polP}). Certain helicity configurations lead to identical expressions for $s_{23} \,{\cal C}_{1^-|2^-3^+ 4^+}$ and ${\cal P}_{1^-2^-|3^+4^+} $ -- see (\ref{C_mmpp}) and (\ref{4ch}) --- but the vanishing of the MHV component ${\cal P}_{1^-2^+|3^-4^+} $ shows that there can be no helicity-independent proportionality between 
${\cal C}_{1|234}$ in (\ref{polC}) and ${\cal P}_{12|34}$ in  (\ref{polP}).


\subsection{Disentangling the supermultiplets in the loop}
\label{sec:run}

As emphasized below (\ref{BG27simp}),  in section \ref{sect34} we constructed a particular 4-point solution to locality and gauge invariance in the presence of bubble and triangle diagrams, but 
this particular solution admits the freedom of adding
 the maximally supersymmetric 1-loop amplitude ${A}_{{\mathcal N}=4}^{\te{1-loop}}$ with arbitrary coefficient.
With this in mind, we want to compare our results to the well-known ${\mathcal N}=2$ supersymmetric 1-loop amplitudes in $D=4$ field theory (see for example \cite{Bern:1994cg,Bern:1995db}). We focus on the recent work \cite{Johansson:2014zca}, that
discusses the contribution of the hypermultiplet running in the loop. In general, the single-trace part of the color-ordered 1-loop amplitude in an ${\cal N}=2$ theory can be written schematically as
\bea  
\tilde{A}_{{\mathcal N}=2}^{\te{1-loop}}=c_{\rm vec}A_{\rm {\cal N}=2}^{\rm vec}+c_{\rm hyp}A_{\rm {\cal N}=2}^{\rm hyp}\,,
\label{original}
\eea
where $A_{\rm {\cal N}=2}^{\rm vec}$ and $A_{\rm {\cal N}=2}^{\rm hyp}$ denote the contributions 
from one ${\cal N}=2$ vector multiplet and one hypermultiplet running in the loop, respectively. The model-dependent coefficients $c_{\rm vec}$ and $c_{\rm hyp}$ depend on the gauge groups and the numbers of hypermultiplets.
Then, the on-shell helicity content $(+1,+\frac{1}{2},0,-\frac{1}{2},-1)$ of the supersymmetry multiplets\footnote{A hypermultiplet has two complex scalars. Following the references we are using in this section,
we write $2({\mathcal N}=2)_{{\rm hyp}}$, where the factor of 2 means that what we call ``hypermultiplet'' here is actually a ``half-hypermultiplet'' with two real scalars. 
See also the conclusions for comments on supersymmetry decomposition with other multiplets.}
\begin{align}
({\mathcal N}=2)_{{\rm vec}} &\leftrightarrow  (1,2,2,2,1) \ , \ \ \ \ \ \ ({\mathcal N}=2)_{{\rm hyp}} \leftrightarrow  (0,1,2,1,0)
\notag \\
({\mathcal N}=4) &\leftrightarrow (1,4,6,4,1)=
    ({\mathcal N}=2)_{{\rm vec}}  \oplus 2({\mathcal N}=2)_{{\rm hyp}} 
\end{align}
translates into the following amplitude contributions from these multiplets in the loop:
\be \label{decomp}
A_{{\mathcal N}=2}^{\rm vec} = 
A_{{\mathcal N}=4}^{\te{1-loop}} - 
2 A_{{\mathcal N}=2}^{\rm hyp} \,,
\ee 
which is meaningful in both $D=6$ and $D=4$. Hence,
one can write the generic one-loop amplitude (\ref{original}) as
\bea  \label{original2}
\tilde{A}_{{\mathcal N}=2}^{\te{1-loop}}=c_{\rm vec}A_{\rm {\cal N}=2}^{\rm vec}+c_{\rm hyp}A_{\rm {\cal N}=2}^{\rm hyp}
=c_{\rm vec}A_{\mathcal N=4}^{\te{1-loop}}+(c_{\rm hyp}-2c_{\rm vec})A_{\rm {\cal N}=2}^{\rm hyp}\,,
\eea
and normalizing
the hypermultiplet contribution to unity, we have
\bea  \label{hyp1}
{A}_{{\mathcal N}=2}^{\te{1-loop}}=
(c_{\rm hyp}-2c_{\rm vec})^{-1} \tilde{A}_{{\mathcal N}=2}^{\te{1-loop}}={c_{\rm vec} \over c_{\rm hyp}-2c_{\rm vec}}A_{\mathcal N=4}^{\te{1-loop}}+A_{{\cal N}=2}^{\rm \rm hyp} \; .
\eea
It is now evident that if we can match $A_{{\cal N}=2}^{\rm hyp}$ to \cite{Johansson:2014zca}
with unit normalization, the detailed 
model-dependence 
of the original \eqref{original} (for example, on the field content) is confined to the coefficient
of the scalar-box contribution $A_{\mathcal N=4}^{\te{1-loop}}$. We need to compare the generic form (\ref{hyp1}) with
the 4-point amplitude \eqref{BG27simp} shifted by any multiple $c$ of $A^{\te{1-loop}}_{{\cal N}=4}(1,2,3,4)$, 
\begin{align}
A^{\te{1-loop}}_{{\cal N}=2}(1,2,3,4;c) &\equiv 2 \int \frac{ \dd^D\ell }{(2\pi)^D} \, \bigg\{ \frac{ C_{1|234} }{2 \ell^2 \ell_{1}^2} + \frac{ \ell_m C^m_{1|23,4} }{\ell^2 \ell_{1}^2 \ell_{123}^2} + \frac{ \ell_m C^m_{1|2,34} }{\ell^2 \ell_{1}^2 \ell_{12}^2}  - \frac{ P_{1|4|2,3} }{\ell_{1}^2 \ell_{12}^2 \ell_{123}^2}
&\label{A4-4dim} \\
&\! \! \! \! \! \! \! \! \! \! \! \! \! \! \! \! \! \! \! \!  \! \! \! \!  + \frac{ \ell_m \ell_n C^{mn}_{1,2,3,4} + \ell_m (s_{23} C^m_{1|23,4}  +s_{24} C^m_{1|24,3}  +s_{34} C^m_{1|2,34}   ) -(c  + \frac{1}{3})t_8(1,2,3,4)}{\ell^2 \ell_{1}^2 \ell_{12}^2 \ell_{123}^2}  \bigg\}   \ ,
\notag
\end{align}
where we used the explicit form of the maximally supersymmetric contribution
\be\label{Amax}
 A^{\te{1-loop}}_{{\cal N}=4}(1,2,3,4)= -2
 \int \frac{ \dd^D\ell }{(2\pi)^D}  \frac{ t_8(1,2,3,4) }{\ell^2 \ell_{1}^2 \ell_{12}^2 \ell_{123}^2}
 = - \int \frac{ \dd^D\ell }{(2\pi)^D}  \frac{ s_{12}s_{23} A^{\rm tree}(1,2,3,4) }{\ell^2 \ell_{1}^2 \ell_{12}^2 \ell_{123}^2}\,,
\ee
and the representation of the scalar box as $s_{12}s_{23}C_{1|234}=2t_8(1,2,3,4)$. As we will see in section \ref{sect6}, the field-theory limit of string amplitudes in 4- and 6-dimensional orbifold compactifications identifies the hypermultiplet contribution with  (\ref{A4-4dim}) at $c=-\frac{1}{3}$ where $t_8(1,2,3,4)$ drops out,
\beq
A_{{\cal N}=2}^{\rm \rm hyp}(1,2,3,4)= -\frac{1}{4}
A^{\te{1-loop}}_{{\cal N}=2}\big(1,2,3,4;c=-\tfrac{1}{3}\big) \ .
\label{hyphyp}
\eeq
In the next section, we will check that this recreates the $D=4$ expressions of \cite{Johansson:2014zca}.


\subsection{Matching 4-dimensional spinor-helicity expressions}
\label{sec:spinhel}

Now we use the 4-dimensional MHV components
of section \ref{sect5_1} to  simplify the hypermultiplet contribution to the 4-point amplitude, and match our result to \cite{Johansson:2014zca}. In $D=4-2\epsilon$ dimensions\footnote{The expressions in this section are valid for 4-dimensional external  gluons. For other external states, we would use $D$-dimensional expressions to resolve subtleties with dimensional regularization.}, performing the integrals in dimensional regularization, from eq.\ \eqref{hyphyp}  and  \eqref{A4-4dim} we have 
\begin{align}
A_{{\cal N}=2}^{\rm \rm hyp}&(1,2,3,4) \Big|_{D=4-2\epsilon}= 
-\frac{i}{4(4\pi)^2}  \Bigg\{  \frac{I_2(s_{23})}{s_{23}} {\cal C}^m_{1|23,4}  \big(k_{23}^m +2k_{4}^m \big)  +\frac{I_2(s_{12})}{s_{12}} {\cal C}^m_{1|2,34}  \big({k_{1}^m\over \epsilon} +k_{34}^m \big)
-2{\cal P}_{14|23} I_3(s_{14}) 
 \notag\\
& \hspace{-1cm}
-2 (s_{23} {\cal C}^m_{1|23,4}  {+}s_{24} {\cal C}^m_{1|24,3}  {+}s_{34} {\cal C}^m_{1|2,34}   ) \sum_{i=2}^4 I_4[ {\scriptstyle{\sum}_{j=2}^ia_j}] k_i^m
+2  {\cal C}^{mn}_{1|2,3,4} 
\bigg( \sum_{i,j=2}^4 I_4[ {\scriptstyle {\sum}_{k=2}^i a_k {\sum}_{l=2}^j a_l  } ] k_i^m k_j^n - \frac{I_4^{\scriptscriptstyle D=6}\,\eta^{mn}}{2} \bigg)  \Bigg\}\,,
\label{calc}
\end{align}
where the explicit expressions for the integrals $I_2(s_{ij})$, 
$I_3(s_{ij})$,
$I_4^{\scriptscriptstyle D=6}$, $I_4[ a_i]$ and $I_4[ a_i a_j]$ (with Feynman parameters $a_j$) are given in appendix \ref{sec:basis}.
Then, 
in the helicity configuration $(1^-,2^-,3^+,4^+)$, we find the nonvanishing kinematic factor to be proportional to $\la 12\ra^2 [34]^2 \sim t_8(1^-,2^-,3^+,4^+)$,
\begin{align}
A_{{\cal N}=2}^{\rm \rm hyp}(1^-,2^-,3^+,4^+) =&\, 
 \frac{i}{(4\pi)^{2}} \frac{\la 12\ra^2 [34]^2}{4}
 \bigg\{
 \frac{ \,I_2(s_{23})}{2s_{12}s_{23}}\, -I_4[ a_2] + I_4[ {\scriptstyle{\sum}_{j=2}^3a_j}] + I_4[ {\scriptstyle{\sum}_{j=2}^4a_j}] \\
& 
-\sum_{i=2}^4  I_4[ {\scriptstyle {\sum}_{k=2}^i a_k {\sum}_{l=2}^i a_l}] 
-\frac{2}{s_{12}}\sum_{i=3}^4 s_{1i} I_4[ {\scriptstyle {\sum}_{k=2}^{(i-1)} a_k {\sum}_{l=2}^{i} a_l}]-
\frac{I_4^{\scriptscriptstyle D=6}}{s_{12}}
\bigg\}\,.
\notag
\end{align}
Up to and including order $\epsilon^0$, we arrive at the simple result
\be\label{A4mmpp_hyp}
A_{{\cal N}=2}^{\rm \rm hyp}( 1^-,2^-,3^+,4^+ )=-
\frac{i}{(4\pi)^{2}}\frac{\la 12\ra^2 [34]^2}{2}
  { I_2(s_{23})\over 4s_{12}s_{23}}\,.
\ee 
Analogously, in the $(1^-,2^+,3^-,4^+)$ helicity configuration we can factor out $\la 13\ra^2 [24]^2$, and find
\be
A_{\rm {\cal N}=2}^{\rm hyp}(1^-,2^+,3^-,4^+)=
\frac{i}{(4\pi)^{2}}
\frac{\la 13\ra^2 [24]^2}{2}\left(
{I_2(s_{12}) \over 4s_{12}s_{13}}+{I_2(s_{23}) \over 4s_{23}s_{13}} -\frac{I_4^{D=6}}{2s_{13}} \right)\,.
\label{A4mpmp_hyp}
\ee
The amplitudes  (\ref{A4mmpp_hyp}) and (\ref{A4mpmp_hyp}) perfectly match known results  \cite{Bern:1994cg,Bern:1995db,Johansson:2014zca}.\footnote{Note that these references normalize
their Mandelstam invariants differently, we have $s_{ij}=k_i\cdot k_j$.}

As is well-known, the maximally supersymmetric 4-point amplitude (\ref{Amax}) is proportional to the corresponding
tree amplitude, by a supersymmetry Ward identity. 
Here in half-maximal supersymmetry, we have the additional 8-supercharge
tensor structure ${\cal P}_{12|34}$ in (\ref{polP}).
This leads to the situation that
 even though the two helicity components (\ref{A4mmpp_hyp}) and (\ref{A4mpmp_hyp}) are separately proportional to the corresponding tree $-2s_{12}s_{23}A^{\rm tree}(1^-,2^+,3^-,4^+) = \la 13\ra^2 [24]^2$ and $-2s_{12}s_{23}A^{\rm tree}(1^-,2^-,3^+,4^+) = \la 12\ra^2 [34]^2$, we cannot relate $A_{\rm {\cal N}=2}^{\rm hyp}(1,2,3,4)$ to $A^{\rm tree}(1,2,3,4)$ with a helicity-independent prefactor. 
 
Finally, a comment about anomalies. We have dimensionally
reduced from $D=6$ to $D=4$ on a torus, maintaining half-maximal supersymmetry,
leaving no parity-odd sector at all.
If we would instead consider
 minimal supersymmetry in $D=4$, one  expects to see the triangle anomaly in the 3-point function.  For a nice general discussion, including ${\cal N}=2$ subsectors of ${\cal N}=1$ gauge theories in $D=4$, and the triangle anomaly from a minahaned 3-point function, see \cite{Anastasopoulos:2006cz}.

\section{1-loop SYM amplitudes from orbifolds of the superstring}
\label{sect6}

In this section we reproduce the above 1-loop SYM amplitudes from the field-theory limit of 
open-string amplitudes in  orbifold compactifications of the 10-dimensional RNS string. 
Denoting $d$-dimensional Minkowski spacetimes and tori by $M_d$ and $T^d$, respectively, we consider toroidal orbifold compactifications of the form $M_6\times (T^2\times T^2)/\mathbb{Z}_N$ and  
 $M_4\times T^2\times (T^2\times T^2)/\mathbb{Z}_N$, yielding effective theories with half-maximal supersymmetry in 6 and 4 dimensions, respectively
 (see e.g.\ the textbooks
 \cite{Blumenhagen:2013fgp,Ibanez:2012zz}  or the review \cite{Angelantonj:2002ct}). 
It is convenient to use complexified coordinates on the two twisted 2-tori, i.e.\ $z_j=x^{2j+4}+U_j x^{2j+5}$, with $j=1,2$ and $U_j$ denoting the complex structure
of the $j^{\te{th}}$ 2-torus. The generator $\Theta$ of the cyclic group $\mathbb{Z}_N$ acts on the tori (and on the string spectrum) as a discrete rotation, 
\be 
\Theta^k z_j=e^{2\pi i k v_j}z_j\,, \hspace{2cm} k\in\{0,1,\dots N-1\}, \hspace{1cm} j\in\{1,2\}\,.
\label{orbtw}
\ee
The rational numbers $v_j$ are chosen such that $\Theta^N =1$ and collected in the {\it twist vector} $\vec{v}_k=k(v_1,v_2)$ with $k=0,1,\ldots,N{-}1$. For the half-maximal models we consider we have\footnote{We consider only $\mathbb{Z}^A_N$ models, in the language of  \cite{Gimon:1996ay}, 
except that we will not implement worldsheet parity constraints, see below and appendix \ref{appE}.} $\vec{v}_k=(k/N,-k/N)$.
The rank $N$ of the cyclic group determines the gauge groups and the number of hypermultiplets of the $D$-dimensional effective theory.

Strictly speaking, toroidal orbifold compactifications lead to fully consistent open-string models only when the string spectrum is quotiented out by worldsheet parity,
to create an orientifold (see e.g.\ \cite{Blumenhagen:2013fgp,Ibanez:2012zz}), where one
also computes unoriented string diagrams like the M\"obius strip. 
Although doing so is straightforward, it is possible to extract partial amplitudes for a large class of gauge theories by performing the field-theory limit of just the oriented-string worldsheet topologies. (We should warn the reader that such truncated theories
may contain spurious divergences that are cancelled in the complete  theory, but we have not seen any indication
that this should affect the discussion here.)

We will now reproduce the 1-loop color-ordered single-trace amplitudes computed in the previous sections from the field-theory limit of the planar cylinder diagram. 

\subsection{1-loop open-string amplitudes with half-maximal supersymmetry}
\label{sect61}

Extending previous work in \cite{Bianchi:2006nf, Tourkine:2012vx, Ochirov:2013xba, Bianchi:2015vsa}, 1-loop open-string amplitudes with external gluons in the context of orbifold compactifications were studied and simplified in \cite{Berg:2016wux}. In particular, toroidal orbifolds preserving half-maximal supersymmetry give rise to the following decomposition of the planar cylinder diagram for $n$-point 1-loop amplitudes:
\be  
\label{AN2}
{\cal A}_{1/2}^D(1,2,\ldots,n) = \int \dd \mu_{12\ldots n}^{D} \, \bigg\{ \Gamma_{\cal C}^{(10-D)} c_0\, {\cal I}_{n,\te{max}}\, + \, 
\Gamma_{\cal C}^{(6-D)} \sum_{k=1}^{N-1} c_k \, \hat \chi_k  \, {\cal I}_{n,1/2}(\vec{v}_k) 
\bigg\} \ .
\ee 
In this expression, $D$ is the number of non-compact spacetime dimensions (here $D=6$ or $D=4$),  the $\Gamma_{\cal C}^{(\ldots)}$ denote lattice sums over momenta in the remaining compact  dimensions, and $c_0$ and $c_k$ are constants determined by the action of the orbifold group on the Chan--Paton (gauge-group) factors. The constants
\beq
\hat \chi_k\equiv   - \left( \frac{ \sin(\pi k v)}{\pi} \right)^2
\eeq
depend solely on the orbifold sector labelled by $k=1,2,\ldots,N{-}1$, and the integrands $ {\cal I}_{n,\te{max}}$ and ${\cal I}_{n,1/2}(\vec{v}_k)$ are determined by the external states and the partition function. The subscripts ``max'' or ``1/2'' distinguish orbifold sectors that preserve all or half the supersymmetries, respectively. While the maximally supersymmetric integrand $ {\cal I}_{n,\te{max}}$ is parity-even and independent of the dimension $D$, the half-maximal integrand ${\cal I}_{n,1/2}(\vec{v}_k)$ 
has both parity-even and parity-odd parts. The parity-odd part vanishes in $D<6$. Finally, the integration measure in eq.\ \eqref{AN2} is given by
\be
\int \dd \mu_{12\ldots n}^D \equiv {(\alpha')^n V_D \over 8N} \int^{\infty}_0 \frac{ \dd \tau_2 }{(8 \pi^2 \alpha' \tau_2)^{D/2}} \! \! \! \! \!  \int \limits_{0 \leq \Im(z_1) \leq \Im(z_2) \leq \ldots \leq \Im(z_n) \leq \tau_2} \! \! \! \! \! \! \! \! \! \! \! \! \! \! \! \! \! \!  \dd z_1 \, \dd z_2 \, \ldots \, \dd z_n \, \delta(z_1) \,\Pi_n \ ,
\label{openmeas}
\ee
where $\tau_2$ is the modular parameter of the cylinder worldsheet whose non-empty boundary is parametrized by purely imaginary coordinates $z_i$ with $0 \leq \Im (z_i) \leq \tau_2$. 
In the measure \eqref{openmeas} we have also incorporated the regularized external volume $V_D$, the order $N$ of the orbifold ${\mathbb Z}_N$, as well as the ubiquitous Koba-Nielsen  factor $\Pi_n$ which arises from the plane-wave factors of the vertex operators; with the conventions of reference \cite{Berg:2016wux} we have
\be\label{G_string}
\Pi_n  \equiv \prod_{1\leq i<j}^n e^{s_{ij} G_{ij}}\,, \hspace{1.5cm}  G_{ij} \equiv G(z_{ij},\tau) = \frac{\alpha'}{2}\Big\{\log \left| \frac{ \theta_1(z_{ij},\tau)}{\theta_1'(0,\tau)} \right|^2 - \frac{2 \pi}{\Im(\tau)} \Im^2(z_{ij})\Big\} \,,
\ee
with $z_{ij}\equiv z_i - z_j$, the Jacobi theta function $\vartheta_1$,
and $G(z,\tau)$ denotes the bosonic Green's function on a genus-one worldsheet with modular parameter $\tau$.

\subsubsection{The worldsheet integrands}

The maximally supersymmetric integrands in the 3- and 4-point amplitudes (\ref{AN2}) are well-known \cite{Green:1982sw},
\beq
{\cal I}_{3,\te{max}} = 0 \ , \ \ \ \ \ \ 
{\cal I}_{4,\te{max}} = - 2 t_8(1,2,3,4) \, \label{maxint} \ ,
\eeq
see (\ref{t8}) for the $t_8$-tensor. The half-maximal integrands ${\cal I}_{n,1/2}(\vec{v}_k)$ at $n=3,4$ were expressed in terms of a basis of worldsheet functions\footnote{A similar reduction to a basis of integrals was performed in \cite{Bianchi:2015vsa}. The 4-dimensional spinor-helicity setup of this reference departs from the infrared regularization of this work (section \ref{sect32}), leading to different final expressions for the simplified integrands. In \cite{Bianchi:2015vsa},
the field-theory limit  was performed for the integrands of particular orientifold string models, and it would be interesting
to compare those results in greater detail to the results in this section.} in \cite{Berg:2016wux},
\begin{align}
{\cal I}_{3,1/2}(\vec{v}_k)& = s_{23} f^{(1)}_{23} C_{1|23}\,,
\label{new3pt}
\\
{\cal I}_{4,1/2}(\vec{v}_k)&= -2  F_{1/2}^{(2)}(kv)  t_8(1,2,3,4)
+ \big[ s_{12}(  f^{(2)}_{12}+ f^{(2)}_{34}) P_{1|2|3,4} +(2\leftrightarrow 3,4) \big]
\notag \\
& \ \ \ \ \ \ \  \ \ \   + \big[ s_{23}  f^{(1)}_{23}(s_{24}  f^{(1)}_{24} + s_{34} f^{(1)}_{34})
 C_{1|234} + (3\leftrightarrow 4) \big]   \,.
\label{new4pt}
\end{align}
The underlying correlation functions of RNS vertex operators have been organized in terms of Berends--Giele currents\footnote{See appendix \ref{appA} for an intuitive motivation.} introduced in section \ref{sect2}, and total derivatives with respect to the worldsheet variables have been discarded to render all the kinematic factors manifestly gauge-invariant.
The worldsheet coordinates enter through the modular functions $f^{(n)}$ of weight $n$ \cite{Broedel:2014vla},
\begin{align}
 f^{(1)}_{ij}\equiv f^{(1)}(z_{ij},\tau) &=   \partial \ln \theta_1(z_{ij},\tau) +  2\pi i \, \frac{ \Im (z_{ij}) }{\Im (\tau)}\,,
\label{red12}
\\
f^{(2)}_{ij}\equiv f^{(2)}(z_{ij},\tau) &=  \frac{1}{2} \Big\{  \Big( \partial \ln \theta_1(z_{ij},\tau) +  2\pi i \, \frac{ \Im (z_{ij}) }{\Im (\tau)} \Big)^2 + \partial^2 \ln \theta_1(z_{ij},\tau) - \frac{\theta_1'''(0,\tau)}{ 3\theta_1'(0,\tau)} \Big\} \,,
\label{red13}
\end{align}
which are evaluated at purely imaginary coordinates $z_j$ and purely imaginary
modular parameters of the cylinder, $\tau\equiv i\tau_2$ with $\tau_2$ real. Note that $f^{(1)}(z,\tau)$ has a simple pole at the origin $z\rightarrow 0$, and the combination
\be
F_{1/2}^{(2)}(kv, \tau) \equiv 2 f^{(2)}(kv,\tau) -f^{(1)}(kv,\tau)^2
\ee
is the Weierstrass $\wp$-function, that
encodes the entire dependence of (\ref{new4pt}) on the orbifold twists (\ref{orbtw}).
 (See also \cite{Bianchi:2006nf} for the role of $\wp$ in string-theory setups with half- and quarter maximal supersymmetry.)

\subsection{The field-theory limit of the open-string amplitudes}
\label{sect62}

It is well-known how to perform the field-theory limit of 1-loop string amplitudes  to obtain the corresponding Feynman diagrams in  Schwinger parametrization, see e.g.\ 
\cite{Green:1982sw, Bern:1991aq, Schubert:2001he, BjerrumBohr:2008vc} and references therein. 
The limit is $\alpha'\rightarrow 0$ while simultaneously degenerating the genus-one surface to a worldline by sending $\tau_2\rightarrow \infty$. The limits are taken such that $\alpha'\tau_2$ is kept finite and reduces to the  worldline length $\alpha' \tau_2 \rightarrow t$ in the Schwinger parametrization of the corresponding field-theory diagrams. The finite parts $\nu$ of the cylinder coordinates $ {\rm Im}(z)= \tau_2 \nu $ are then identified with proper times on the worldline. In this limit, the bosonic Green's function in eq.\ \eqref{G_string} reduces to
\be
G_{ij} \longrightarrow  -\pi t (\nu^2_{ij} - |\nu_{ij}| )\, .
\ee 
Then, the limit of the Koba-Nielsen factor is
\be 
\Pi_n \longrightarrow e^{-\pi t Q_n[k_1, k_2, \dots, k_n]},\hspace{1cm}  
Q_n[k_{A_1}, k_{A_2}, \dots, k_{A_n}] \equiv \sum_{i < j}^n(k_{A_i}\cdot k_{A_j})(\nu^2_{ij} - |\nu_{ij}| ) \ ,
\label{defQn}
\ee 
and the  measure in eq.\ \eqref{AN2} reduces to\!
\footnote{On the right-hand side of eq.\ \eqref{mu_ftl} we have suppressed the overall constant factor ${V_D \over {8N (8 \pi^2  )^{D/2}}} $ from eq.\ \eqref{openmeas}. In this section, 
we will suppress all such overall prefactors that do not depend on the kinematic variables.}
\be
\int \dd \mu_{12\ldots n}^D \longrightarrow  
\int^{\infty}_0 \frac{ \dd t }{t} \,t^{-D/2+n} \,\,\,\,i^{n-1} \!\!\!\!\!\!\!\!\!\!\!\!\!  \int \limits_{0 \leq \nu_2 \leq \nu_3 \leq \ldots \leq \nu_n \leq 1} \! \! \! \! \! \! \! \! \! \! \! \! \! \! \!   \dd \nu_2 \, \ldots \, \dd \nu_n \,\, e^{-\pi t Q_n[k_1,k_2,\dots k_n]} \, \Big|_{\nu_1=0}\, ,
\label{mu_ftl}
\ee 
see \cite{Tourkine:2013rda} for an extension of these techniques to higher loops. Note that the definition of $Q_n$ in (\ref{defQn}) admits massive external legs with composite momenta  $k_{A}=k_{a_1} +k_{a_2} + \ldots + k_{a_m}$ for $A=a_1a_2\ldots a_m$.

In the above limit, the worldsheet functions (\ref{red12}) and (\ref{red13}) involved in the half-maximal integrands of the  3- and 4-point functions reduce to 
\beq
f^{(1)}_{ij}\longrightarrow   2\pi i \left[\nu_{ij} -\frac{1}{2} {\rm sgn}(\nu_{ij})\right]\,,\ \ \ \ \ \ 
f^{(2)}_{ij}\longrightarrow  (2\pi i)^2 \left[\frac{1}{2} \nu_{ij}^2-\frac{1}{2} \nu_{ij}{\rm sgn}(\nu_{ij})  +\frac{1}{12}\right]\,,
\label{degfs}
\eeq
while the function $F^{(2)}_{1/2}( k v)$ of the orbifold twists in the 4-point integrand degenerates as follows:
\be\label{ft_F2}
F^{(2)}_{1/2}( k v)\longrightarrow  \hat{F}^{(2)}_{1/2}( k v)\equiv {\pi^2}\left[\frac{1}{3}-\frac{1}{\sin^2 (\pi k v)}\right]\,.
\ee
We would like to highlight the following two additional aspects in performing the field-theory limit of string amplitudes:
\begin{itemize}
\item The quantized momenta in the internal dimensions need to decouple, this is obtained by shrinking the size of the internal dimensions at the same rate as the string-length goes to zero \cite{Green:1982sw, BjerrumBohr:2008vc} so that
 lattice
sums become unity\footnote{For toroidal compactifications, given the K{\"a}hler moduli  $T_2^i$, the internal momenta decouple when $T_2^i\rightarrow 0$, such that  $T_2^i/{{\alpha'}}$ stays finite. In this limit, with the conventions adopted in \cite{Berg:2016wux}, we have
\[
\Gamma_{\cal C}^{(2m)} = 
\prod_{i=1}^m \bigg\{\frac{T^i_2}{\alpha'\tau_2} \sum_{n_1,n_2\in \mathbb{Z}} {\rm exp}\bigg({\frac{-\pi T^i_2}{\alpha'\tau_2}\frac{|n_1+n_2 U^i|^2}{U_2^i}}\bigg) \bigg\}
\rightarrow \,\,1 \; . 
\]
Often in the literature on maximal supersymmetry, e.g.\ in \cite{BjerrumBohr:2008vc}, 
the $\Gamma_{\cal C}^{(2m)}$ differ from ours by a factor of $(\alpha'\tau_2)^m$. This is because in these papers  $\Gamma_{\cal C}^{(2m)}$ are prefactors in non-compact amplitudes, where the integration measure has an overall factor $1/(\alpha'\tau_2)^5$, unlike our $1/(\alpha'\tau_2)^{D/2}$.  }
\be 
\Gamma_{\cal C}^{(\ldots )} 
\rightarrow \,\,1 \; .   
\ee
\item The worldsheet functions $f^{(1)}_{ij}$ have simple poles when vertex operators collide, $ f^{(1)}_{ij} \sim 1/ z_{ij}$ as $z_i \rightarrow z_j$. These regions of the integration domain require  separate treatment and yield kinematic poles through interplay of the singular functions $f^{(1)}_{ij}$ with the Koba-Nielsen factor (\ref{G_string}):
\be 
f^{(1)}_{ij}\,\Pi_n \sim  \frac{1}{z_{ij}^{1-\alpha's_{ij}}}\,\bigg( \prod_{\underset{\{kl\}\neq \{ij\}}{1\leq k<l}}^n e^{s_{kl} G_{kl}}\Big|_{z_i=z_j} \bigg) \,.
\ee
For example, the regime of $z_2\rightarrow z_3$ in the 3-point integrand (\ref{new3pt}) gives rise to
\be
\int_0^1 \dd z_3 \int_0^{z_3} \! \dd z_2 \int_0^{z_2} \! \dd z_1\,\delta(z_1)\, f_{23}^{(1)} \,\Pi_3 = -\frac{1}{\alpha's_{23}} \int_0^1 \dd z_3 \,\,e^{(s_{12}+ s_{13})G_{13}}\Big|_{z_1=0} + {\cal O}(\ap^0)\,.
\label{pol23}
\ee 
The ubiquitous combinations $s_{ij}f^{(1)}_{ij}$ which accompany the gauge-invariant kinematic factors in the half-maximal correlators of section \ref{sect61} cancel these kinematic poles. For example, the factor of $s_{23}^{-1}$ on the right-hand side of (\ref{pol23}) (due to the integration region where $z_2\rightarrow z_3$) cancels in
\be 
\lim_{z_2\rightarrow z_3}\int \dd \mu_{12\ldots n}^D \,s_{23} f^{(1)}_{23} \longrightarrow  
i^n \int^{\infty}_0 \frac{ \dd t }{t} \,t^{-D/2+n-1} \,\, \! \!\! \! \! \! \! \! \! \! \! \! \!  \! \!
\int \limits_{0  \leq \nu_2 \leq \ldots \leq \nu_{n-1} \leq 1} \! \! \! \! \! \! \! \! \! \! \!  \! \!
 \dd \nu_2 \, \ldots \, \dd \nu_{n-1} \, e^{-\pi t Q_{n-1}[k_{1},k_{23},k_4,\dots, k_n]}\, \Big|_{\nu_1=0} \, .
\label{}
\ee 
\end{itemize}
To make contact with the momentum-space construction in section \ref{sect3}, it remains to translate the resulting Schwinger parametrization
of field-theory amplitudes into Feynman integrals. The following map will be sufficient for the 3- and 4-point amplitudes to
trade the integration over worldlines with length $t$ and proper times $\nu_j$ for the integration over the loop momentum $\ell$:
\begin{align}\label{S_to_F}
\int &\frac{\dd^D \ell \ (\alpha + \beta_m \ell^m + \gamma_{mp} \ell^m \ell^p) }{\ell^2 (\ell-k_1)^2 (\ell-k_{12})^2 \ldots (\ell-k_{12\ldots n-1})^2} = \pi^n \int_0^\infty \frac{ \dd t }{t} \,t^{n-D/2}  \\
&\times \! \! \!   \! \! \!  \int \limits_{0 \leq \nu_2\leq \nu_3 \leq \ldots \leq \nu_n \leq 1} \! \! \!   \! \! \!   \dd \nu_2 \, \dd \nu_3\, \ldots \, \dd \nu_n \, \Big( \alpha + \beta_m L^m + \gamma_{mp} \Big[ L^m L^p + \frac{ \eta^{mp} }{2\pi t} \Big] \Big) e^{-\pi t Q_n[k_1,k_2,\ldots,k_n]} \, \Big|_{\nu_1=0} \ ,
\notag
\end{align}
where $\alpha, \beta_m$ and $\gamma_{mp}$ are arbitrary scalars, vectors and tensors, respectively, and
\be\label{Lm}
L^m\equiv -\sum_{i=1}^n k_i^m \nu_i\,.
\ee
In the following, we will apply these techniques to compute the field-theory limit of the above 3- and 4-point functions, taking the manifestly gauge-invariant integrands in \eqref{new3pt} and \eqref{new4pt} as a starting point.

\subsubsection{The 3-point amplitude}

At 3 points, the maximally supersymmetric integrand vanishes,  
and the half-maximal open-string amplitude reads
 \be
{\cal A}_{1/2}^D(1,2,3) =
\int \dd \mu_{123}^{D} \,\Gamma_{\cal C}^{(6-D)} \sum_{k=1}^{N-1} c_k \,\hat \chi_k  \, {\cal I}_{3,1/2}(\vec{v}_k)\,. 
 \ee
Starting from the half-maximal integrand in \eqref{new3pt} and formally equating the vanishing expressions $k_2^m C^m_{1|2,3}=-s_{23} C_{1|23}$, we have (up to irrelevant global prefactors)
\begin{align}\label{ft_3pt_a}
\hspace{-0.5cm}{\cal A}_{1/2}^D(1,2,3)\longrightarrow&
\int_{0}^{\infty} \frac{\dd t}{t}\,\,\,\, t^{2-D/2} \int_0^1 \! \dd\nu_2 \,
\Big\{C_{1|23} e^{-\pi t Q_2[k_{1},k_{23}]}\Big\}\\
&+\pi\,\int_{0}^{\infty} \frac{\dd t}{t}\,\, t^{3-D/2} \!\!\!\! \int \limits_{0 \leq \nu_2 \leq \nu_3 \leq 1}\!\!\!\!\!\!\!\! 
\dd\nu_2 \dd\nu_3 \Big\{s_{23}C_{1|23} +2L_mC^m_{1|2,3}\Big\}e^{-\pi t Q_3[k_1,k_2,k_3]}\,.\notag
\end{align}
By virtue of \eqref{S_to_F}, this is easily seen to reproduce \eqref{BG30a}.

\subsubsection{The 4-point amplitude}

The 4-point string amplitude reads
\be  
\label{all4pt}
{\cal A}_{1/2}^D(1,2,3,4) = \int \dd \mu_{1234}^{D} \, \left\{ \Gamma_{\cal C}^{(10-D)} c_0\, {\cal I}_{4,\te{max}}\, + \, 
\Gamma_{\cal C}^{(6-D)} \sum_{k=1}^{N-1} c_k \, \hat \chi_k  \, {\cal I}_{4,1/2}(\vec{v}_k) 
\right\} \ ,
\ee 
with the gauge-invariant form of the integrand ${\cal I}_{4,1/2}(\vec{v}_k) $ given in eq.\ \eqref{new4pt}. It is convenient to organize the field-theory limit according to the configurations of colliding vertex operators: 
\begin{itemize}
\item Simultaneous contact of 3 vertex operators occurs for products of two $f^{(1)}_{ij}$,
\begin{align}\label{ft_4pt_1}
{\cal A}_{1/2}^D(1,2,3,4)\bigg|_{z_2 \leftrightarrow z_3\leftrightarrow z_4}\!\!\!\!\!\!\longrightarrow
-\sum_{k=1}^{N-1} c_k \, \hat \chi_k \int_{0}^{\infty} \frac{\dd t}{t}\,\,\,\, t^{2-D/2} \int_0^1  \! \dd\nu_2 \,
C_{1|234}e^{-\pi t Q_2[k_1,k_{234}]}\,,
\end{align}
reproducing the bubble integral in \eqref{BG27simp}. We take the compactification dependent coefficients $\sum_{k=1}^{N-1} c_k \, \hat \chi_k$ into account to  combine the resulting contributions with the maximally supersymmetric sector of (\ref{all4pt}).
\item From worldsheet regions where only one pair of $z_i$ and $z_{i+1}$ collides, we instead have
\begin{align}\label{ft_4pt_2}
{\cal A}_{1/2}^D(1,2,3,4)\bigg|_{\hspace{-0.2cm}\begin{array}{>{\scriptstyle}l}z_2 \leftrightarrow z_3\vspace{-0.35cm}\\z_3 \leftrightarrow z_4\end{array}}
\!\!\!\!\!\!&\longrightarrow
-2\pi \sum_{k=1}^{N-1} c_k \, \hat \chi_k \,\int_{0}^{\infty} \frac{\dd t}{t}\,\, t^{3-D/2} \!\!\!\! \int \limits_{0 \leq \nu_2 \leq \nu_3 \leq 1}\!\!\!\!\!\! 
\dd\nu_2 \dd\nu_3 \Big[\nu_{23}+\frac{1}{2}\Big]  \\
&\! \! \! \! \! \! \! \! \! \! \! \! \! \! \!\! \! \! \! \! \times \Big[\big(s_{34}C_{1|234}-s_{24}C_{1|324}\big)e^{-\pi tQ_3[k_1,k_{23}, k_4]}
+\big(s_{23}C_{1|234}-s_{24}C_{1|243}\big)e^{-\pi tQ_3[k_1,k_2, k_{34}]} \Big]\notag\,.
\end{align}
The vanishing kinematic combinations in the second line translate into the spurious scalar triangles in eq.\ \eqref{BG27}, and the vector triangles in \eqref{BG27simp} will be traced back to the irreducible part of the worldsheet integrand,
that we consider next. 
\item In the absence of collisions among vertex operators, the worldsheet functions $f^{(n)}_{ij}$ may be replaced by their worldline degenerations (\ref{degfs}), giving the following irreducible contributions to the field-theory limit: 
\begin{align}
{\cal A}_{1/2}^D&(1,2,3,4)\bigg|_{\rm irred}\!\!\!\!\!\!\longrightarrow
\int_{0}^{\infty} \frac{\dd t}{t}\,\, t^{4-D/2} \!\!\!\!\!\!\! \int \limits_{0 \leq \nu_2 \leq  \nu_3 \leq \nu_4 \leq 1}\!\!\!\!\!\!\!\!\!\!\! 
\dd\nu_2 \dd\nu_3 \dd\nu_4 \, \bigg\{ 
\!-2t_8(1,2,3,4) \bigg[ c_0 + \sum_{k=1}^{N-1} c_k \hat{\chi}_k \hat{F}_{1/2}^{(2)}(kv) \bigg]  \notag \\
&
-\pi^2\sum_{k=1}^{N-1} c_k \hat{\chi}_k\bigg[
C^{\rm scal}_{1|2|3|4}+2 L_m \big(s_{23}C^m_{1|23,4} + s_{24}C^m_{1|24,3} + s_{34}C^m_{1|34,2}\big)
+2\big(L_mL_n+\frac{\eta_{mn}}{2\pi t} \big)C^{mn}_{1|2,3,4}
\notag\\
&
\hspace{2.75cm}-4\, \Big( P_{1|4|2,3}\nu_{14} \big( L\cdot k_4 + \frac{1}{2\pi t \nu_{14}}\big)+
           P_{1|3|2,4} {\nu_{13}} \big( L\cdot k_3 +s_{34} + \frac{1}{2\pi t \nu_{13}}\big) 
\label{ft_4pt_3}\\
&\hspace{3.75cm}
        +  P_{1|2|3,4}\nu_{12} \big( L\cdot k_2 +s_{23}+ s_{24} + \frac{1}{2\pi t \nu_{12}}\big)\Big)\bigg]\bigg\}
 e^{-\pi t Q_4[k_1,k_2,k_3,k_4]}       \notag
\end{align}
The coefficient $C^{\rm scal}_{1|2|3|4}=-\frac{2}{3}t_8(1,2,3,4)$ is obtained by simplifying a collection of scalar blocks 
using appendix \ref{sectB} and the BCJ relations among permutations of $C_{1|234}$. 
Using \eqref{S_to_F}, we see that the second line in \eqref{ft_4pt_3} reproduces the box numerator in \eqref{BG27}. Recalling that $\hat{F}^{(2)}_{1/2}( k v) \sim
 \frac{1}{3}-\frac{1}{\sin^2 (\pi k v)}$, we see that the scalar coefficient $C^{\rm scal}_{1|2|3|4}$ cancels due to the $1/3$ in $\hat{F}^{(2)}_{1/2}( k v)$. 
This cancellation is essentially the reason why the field-theory amplitude (\ref{BG27simp}) constructed from locality and gauge invariance is not a pure hypermultiplet contribution $A_{{\cal N}=2}^{\rm hyp}$. 
\item The integrands in last two lines of \eqref{ft_4pt_3} can be identified as total derivatives in $\nu_j$ and can be easily evaluated as boundary terms\! 
\footnote{To identify the total derivatives note that
\be\nonumber
\nu_{1i}\big( L\cdot k_i + \sum_{j=i+1}^{4}s_{ij} + \frac{1}{2\pi t \nu_{1i}}  \big)e^{-\pi t Q_4[k_1,k_2,k_3,k_4]}=
-\frac{1}{2\pi t}\partial_{\nu_i}\big(\nu_{1i}e^{-\pi t Q_4 [k_1,k_2,k_3,k_4]}\big) \ , \ \ \ \ \ \ 
i=2,3,4\, .
\ee},
\begin{align}
\!\!\!\!\!\!\! \int \limits_{0 \leq \nu_2 \leq  \nu_3 \leq \nu_4 \leq 1}\!\!\!\!\!\!\!\!\!\!\! \dd\nu_2 \dd\nu_3 \dd\nu_4 &
\Big\{ P_{1|4|2,3}\nu_{14} \big( L\cdot k_4 + \frac{1}{2\pi t \nu_{14}}\big)+
         P_{1|3|2,4}\nu_{13} \big( L\cdot k_3 +s_{34} + \frac{1}{2\pi t \nu_{13}}\big) \notag \\
& +   P_{1|2|3,4}\nu_{12} \big( L\cdot k_2 +s_{23}+ s_{24} + \frac{1}{2\pi t \nu_{12}}\big)\Big\} e^{-\pi t Q_4[k_1,k_2,k_3,k_4]}   \label{triags}\\
=- \frac{1}{2\pi  t}\!\!\!\!\int \limits_{0 \leq \nu_2 \leq \nu_3 \leq 1}\!\!\!\!\!\!\!\! \dd\nu_2 \dd\nu_3&
\Big\{  - P_{1|4|2,3} e^{-\pi t Q_3[k_{41},k_{2},k_{3}]}+ 
\nu_{13} ( P_{1|3|2,4} - P_{1|4|2,3})e^{-\pi t Q_3[k_1,k_{2},k_{34}]} \notag\\
&+\nu_{12} ( P_{1|2|3,4} - P_{1|3|2,4})e^{-\pi t Q_3[k_1,k_{23},k_4]} \Big\}\,, \notag
\end{align}
where the relations in eq. \!\eqref{B3}  yield
\begin{align}\label{P2_to_C}
\nu_{13}(P_{1|3|2,4} - P_{1|4|2,3})&=L^m C^m_{1|34,2} \, \big|_{\nu_3=\nu_4}  \ , \ \ \ \ \ \
\nu_{12}(P_{1|2|3,4} - P_{1|3|2,4})&=L^m C^m_{1|23,4} \, \big|_{\nu_2=\nu_3}\,.
\end{align}
Therefore, by use of \eqref{triags} and \eqref{P2_to_C} as well as the massive-leg generalization of \eqref{S_to_F}
\be\label{S_to_F_g}
\int \frac{\dd^D \ell \  \beta_m \ell^m  }{\ell^2 (\ell-k_A)^2 (\ell-k_A-k_B)^2 } = 
\pi^3 \int_0^\infty \frac{ \dd t }{t} \,t^{3-D/2} 
\! \! \! \! \! \! \! \! \!\int \limits_{0 \leq \nu_2\leq \nu_3  \leq 1}  \! \! \!\! \! \! \! \! \! \dd \nu_2 \, \dd \nu_3\,
 \beta_m ( \nu_{12} k_B^m + \nu_{13} k_C^m  ) 
e^{-\pi t Q_3[k_{A},k_{B}, k_{C}]} \, \Big|_{\nu_1=0} \ ,
\ee
we see  that the last two lines of \eqref{ft_4pt_3} reproduce the triangles in eq.\ \eqref{BG27simp}. In eq.\ \eqref{S_to_F_g}, the combination $\nu_{12} k_B^m + \nu_{13} k_C^m$ generalizes $L^m$ in (\ref{Lm}) to incorporate massive legs.
\end{itemize}

\subsubsection{Streamlining the model dependence in the 4-point amplitude}

Collecting the above results, and using the expression (\ref{ft_F2}) for $\hat{F}^{(2)}_{1/2}( k v)$ as well as 
$\hat\chi_k=-(\sin(\pi k v)/\pi)^2$,
we can finally write the field-theory limit of the 4-point amplitude as
\be\label{ft_4pt_f}
{\cal A}_{1/2}^D(1,2,3,4)\longrightarrow \Big\{c_o +  \sum_{k=1}^{N-1}  \,c_k\,  \Big\} A^{\te{1-loop}}_{{\cal N}=4 } -4 
\sum_{k=1}^{N-1}  \,c_k\, \sin(\pi kv)^2  A_{{\cal N}=2}^{ \rm hyp} 
\ee 
with $A^{\te{1-loop}}_{{\cal N}=4 }$ and $A_{{\cal N}=2}^{ \rm hyp }$ given by (\ref{Amax}) and (\ref{hyphyp}), respectively.
In eq.\ \eqref{ft_4pt_f}, the unambiguous identification of the hypermultiplet contribution (and of its normalization relative to 
the  ${\cal N}=4$ multiplet) is possible because the coefficients arising from Chan--Paton factors, $c_0$ and $c_k$, are related to the numbers of multiplets, $c_{\rm vec}, \,c_{\rm hyp}$, as
\be \label{ck_vs_cvh}
\frac{c_0 +  \sum_{k=1}^{N-1}c_k}{-4\sum_{k=1}^{N-1}  \,c_k\, \sin(\pi kv)^2 } = \frac{c_{\rm vec}}{c_{\rm hyp} - 2 c_{\rm vec}}\,,
\ee
see \eqref{hyp1} and \eqref{ft_4pt_f}. In appendix \ref{appE} we explicitly check this relation in simple 4-dimensional models arising from orbifold compactifications of oriented open strings.


\section{Conclusions}

In this work, we have presented a method to construct 1-loop amplitudes of 6-dimensional gauge theories with half-maximal supersymmetry from first principles: starting from a string-theory inspired ansatz of kinematic building blocks, imposing locality and gauge invariance led us to unique answers for 3 and 4 external gauge bosons, respectively, see (\ref{BG30b}) and (\ref{BG27simp}) for the manifestly gauge-invariant results. We have checked that,
when dimensionally reducing the integrands to $D=4$, these expressions
 integrate to results known from the field-theory literature, and also
emerge from the field-theory limit of the corresponding open-string amplitudes \cite{Berg:2016wux} (also see \cite{Bianchi:2006nf, Tourkine:2012vx, Ochirov:2013xba, Bianchi:2015vsa} for earlier work). Most of the building blocks are introduced in generality for any number of legs, so it appears feasible to construct higher-multiplicity amplitudes along the same lines.

A similar strategy has been applied to tree and loop amplitudes of 10-dimensional SYM \cite{Mafra:2010jq, Mafra:2014gja, Mafra:2015mja} where supersymmetry is imposed along with gauge invariance and the kinematic ansatz is inspired by the pure-spinor superstring \cite{Berkovits:2000fe}. Accordingly, the 1-loop amplitudes of this work inherit their structure from their maximally supersymmetric counterparts in pure-spinor superspace \cite{Mafra:2014gja} with two additional legs. It would be desirable to reexpress and supersymmetrize the present results in a comparable superspace setup and to derive their superstring ancestors from the hybrid formalism \cite{Berkovits:1994wr, Berkovits:1999im}.

The structural similarities between loop amplitudes with $n$ legs and  16 supercharges and $n{+}2$ legs and 8 supercharges naturally lead to questions about further supersymmetry breaking to 4 supercharges. For ${\cal N}=1$ supersymmetric amplitudes in 4 dimensions, the parity-even part is a straightforward dimensional reduction of the parity-even ${\cal N}=2$ contributions in our results. This is due to the enhanced supersymmetry when summing over the fundamental and antifundamental chiral multiplets in the loop \cite{Johansson:2014zca}. However, it remains an open challenge to construct parity-odd parts of SYM amplitudes with quarter-maximal supersymmetry and pure Yang--Mills amplitudes from the principles we used here. It would  be equally interesting to extend this to minimal supergravity; as reviewed in  \cite{Berg:2016wux}, only minimal supersymmetry allows for 1-loop corrections to the Einstein-Hilbert term in the low-energy effective action. There can also be interesting couplings of open- and closed-string sectors, for example one could investigate loop amplitudes of gauge bosons coupled to 6-dimensional tensor multiplets,
that we did not include in our decomposition \eqref{decomp} in the context of that section.

While the 3-point amplitude of this work obeys the BCJ duality between color and kinematics and thereby yields the half-maximal supergravity amplitude (\ref{sug3c}) as a byproduct, we encounter an obstacle at the 4-point level to satisfy kinematic Jacobi identities. This ties in with the findings of \cite{Mafra:2014gja} on maximally supersymmetric 5- and 6-point amplitudes. It is a particularly burning question how to reconcile the present approach to $D$-dimensional 1-loop $n$-point amplitudes with the BCJ duality and the double-copy construction\footnote{We are grateful to Yu-tin Huang and Henrik Johansson for informing us about their unpublished BCJ form of the 4-point 1-loop amplitude in $D=6$ half-maximal SYM, including parity-odd terms \cite{WIP1}. We recall that 4-dimensional BCJ representations are available in \cite{Carrasco:2012ca, Johansson:2014zca}.}. As a rewarding first step forward, it would be helpful to directly compute the 6-dimensional 1-loop supergravity amplitudes beyond the current reach of the double copy, based on the field-theory limit of the closed-string amplitudes in \cite{maxsusy, Berg:2016wux} and comparison with the 4-dimensional BCJ analysis of \cite{Ochirov:2013xba}.


\section*{Acknowledgements}

We would like to thank John Joseph Carrasco, Alexander Ochirov, Rodolfo Russo and Bo Sundborg
for inspiring discussions on topics related to those in this paper.
We are grateful to Massimo Bianchi, Dario Consoli, Michael Haack, Henrik Johansson and Piotr Tourkine for enlightening discussions and valuable comments on a draft of the manuscript. 
We are particularly indebted to Erik Panzer for invaluable help with Feynman integral calculations. 
OS is grateful to Carlos Mafra for collaboration on related topics and to Karlstad University for financial support and kind hospitality during early stages of this project. MB and OS are indebted to Nordita and in particular Paolo Di Vecchia and Henrik Johansson for providing stimulating atmosphere, support and hospitality through the ``Aspects of Amplitudes'' program.
 
\newpage{}

\appendix

\section{RNS approach to Berends--Giele currents}
\label{appA}

In units $2\alpha'=1$, the color-stripped vertex operator for the gluon in the RNS formalism is given by
\beq
V(e,k,z)= (e^m \partial X_m(z) + \frac{1}{2} \, f^{mn}  \psi_m(z) \psi_n(z)) e^{k\cdot x(z)} \ .
\label{rns1}
\eeq
We have chosen its integrated representative of conformal weight $h=1$ in the superghost picture zero. The embedding coordinates $X^m(z)$ of the superstring and their worldsheet superpartners $\psi^m(z)$ depend on the worldsheet coordinate $z$, and their short-distance behaviour is captured by the operator product expansion (OPE)
\beq
\partial X^m(z) X^n(0) \sim \frac{ \eta^{mn}}{z} + {\cal O}(z^0) \co
\psi^m(z) \psi^n(0) \sim \frac{\eta^{mn}}{z} + {\cal O}(z^0)
\label{rns2}
\eeq
with no mutual singularities between $X^m$ and $\psi^n$. When assembling all contractions of the form (\ref{rns2}), the OPE of two vertex operators (\ref{rns1}) yields
\begin{align}
V(e_1,k_1,z_1) &V(e_2,k_2,z_2) \sim z_{12}^{s_{12}} \Big\{ \frac{1-s_{12}}{z_{12}^2}  (e_1\cdot e_2) + \frac{\partial X_m(z_2)}{z_{12}} \big[ e_2^m (e_1 \cdot k_2) - e_1^m( e_2 \cdot k_1) \big] \label{rns3}\\
&+ \frac{ \psi_m \psi_n(z_2)}{2 z_{12}} \big[ (e_1 \cdot k_{2}) k_{12}^m e_2^n - (e_2 \cdot k_1) k_{12}^m e_1^n - (e_1 \cdot e_2) k_1^m k_2^n - (k_1 \cdot k_2) e_1^m e_2^n - (m\leftrightarrow n)\big]
\Big\} \ ,
\notag
\end{align}
where the overall power of $z_{12}^{s_{12}}$ stems from the plane waves $e^{k_1\cdot x(z_1)} e^{k_2\cdot x(z_2)} \sim e^{k_{12} \cdot x(z_2)} z_{12}^{s_{12}} (1 + {\cal O}(z_{12}))$. After absorbing the double pole in the first line of (\ref{rns3}) into total derivatives such as
\beq
\frac{\partial }{\partial z_2} (z_{12}^{s_{12}-1} e^{k_2 \cdot x(z_2)}) = z_{12}^{s_{12}-1} e^{k_2 \cdot x(z_2)} \Big\{ k_2 \cdot \partial X(z_2) + \frac{1-s_{12}}{z_{12}} \Big\} \ ,
\label{rns4}
\eeq
one can identify the 2-particle polarizations $e_{12}^m$ and $f_{12}^{mn}$ in (\ref{BG31}) and (\ref{BG32}) along with $\partial X_m$ and $\psi_m \psi_n$,
\begin{align}
V(e_1,k_1,z_1) V(e_2,k_2,z_2) &\sim z_{12}^{s_{12}-1} \Big(  e^m_{12} \partial X_m(z_2) + \frac{1}{2}  \, f^{mn}_{12}\psi_m \psi_n(z_2)  \Big) e^{k_{12}\cdot x(z_2)}\label{rns5}\\
&\ \ \ \  + \frac{1}{2} (e_1 \cdot e_2) \left( \frac{\partial}{\partial z_2} - \frac{\partial}{\partial z_1} \right) z_{12}^{s_{12}-1} e^{k_1 \cdot x(z_1)} e^{k_2 \cdot x(z_2)}
 \ ,
\notag
\end{align}
in formal analogy to (\ref{rns1}).
A supersymmetric identification of Berends--Giele currents in the OPE of integrated vertex operators has been performed in the pure-spinor formalism \cite{Mafra:2014oia}, also see \cite{Mafra:2010ir, Mafra:2010gj} for their emergence from the OPE between an integrated vertex and an unintegrated one.

Since string amplitudes at any genus are obtained from integrating correlation functions of vertex operators over $z_j$, the contributions from the total derivatives in (\ref{rns5}) ultimately drop out. The overall correlator is largely\footnote{Starting from genus one, multiparticle correlators involve additional worldsheet functions without any singularities as $z_i \rightarrow z_j$, see e.g.\ the function $f^{(2)}(z)$ defined by (\ref{red13}) in the 6- and 4-point amplitudes of \cite{maxsusy} and \cite{Berg:2016wux}. The analysis of the OPE in this section does not constrain the kinematics along with such non-singular functions.} determined by summing OPEs among the vertex operators, so the appearance of the 2-particle currents $\efrak_{12}^m$ and $\ffrak_{12}^{mn}$ in (\ref{rns5}) identifies Berends--Giele currents as a suitable language to describe the kinematic dependence of string amplitudes. In the same way as multiparticle correlators give rise to nested OPEs with the same 2-point contractions among $X^m$ and $\psi^m$ at each step, the iteration of the Berends--Giele recursion (\ref{BG01}) and (\ref{BG02}) yields currents $\efrak_{P}^m$ and $\ffrak_{P}^{mn}$ of arbitrary multiplicity.


\section{Spinor-helicity conventions}
\label{sectY}

In this appendix we summarize the spinor-helicity conventions used in this paper. Given solutions to the Weyl equations 
\begin{align} \label{B1}
\sigma^m k_m v_{-}(k)&=0\hspace{1cm} \bar{\sigma}^m k_m v_{+}(k)=0\,\\
\bar{u}_{-}(k)\sigma^m k_m&=0 \hspace{1cm} \bar{u}_{+}(k) \bar{\sigma}^m k_m=0 \ ,\notag
\end{align}
where $\sigma^m\equiv(\mathds{1},\sigma^i)$ and $\bar{\sigma}^m\equiv(\mathds{1},-\sigma^i)$, with Pauli matrices $\sigma^i$ such that 
\begin{align}
\sigma^m\bar{\sigma}^n+\sigma^n\bar{\sigma}^m &=-2\eta^{mn}  \mathds{1}  \ , \ \ \ \ \ \ 
\bar{\sigma}^m{\sigma}^n+\bar{\sigma}^n \sigma^m =-2\eta^{mn}  \mathds{1} \ ,\notag
\end{align}
we use the following conventions and notation
\begin{align} \label{B2}
|i]\equiv v_+(k_i)=u_{-}(k_i) &\hspace{1cm} [i|\equiv \bar{u}_+(k_i)=\bar{v}_{-}(k_i)\\
|i\rangle\equiv v_{-}(k_i)=u_+(k_i) &\hspace{1cm} \langle i|\equiv \bar{u}_-(k_i)=\bar{v}_{+}(k_i)\ .\notag
\end{align}
The computations in section \ref{sect5_1} are performed by choosing the following expressions for the polarization vectors $e^m_{\pm}$ of gluons with helicity $\pm 1$,
\be  \label{B3}
e^m_{i+}\equiv-\frac{ [i|\sigma^m |q_i \rangle}{\sqrt{2}\langle q_i i \rangle}\hspace{1cm}
e^m_{i-}\equiv-\frac{[ q_i|\sigma^m|i\rangle}{\sqrt{2}[ q_i i ]}\,,
\ee
where $q_i$ are arbitrary massless reference momenta that cancel in gauge-invariant expressions.

With the above choices, by use of completeness 
\be \label{B4}
k^m_i \sigma_m =- |i]\langle i|\,,  \,\hspace{1cm}k^m_i \bar{\sigma}_m =- | i\rangle[i |\,,
\ee
Fierz identities
\be
[i|\sigma^m|j\rangle\langle k|\bar{\sigma}_m|l] =2[il]\langle jk\rangle\,,
\ee
as well as
\be
k^m_i =\frac{1}{2}[ i |\sigma^m| i\rangle\,,
\ee
we have the following useful formulae 
\begin{flalign} \label{B5}
\ek{i^+}{j}&=\frac{\la q_i j\ra [j i]}{\sqrt{2}\la q_i i\ra} \ , \hspace{1cm}  \ \  \ \ek{i^-}{j}=\frac{[ q_i j] \la j i\ra}{\sqrt{2}[ q_i i]}\\
\notag\\
\eeij{i^+}{i^+}&=\frac{\la q_i q_j\ra [i j]}{\la q_i i\ra \la q_j j \ra}  \ , \hspace{1cm}   
\eeij{i^-}{j^-}=\frac{\la i j\ra [q_i q_j]}{[q_i i][ q_j j] } \ , \hspace{1cm}
\eeij{i^+}{j^-}=\frac{\la q_i j\ra [i q_j]}{\la q_i i\ra [q_j j]}\,, \label{B6}
\end{flalign}
these are used to derive the spinor-helicity expressions in section \ref{sect5_1}.


\section{Relations for the kinematic building blocks}
\label{sectB}

In this appendix we summarize a series of on-shell identities relating the scalar blocks to the vector and tensor blocks introduced in section \ref{sect2}. 
These identities are used in the main text, in particular to make contact between the Schwinger parametrization of the field-theory amplitudes, which naturally comes in terms of just scalar blocks, and the corresponding Feynman-integral representations. While contractions with external momenta give rise to 
\begin{align}
k_1^m C^m_{1|23,4} &= P_{1|3|2,4} - P_{1|2|3,4} \notag \\
k_2^m C^m_{1|23,4} &= P_{1|2|3,4} - P_{1|4|2,3} + s_{24} C_{1|324} \notag  \\
k_4^m C^m_{1|23,4} &= s_{34} C_{1|234}  - s_{24} C_{1|324}  =0  \label{B3} \\
k_1^m C^{mn}_{1|2,3,4} &= -\big[ k_2^n  P_{1|2|3,4} + (2\leftrightarrow 3,4) \big]  \notag\\
k_2^m C^{mn}_{1|2,3,4} &= k_2^n  P_{1|2|3,4} - s_{23} C_{1|23,4}^n - s_{24} C_{1|24,3}^n \,,\notag
\end{align}
the trace of the tensor pseudo-invariant can be expanded as
\be
\eta_{mn} C^{mn}_{1|2,3,4}  = 2 ( P_{1|2|3,4}+P_{1|3|2,4}+P_{1|4|2,3}) \ .
\label{BG4}
\ee
In addition, the scalar pseudo-invariants $P_{1|2|3,4}$ satisfy
\be\label{B5}
  s_{12}{P}_{1|2|3,4} + s_{13}{P}_{1|3|2,4} +s_{14}{P}_{1|4|2,3}=s_{23} s_{34} C_{1|234}\,,
\ee
that is, this combination is gauge-invariant because the parity-odd part cancels.

The following corollaries of (\ref{B3}) and (\ref{B5}) are useful to simplify the expressions after solving the integrals in section \ref{sec:spinhel}:
\begin{align}
k_{23}^m C^m_{1|23,4}&= {P}_{1|2|3,4}  - {P}_{1|3|2,4} \notag\\
k_2^m (s_{23} C^m_{1|23,4}+s_{24} C^m_{1|24,3}+s_{34} C^m_{1|34,2}) &=  s_{23} s_{24} C_{1|324} \notag \\
k_3^m (s_{23} C^m_{1|23,4}+s_{24} C^m_{1|24,3}+s_{34} C^m_{1|34,2}) &= {  s_{14} {P}_{1|4|2,3} - s_{12}{P}_{1|2|3,4} + 
( s_{12}-s_{14}) {P}_{1|3|2,4} }  \\
k_2^m k_2^m C^{mn}_{1|2,3,4} &= - s_{23} s_{24} C_{1|324} \notag\\
k_2^m k_3^m C^{mn}_{1|2,3,4} &=   - s_{13} {P}_{1|2|3,4} - s_{12} {P}_{1|3|2,4}\,.\notag
\end{align}

\linespread{1.1}

\section{Feynman integral ``basis''}
\label{sec:basis}

We use the following ``basis'' of Feynman integrals 
for the computations in $D=4-2\epsilon$ dimensions in section \ref{sec:spinhel}
(though the result will ultimately be expressed only in terms of $I_2(s_{ij})$ and $I_4^{D=6}$):
\bea
I_2(s_{ij}) &=& {r_{\Gamma} \over \epsilon(1-2\epsilon)}(-2s_{ij})^{-\epsilon} \\
I_3(s_{ij}) &=& {r_{\Gamma} \over \epsilon^2  }(-2s_{ij})^{-1-\epsilon} \\
I_4^{D=6}&=&{r_\Gamma \over 4s_{13}}\left(\ln\left({-s_{12}\over -s_{23}}\right)^2+\pi^2\right) \\
L &=& r_\Gamma \ln \left({-s_{12} \over -s_{23}}\right)
\eea
where
\be
r_\Gamma \equiv \frac{\Gamma(1+\epsilon)\Gamma^2(1-\epsilon)}{\Gamma(1-2\epsilon)}\,.
\ee
We express our generic integrals  in terms of these basis integrals as follows. First, integrals $I_4[a_i]$
with a single Feynman parameter in the numerator decompose as
\bea
I_4[a_2]&=&{1-2\epsilon \over \epsilon} {I_{2}(s_{12})\over 4s_{12}s_{23}} +{1 \over 2s_{23}}I_4^{D=6}\\
I_4[a_3]&=&{1-2\epsilon \over \epsilon} {I_{2}(s_{23})\over 4s_{12}s_{23}}+{1 \over 2s_{12}}I_4^{D=6} \ .
\eea 
Due to the symmetry $1\leftrightarrow 3$ and $2\leftrightarrow 4$,
we only give two out of the four $I_4[a_i]$ integrals.
As for the integrals $I_4[a_ia_j]$
with two Feynman parameters in the numerator, using the symmetry $I_4[a_ia_j]=I_4[a_ja_i]$ and again $1\leftrightarrow 3$ and $2\leftrightarrow 4$, 
we only give five of the $I_4[a_ia_j]$,
\bea
I_4[a_2^2] &=& {1 \over 4\epsilon s_{12}s_{23} } {I_{2}(s_{12}) } -{s_{12} \over 2s_{13}s_{23}}I_4^{D=6}-{1 \over 4s_{13}s_{23}}L \\ 
I_4[a_2 a_3] &=& -{1 \over 4\epsilon s_{12}s_{23}} {I_{2}(s_{12})} -{1 \over 2s_{13}}I_4^{D=6}+{1 \over  4s_{13}s_{23}}L \\
I_4[a_2 a_4] &=& -{1 \over 2s_{13}}I_4^{D=6}-{1 \over 4s_{13}s_{23}}L \\
I_4[a_3^2] &=& {1 \over 4\epsilon s_{12}s_{23}} {I_{2}(s_{23})} - {s_{23} \over 2s_{12}s_{13}}I_4^{D=6}+{1 \over 4s_{12}s_{13}}L \\
I_4[a_3 a_4] &=& -{1 \over  4s_{12}s_{23}}I_2(s_{12})-{1 \over 2s_{23}}I_4^{D=6}-{1 \over 4s_{12}s_{13}}L  \ .
\eea
This is sufficient to perform the check in the main text. As we see there,
the ``bare'' logarithms $L$ all cancel. In  intermediate steps,
they are required by the symmetries of the Feynman parameter integrals. 

\section{Explicit 4-dimensional models from oriented-string orbifolds}
\label{appE}

In this appendix we show that eq.\ \eqref{ck_vs_cvh}, which relates Chan--Paton traces in string theory to the gauge
group and supermultiplet content of the effective theory, is satisfied in explicit string models.
As emphasized in the main text, the oriented-string orbifold models we consider here
are not full-fledged string models, but if taken by themselves suffer from inconsistencies,
especially when coupled to gravity. It would be straightforward to generalize them
to consistent string models, e.g.\ as orientifolds, 
but they suffice as they are for our purposes here: to compare specific coefficients in specific amplitudes.

The spectrum of effective gauge theories with half-maximal supersymmetry arising from $\mathbb{Z}_N$ orbifold compactifications of the RNS string can be obtained by projecting the spectrum of the maximally supersymmetric open string as follows\!
\footnote{As in the main text, we consider the class of  $\mathbb{Z}^A_N$ models in the language of \cite{Gimon:1996ay} but without imposing the worldsheet parity projection and considering only D9-branes.}
\be\label{E1}
P_\Theta \big( |w\rangle\otimes \lambda \big)    \equiv
 \frac{1}{N}\sum_{k=0}^{N-1}\Theta^k\big( |w\rangle\otimes \lambda \big) \equiv
 \frac{1}{N}\sum_{k=0}^{N-1} e^{2\pi i k v( J_1 - J_2)}|w\rangle \otimes \gamma^k {\lambda} (\gamma^k)^{-1}\,.
\ee 
We denote color-stripped string states by $|w\rangle$ and Chan--Paton matrices by $\lambda$.
The rotation operators $ J_1 ,J_2$ act on the twisted tori $T^2_1$ and $T^2_2$, and the matrices
$\gamma^k$ for $k=0,1,\ldots,N{-}1$ represent the orbifold group $\mathbb{Z}_N$
on Chan--Paton factors. If we wanted to build fully consistent string models, e.g.\ free of closed-string tadpoles and gauge anomalies, certain constraints would arise on the number of Chan--Paton factors and the allowed representations spanned by the $\gamma^k$ matrices (see e.g.\ \cite{Gimon:1996rq, Gimon:1996ay}).

\subsection{Simplest models without hypermultiplets}

For our purposes, the simplest effective models that can be built by orbifold projection of the maximally supersymmetric open string are the ones where the $\gamma^k$ are in the trivial representation of $\mathbb{Z}_N$. That is, given open strings with $M \times M$ Chan--Paton factors, we obtain simple gauge-theory models by setting $\gamma^k\equiv{\bold 1}_{M\times M}$ for all $k\in \{0,1,\dots N{-}1\}$. In this case, all the massless excitations in untwisted directions ($m=0,1,\ldots,5$) 
are kept in the effective theory, while all the massless excitations in internal (compact) twisted directions ($i=6,\ldots ,9$) 
are projected out. For instance, from eq.\ \eqref{E1}  for the massless NS states, we have
\be\label{E2}
P_\Theta \big( \psi^m_{-\frac{1}{2}}|0\rangle\otimes \lambda \big) =  \psi^m_{-\frac{1}{2}}|0\rangle\otimes \lambda\,,
\hspace{1cm}
P_\Theta \big( \psi^i_{-\frac{1}{2}}|0\rangle\otimes \lambda \big)=0\,.
\ee
More generally, the spectrum of the $D$-dimensional effective theory of these simple string models consists of a half-maximal 
vector multiplet with $U(M)$ gauge symmetry, and no  hypermultiplets. Referring  to eq.\ \eqref{hyp1}, in these models we then have $c_{\rm vec}=M$ and $c_{\rm hyp}=0$. We now wish to verify eq. \eqref{ck_vs_cvh}, i.e.\ we want to show that in this class of models (where $v= 1/N$),
\be\label{E3}
\frac{c_0 +  \sum_{k=1}^{N-1}c_k}{-4\sum_{k=1}^{N-1}  \,c_k\, \sin(\pi kv)^2 } = \frac{c_{\rm vec}}{c_{\rm hyp} - 2 c_{\rm vec}}\,=-\frac{1}{2} \ .
\ee
The coefficients $c_0$ and $c_k$ are determined by traces of the $\gamma^k$ matrices acting on the Chan--Paton factors. In models with just one $U(M)$ gauge group, we have\!
\footnote{In models with just one gauge group, the identification $c_k=({\rm tr}\gamma^k)^2$ arises straight from the partition function, see e.g.\ \cite{Gimon:1996ay} and appendix A of \cite{Berg:2016wux}.}
\be\label{E4}
c_k=({\rm tr}\gamma^k)^2 = M^2\,, \hspace{1cm}  k\, \in \{0,1,\dots N{-}1\}\,,
\ee 
i.e.\ $c_k$ clearly does not depend on $k$ when $\gamma^k$ are in the trivial representation, and we obtain
\be\label{E5}
\frac{c_0 +  \sum_{k=1}^{N-1}c_k}{-4\sum_{k=1}^{N-1}  \,c_k\, \sin(\pi kv)^2 } = \frac{M^2N}{ - 2 M^2N}=-\frac{1}{2}
\ee
as desired, where we used 
the elementary finite sum $\sum_{k=1}^{N-1}\sin(\pi k v)^2=\sum_{k=1}^{N-1}\sin(\pi k /N)^2=N/2$. 

\subsection{Simplest models with hypermultiplets}

To obtain models with hypermultiplets, the matrices $\gamma^k$ should form a non-trivial representation of $\mathbb{Z}_N$. Without loss of generality, for the $\mathbb{Z}_N$ models under consideration, we can choose a Chan--Paton basis where  
\be\label{E6}
\gamma^1\equiv {\rm diag} ( {\bold 1}_{m_0\times m_0}, \alpha {\bold 1}_{m_1\times m_1}, \dots, \alpha^{N-1} {\bold 1}_{m_{N-1}\times m_{N-1}})\,,\hspace{1cm} \alpha\equiv e^{2\pi i /N}\,,
\ee 
with $m_0+m_1+\dots +m_{N-1}=M$, the total Chan--Paton degeneracy at each string endpoint.
In the $\mathbb{Z}_2$ case,  eq.\ \eqref{E6} specializes to 
\be\label{E7}
\gamma^1\equiv {\rm diag} ( {\bold 1}_{m_0\times m_0}, - {\bold 1}_{m_1\times m_1})\,,
\ee 
and the massless NS states that survive the $\mathbb{Z}_2$ projection have the following block structure
\begin{align}\label{E8}
P_\Theta \big( \psi^m_{-\frac{1}{2}}|0\rangle\otimes \lambda \big) =& \, \psi^m_{-\frac{1}{2}}|0\rangle\otimes 
{\rm diag}\big(\lambda_{m_0\times m_0}, \lambda_{m_1\times m_1} \big)
\\
P_\Theta \big( \psi^i_{-\frac{1}{2}} |0\rangle\otimes \lambda \big)=&\,\psi^i_{-\frac{1}{2}} |0\rangle\otimes
\left(
\begin{array}{cc}
\bold{0} & \lambda_{m_0\times m_1} \\
\lambda_{m_1\times m_0} &\bold{0}
\end{array}
\right),\notag
\end{align}
where $i$ denotes any of the four directions of the twisted tori, 
and the matrices $\lambda_{m_i\times m_j}$ denote sub-blocks of $\lambda \in\, U(M)$ of dimension $m_i\times m_j$.
More generally, the effective spectrum in these $\mathbb{Z}_2$ models consists of two vector multiplets with $U(m_0)$ and  $U(m_1)$ gauge symmetry, and two (half-)hypermultiplets transforming in the bifundamental $[(\square_{m_0}, \overline{\square}_{m_1})+( \overline{\square}_{m_0}, \square_{m_1})]$. Therefore, for external gluons belonging to e.g.\ $U(m_0)$, we have the field-theory parameters
\be\label{E9}
\frac{c_{\rm vec}}{c_{\rm hyp}- 2c_{\rm vec}}=\frac{m_0}{4m_1-2m_0}.
\ee
From the string-theory perspective, for external gluons in the $U(m_0)$ 
we have 
\be\label{E10}
c_k=
\!\!\!\sum_{\underset{(i=0)\vee (j=0)}{i,j=0}}^{1}\!\!\!
{\rm tr}\gamma^k_{[m_i]} {\rm tr}\gamma^{-k}_{[m_j]}= m_0^2 + (-1)^k 2 m_0 m_1\,, \hspace{1cm}  k=0,1\,,
\ee
where $\gamma^k_{[m_i]}$ denotes the $m_i\times m_i$ sub-block of $\gamma^k$; therefore,
\be\label{E11}
\frac{c_0 +  c_1}{-4 \,c_1\, \sin(\pi/2)^2 } =\frac{m_0}{4m_1-2m_0} 
\ee
is equal to \eqref{E9} and confirms eq.\ \eqref{ck_vs_cvh} that we wanted to check. 

The $\mathbb{Z}_N$ models for $N=3,4,6$ can be treated together. In these cases we find vector multiplets with gauge symmetry $U(m_0)\times U(m_1)\times \dots \times U(m_{N-1})$ and one (half-)hypermultiplet for each of the following representations:
$[(\square_{m_0}, \overline{\square}_{m_{N-1}})+( \overline{\square}_{m_0}, \square_{m_{N-1}})]$, 
$[(\square_{m_\alpha}, \overline{\square}_{m_{\alpha+1}})+( \overline{\square}_{m_\alpha}, \square_{m_{\alpha+1}})]$ for $\alpha\in\{0,1,\dots, N{-}1\}$.
Therefore, in $\mathbb{Z}_3$, $\mathbb{Z}_4$ and $\mathbb{Z}_6$ orbifolds, for external gluons belonging to e.g.\ the $U(m_0)$, we have field-theory parameters
\be\label{E12}
\frac{c_{\rm vec}}{c_{\rm hyp}- 2c_{\rm vec}}=\frac{m_0}{2(m_1+m_{N-1})-2m_0} \ .
\ee 
This ties in with the string-theory quantities
\be\label{E13}
c_k=\!\!\!\sum_{\underset{(i=0)\vee (j=0)}{i,j=0}}^{N-1}\!\!\!
{\rm tr}\gamma^k_{[m_i]} {\rm tr}\gamma^{-k}_{[m_j]}=m_0(m_0+(\alpha^{k}+\alpha^{-k}) m_1+ \dots +(\alpha^{(N-1)k}+\alpha^{-(N-1)k}) m_{N-1})\,,
\ee
and we find
\be\label{E14}
\frac{c_0 +  \sum_{k=1}^{N-1}c_k}{-4\sum_{k=1}^{N-1}  \,c_k\, \sin(\pi kv)^2 } =\frac{m_0}{2(m_1+m_{N-1}-m_0)}\,,
\ee
where we used $\sum_{k=0}^{N-1} \alpha^{k}=0$ to see that $\sum_{k=0}^{N-1}c_k=Nm_0^2$. Also note that
$\sum_{k=1}^{N-1} \sin(\pi kv)^2 (\alpha^{kn}+ \alpha^{-kn})$
takes the value $N$ for $n=0$, the  value $-N/2$ for  $n\in\{1, N{-}1\}$ and vanishes
for $1<n<N{-}1$.
The equality of \eqref{E12} and \eqref{E14} completes
our check of eq.\ \eqref{ck_vs_cvh} for these classes of models.

\newpage{\pagestyle{empty}\cleardoublepage}
\end{document}